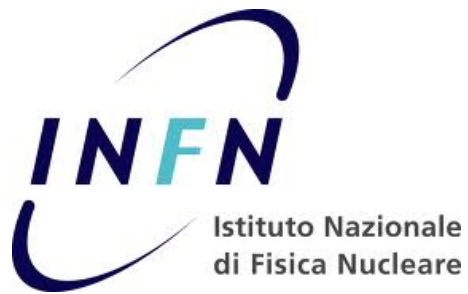

Istituto Nazionale
di Fisica Nucleare

# *Tau/Charm Factory Accelerator Report*

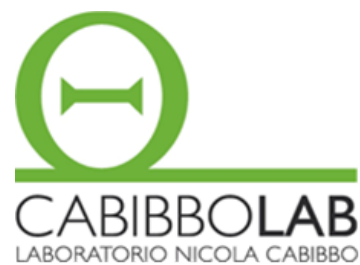

July 2013

**Foreword**

The present Report concerns the current status of the Tau/Charm accelerator project and in particular discusses the issues related to the lattice design, to the accelerators systems and to the associated conventional facilities.

The project aims at realizing a variable energy flavor factory between 1 and 4.6 GeV in the center of mass, and succeds to the SuperB project from which it heritates most of the solutions proposed in this document. The work comes from a cooperation involving the LNF accelerator experts, the young newcomers, mostly engineers, of the Cabibbo Lab consortium and key collaborators from external laboratories.

The result of this effort is impressive, given the little time elapsed since SuperB cancellation, and is due to the enthusiasm of its contributors as well as to the deep and reusable work done for the parent project SuperB, showing the knowledge accumulated in accelerator physics at LNF. In the last section a possible time scale for the construction, as well as the financial load and the personnel requests, are preliminary outlined.

Detector design and specific Physics channels to be studied by such an accelerator will be addressed in a separate document, ready by the end of September. The current work on these topics is concentrating in re-adapting the BaBar detector to a symmetric machine and to more stringent particle identification requirements. The physics case is robust and specific with a few discovery channels, but detector simulations are needed to assess the final potential of the experiment.

A Tau/Charm Factory can provide multiple returns. An immediate economic one, related to the job opportunities of its construction and operation, and to the average presence on the territory of hundreds of physicists, engineers and technicians, most of them from an international community.

The preparation to international tenders for its realization will make the Italian industries more competitive in future tenders of accelerator based infrastructures, including those related to medical physics or light sources. The attraction of young researcher abroad will generate a "brain catch" program.

The project will strongly contribute to the HORIZON 2020 program of excellent science through the development of skills and talents. It will be an incubator of future emerging technologies, anticipated and tested in the High Energy Physics environment (electronics, engineering, web, computing). In particular, detector performances require the development of high technology in 3D electronics devices for the integration of sensors (particle trackers), today one of the major trends in the emerging industrial technologies. Sophisticated software codes are needed to simulate and treat the huge amount of data coming for the experiment, calling for a powerful computational network based on GRID technology. The novel control system developed for the accelerator can be exported to the industrial world.

On the accelerator side, very low emittance rings, such as in Tau/Charm project, will generate skills useful in the development of future linear colliders Damping Rings.

The capability for Italy to host an International laboratory, the Cabibbo Lab, may activate a co-funding process from European countries in a reciprocity scenario with respect to the Italian contributions to major European infrastructures.

Besides a frontier particles detector and collider, the infrastructure aims to host a Free Electron Laser (FEL) facility with Angstrom class resolution, for state of the art material and biophysics studies, and a test area where extracted beam of various type will be available, ensuring to the facility a long exploitation time.

As described briefly at the end of this document, key applications will then be made available for a wider scientific and industrial community.

Roberto Petronzio
Director of the Nicola Cabibbo Laboratory Consortium

**EDITOR:**

M.E. Biagini, INFN-LNF, Frascati, Italy

**CONTRIBUTORS:**

M.E. Biagini, R. Boni, M. Boscolo, A. Chiarucci, R. Cimino, A. Clozza, A. Drago, S. Guiducci,
C. Ligi, G. Mazzitelli, R. Ricci, C. Sanelli, M. Serio, A. Stella, S. Tomassini
INFN – Laboratori Nazionali di Frascati, Italy

S. Bini, F. Cioeta, D. Cittadino, M. D'Agostino, M. Del Franco, A. Delle Piane, E. Di Pasquale,
G. Frascadore, S. Gazzana, R. Gargana, S. Incremona, A. Michelotti, L. Sabbatini
Consorzio Nicola Cabibbo Laboratory, Rome, Italy

G. Schillaci, M. Sedita
INFN – Laboratori Nazionali del Sud, Catania, Italy

P. Raimondi
ESRF, Grenoble, France

R. Petronzio
Rome Tor Vergata University, Rome, Italy

E. Paoloni
Pisa University, Pisa, Italy

S. M. Liuzzo
Rome Tor Vergata University, Rome, Italy, and ESRF, Grenoble, France

N. Carmignani
Pisa University, Pisa, Italy, and ESRF, Grenoble, France

M. Pivi
IMS Nanofabrication, Vienna, Austria

# CONTENTS

PART 1 *Accelerator Complex*

# Contents

# 1    Introduction

The SuperB Flavor Factory is part of the Research Plan as Flagship Project since 2010. The accelerator was supposed to work primarily on the Υ(4S) resonance (center of mass energy 10.58 GeV) but also to be able to reduce the center of mass energy to measure rare decays around the Tau/Charm production threshold (center of mass energy about 4 GeV). This was made possible by a flexible design of the two rings lattice, and by a proper choice of the main beam parameters. Already in both the first [1.1] and the second [1.2] SuperB Conceptual Design Reports this possibility was incorporated in the design. It is then a natural evolution of the project, once established that the budget allocated is not sufficient to entirely cover the SuperB complex construction and operation, to have a transition to a smaller and cheaper, but still frontline, accelerator such as the Tau/Charm.

The principles of operation of such an accelerator are still based on the SuperB ones, like the "crab waist and large Piwinski angle" collision scheme, which has been successfully tested at the Φ-Factory DAΦNE in Frascati [1.3], with small beam emittances and smaller beam sized at the Interaction Point. As a plus the Tau/Charm, due to the lower beam energy, will have very low power consumption and running costs.

The accelerator is designed to have a main operation point at the Tau/Charm threshold, however operation at lower and slightly higher energies is also foreseen. The lower center of mass energy should be at the Φ resonance (1.05 GeV) to complete the data collected at the DAΦNE collider at LNF Frascati. Also 2 GeV in the center of mass will provide interesting data at the threshold of the nucleon antinucleon production, for studies of the nucleon form factors. Upper energies will allow the study of the $\Lambda_s$ resonance at 4.35 GeV. For this reasons the maximum energy is 2.3 GeV. The new design is based on a symmetric beams collision, rather than the asymmetric one planned for SuperB. This is justified by the different processes, which will be studied, and makes the design a lot simpler. For example, for SuperB a large effort was put on the design of the Final Focus (FF) sections, where the two beams are brought into collision, especially on a state-of-the-art design of the first superconducting quadrupole doublet, able to cope with the high gradients required and the small space available. For the Tau/Charm these elements will have more relaxed characteristics, and the symmetric FF will easy the design.

The beam parameters will be similar to those planned for SuperB, yet more relaxed due to the lower beam energy. The experience with SuperB, for which an extensive study on collective effects has been carried out, turns out now to be very useful in the choice on which are the most critical parameters.  The electron beam polarization, a unique feature in the SuperB project with respect to its Japanese competitor SuperKEKB, will also be part of this design. However, as a consequence of the lower energy and the larger spin depolarization time, the design of the "Spin Rotation System" (SRS) will be a lot simpler. Instead of two SRS, one on each side of the FF, just one, with similar characteristics, will be placed in the straight section opposite to the IP. This insertion is called "Siberian Snake" and does not require a specific bending angle value in the Final Focus section as it was for the SuperB SRS.

The injection system will also profit from the design done for SuperB. An adjustment of the complex to the lower beam energy is straightforward. The use of the Linac for a SASE-FEL facility, with the increment in energy to 6 GeV, thanks to the C-band technology under development at LNF, will allow for state of the art studies in material science and biophysics. A Beam Test Facility for detectors tests and other particle sources is also under study and will be briefly mentioned in the last Part of this document.

## 2   Collider Main Rings

### 2.1   Introduction

The Main Rings are the principal component of the collider. Electrons and positrons will circulate in two separate rings, crossing at the Interaction Point (IP) where the events will be collected by the Detector.

The two Rings have similar magnetic structure, called "lattice", the only difference being the presence in the electron ring of a Siberian Snake (SS) used to rotate the transverse spin of injected electrons into a longitudinal spin at the IP. The electron beam polarization is a unique feature of the Tau/Charm accelerator.

To further simplify the design and save money, operation with equal beam energies has been chosen. This has an impact on the processes that can be studied and is still reason of debate, and for the first phase of operation symmetric energies are the baseline. Since tunable beam energies are part of the design, a small boost (< 0.2 at 4 GeV c.m.) could be provided in a second phase.

The collision scheme adopted is the "Large Piwinski Angle and Crab Waist" (LPA and CW) [2.1] scheme already tested at DAΦNE and baseline of the design of the upgraded SuperKEKB B-Factory at KEK in Japan. The characteristics of this scheme are the large crossing angle, which reduces the parasitic crossings and the beam overlap area useful for the luminosity, the extremely low beam emittances and IP beam sizes, and the use of sextupoles to cancel the resonances appearing with the crossing angle. Details on the principles of this scheme can be found in [1.1,1.2,2.1].

### 2.2   Luminosity and Beam Parameters

The beam parameters have been chosen in order to have a peak luminosity of $10^{35}$ cm$^{-2}$ sec$^{-1}$ at the Tau/Charm threshold and upper, as indicated by the Physics case study. At lower center of mass energies a lower luminosity, but still an order of magnitude higher than that of present colliders operating in the same energy range, can be achieved with a suitable choice of parameters.

In Table 2.2.1 is the list of beam parameters relevant to achieve such a luminosity, for the energy of 2 GeV/beam, as an example. The emittance, bunch length and energy spread of such intense bunches are dominated (and increased) at these energies by the Intra Beam Scattering (IBS) mechanism. For this reason the numbers in Table 2.2.1 include an estimation of this effect, as well as the hourglass effect, which reduces the luminosity due to the bunch length longer than the IP $\beta_y$.

It has to be noted that these parameters are not pushed to the limit, so it is possible for example to reach the same luminosity with a larger coupling factor but slightly higher currents, or even foresee a factor of two in the peak luminosity pushing up the beam currents.

**Table 2.2.1 – Beam parameters for running at 4 GeV c.m. at $10^{35}$ cm$^{-2}$ s$^{-1}$**

| Parameter | Units | |
|---|---|---|
| LUMINOSITY | $10^{35}$ cm$^{-2}$ s$^{-1}$ | 1.0 |
| cm Energy | GeV | 4.0 |
| Beam Energy | GeV | 2.0 |
| Circumference | m | 340.7 |
| X-Angle (full) | mrad | 60 |
| Piwinski angle | rad | 10.84 |
| Hourglass reduction factor | | 0.85 |
| Tune shift x | | 0.004 |
| Tune shift y | | 0.089 |
| $\beta_x$ @ IP | cm | 7 |
| $\beta_y$ @ IP | cm | 0.06 |
| $\sigma_x$ @ IP | microns | 18.95 |
| $\sigma_y$ @ IP | microns | 0.088 |
| Coupling (full current) | % | 0.25 |
| Natural emittance x | nm | 2.85 |
| Emittance x (with IBS) | nm | 5.13 |
| Emittance y (with IBS) | pm | 12.8 |
| Natural bunch length | mm | 5 |
| Bunch length (with IBS) | mm | 6.9 |
| Beam current | mA | 1745 |
| Buckets distance | # | 1 |
| Ion gap | % | 2 |
| RF frequency | Hz | 4.76E+08 |
| Number of bunches | # | 530 |
| N. Particle/bunch | # | 2.34E+10 |
| Beam power | MW | 0.16 |
| Transverse damping times (x/y) | msec | 35/49 |

To get high luminosity with a relatively low bunch density, in order to keep under control the IBS emittance growth, it has been chosen to fill all buckets, with a bunch distance of 2.1 nsec, keeping a 2% gap to avoid the ion trapping in the electron ring. These values will require an efficient bunch-by-bunch feedback system, like the one already developed for DAΦNE, PEP-II and SuperB, a proper choice of the RF parameters and very effective mitigations for the e-cloud instability. It is worthwhile to note that the amount of energy losses in the Tau/Charm is very low, being about 15 times lower than in SuperB.

Damping times (different in X and Y due to the presence of a gradient dipole in the ARC cell) are slightly higher than in SuperB. For operation at lower beam energy the use of wiggler magnets is foreseen.

The Crab Waist collision scheme is beneficial for beam-beam effects. Due to effective suppression of beam-beam induced resonances it allows for increasing the value of ξ$_y$ by a factor of about 3 as compared with the ordinary head-on collision. Accordingly, the same factor can be gained in the luminosity. As an example of the large available operation area in the tune space with the LPA and CW scheme, the luminosity contour plot versus the betatron tunes for the BINP (Novosibirsk) C-Tau project parameters (very similar to the Tau/Charm ones) is shown in Figure 2.2.1, in the tune region close to 0.5 where these kind of colliders usually work. For the Tau/charm design beam-beam tune shifts are on the safe side: the horizontal is negligible, due to the features of the LPA scheme, the vertical ξ$_y$ is lower than 0.1 for the baseline parameters, a value much lower than those routinely achieved at the B-Factories PEP-II and KEKB.

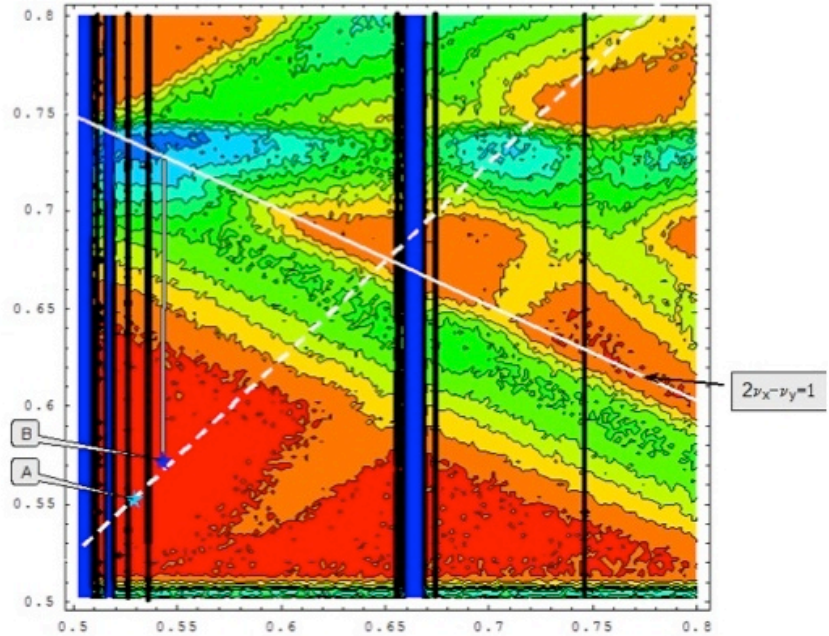

**Figure 2.2.1 – BINP C-Tau project luminosity contour plots as a function of the working point of the betatron tunes (horizontal and vertical axes correspond to the tune fractional part). The red and blue colors show large and small peak luminosity.**

In Table 2.2.2 is a list of beam parameters relevant for the luminosity, for c.m. energies ranging between 4.6 and 2 GeV. At low energy (last column) the insertion of 8 wigglers, of the type installed at DAΦNE, is foreseen to keep the same damping times. The polarization will be maximum around 4 GeV c.m.

**Table 2.2.2 – Beam parameters for different c.m. energies**

| Parameter | Units | | | |
|---|---|---|---|---|
| LUMINOSITY | $10^{35}$ cm$^{-2}$ s$^{-1}$ | 1.0 | 1.0 | 0.2 |
| c.m. Energy | GeV | 4.6 | 4.0 | 2.0 |
| Beam Energy | GeV | 2.3 | 2.0 | 1.0 |
| Circumference | m | 340.7 | 340.7 | 340.7 |
| X-Angle (full) | mrad | 60 | 60 | 60 |
| Piwinski angle | rad | 11.19 | 10.84 | 14.66 |
| Hourglass reduction factor | | 0.86 | 0.85 | 0.83 |
| Tune shift x | | 0.004 | 0.004 | 0.002 |
| Tune shift y | | 0.078 | 0.089 | 0.064 |
| $\beta_x$ @ IP | cm | 7 | 7 | 7 |
| $\beta_y$ @ IP | cm | 0.06 | 0.06 | 0.06 |
| $\sigma_x$ @ IP | microns | 18.50 | 18.95 | 20.67 |
| $\sigma_y$ @ IP | microns | 0.086 | 0.088 | 0.096 |
| Coupling (full current) | % | 0.25 | 0.25 | 0.25 |
| Natural emittance x | nm | 3.76 | 2.85 | 1.42 |
| Emittance x (with IBS) | nm | 4.89 | 5.13 | 6.11 |
| Emittance y (with IBS) | pm | 12.2 | 12.8 | 15.3 |
| Natural bunch length | mm | 6 | 5 | 6 |
| Bunch length (with IBS) | mm | 6.9 | 6.9 | 10.1 |
| Beam current | mA | 1720 | 1745 | 1000 |
| Buckets distance | # | 1 | 1 | 1 |
| Ion gap | % | 2 | 2 | 2 |
| RF frequency | Hz | 4.76E+08 | 4.76E+08 | 4.76E+08 |
| Number of bunches | # | 530 | 530 | 530 |
| N. Particle/bunch | # | 2.3E+10 | 2.3E+10 | 1.3E+10 |
| Beam power | MW | 0.28 | 0.16 | 0.05 |
| Transverse damping times (x/y) | msec | 23/33 | 35/49 | 35/49 |

### 2.3  Main Rings lattice

The magnetic structure of the Main Rings is inspired by the latest work on low emittance lattices developed by the Damping Rings of the Linear Collider and the Synchrotron Light Sources community.

In each ring there are: two ARCs (with 2 arc cells each, plus dispersion suppressor sections), a long Final Focus (FF) section for bringing the beam into collision at the IP, and a long straight section opposite to the IP, used for injection, RF cavities, ring crossing, a Siberian Snake section (in the e⁻ ring only), and some tuning quadrupoles. Some drift spaces are also available in the FF matching section for feedbacks, diagnostics, wigglers, etc. In Figure 2.3.1 is a sketch of one Ring layout.

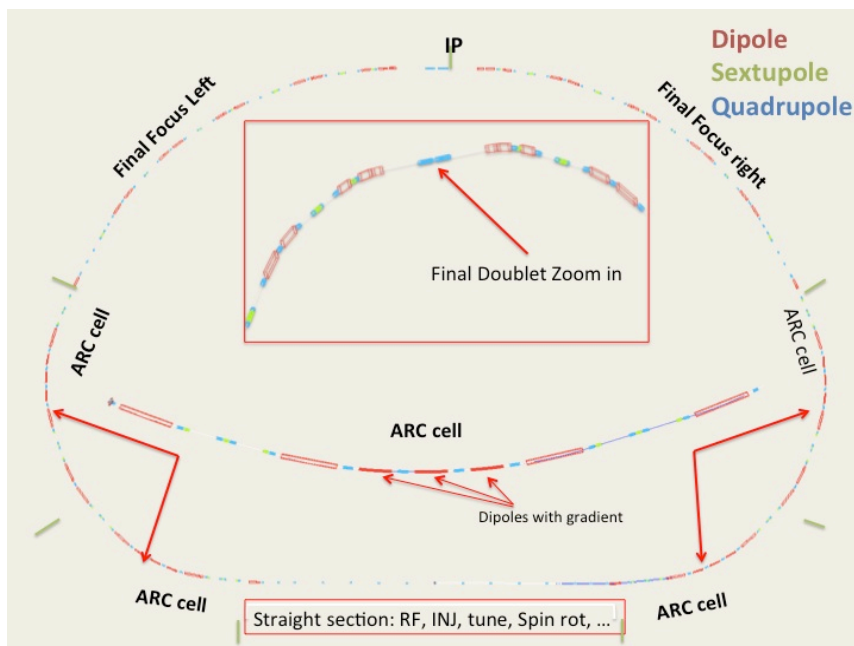

**Figure 2.3.1 – Sketch of one Ring layout.**

The cells lattice is based on a 7-bend achromat scheme: among all the present ring designs, this one has the best ratio between dipole length/total length, and it provides the smallest emittance and has the minimum number of sextupoles with the smaller integrated gradient (because of the large dispersion and betas). The use of a quadrupole gradient (vertical focusing) in some of the cell dipoles reduces the emittance by a factor 1.5 and simultaneously increases the natural bunch length by about a factor 1.25. All the dipoles have a curvature radius of 15m (total dipole length is about 100m), which is the best compromise between damping time and average polarization. The optical functions are shown in Figure 2.3.2 for one ARC. The vertical separation necessary on the opposite side of the IP can be made in different ways. The possibility of tilting slightly the two rings like in the SuperB seems the easiest at the moment. The basic layout, the number and characteristics (in size, field, etc.) of the magnets are not going to change much when going to the TDR phase. So the present lattice allows for a quite precise estimate of the cost and performances of the Tau/charm project.

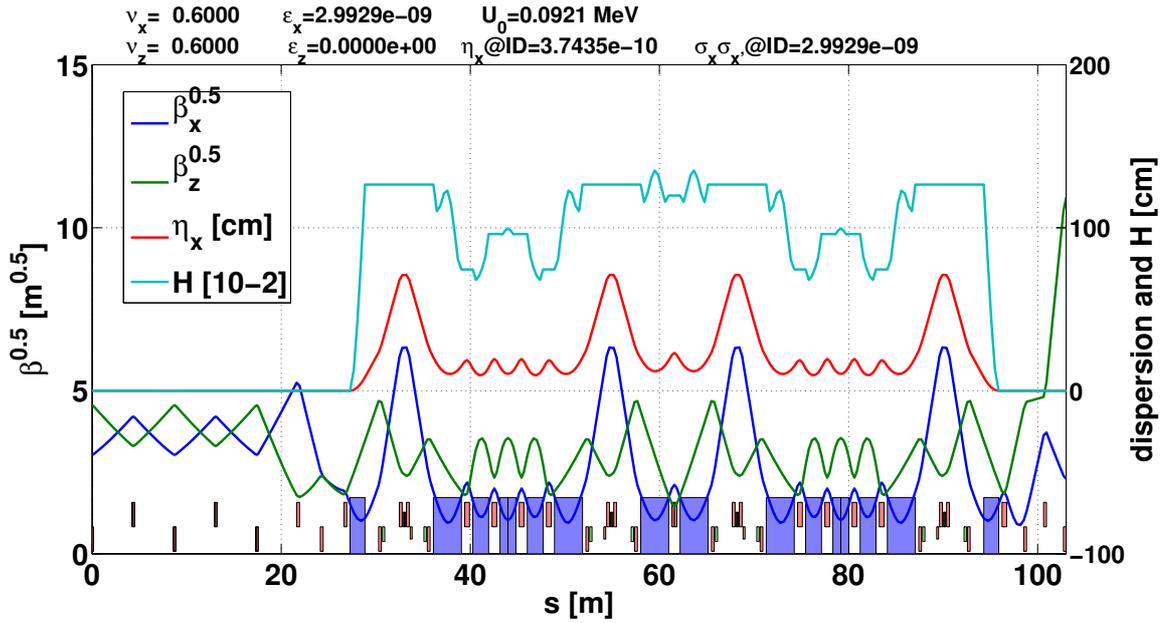

Figure 2.3.2 – Optical functions and H function in one ARC only.

A special care has been devoted to the optimization of the non-linear effects both in the ARCs and in the Final Focus. In the ARC cells 3 families of interleaved sextupoles (2 SD and 1 SF) at about 180 degrees of phase are used to correct for the cells chromaticity (see Figure 2.3.3). Since the sextupoles pairs are interleaved, it generates X and Y tune shift versus $J_x$ and $J_y$ (amplitudes).

A pair of octupoles cancels the X tune shift dependence from $J_x$. The Y tune shift from $J_y$ is canceled by having a proper value of $\alpha_y$ at the X sextupoles (or a proper $R_{43}$ matrix element between them). The cross term is very small and can be zeroed by choosing the proper z-location for the octupoles. As result the ARCs optics is virtually linear for several hundreds beam sigmas (x and y and $\Delta E/E$), as shown in Figure 2.3.4 and 2.3.5, and ARCs dynamic aperture is several times larger than physical aperture, with an energy acceptance larger than ±4%. The ARC sextupoles can also easily cope with the additional chromaticity coming from the straight section hosting the injection, RF and utilities.

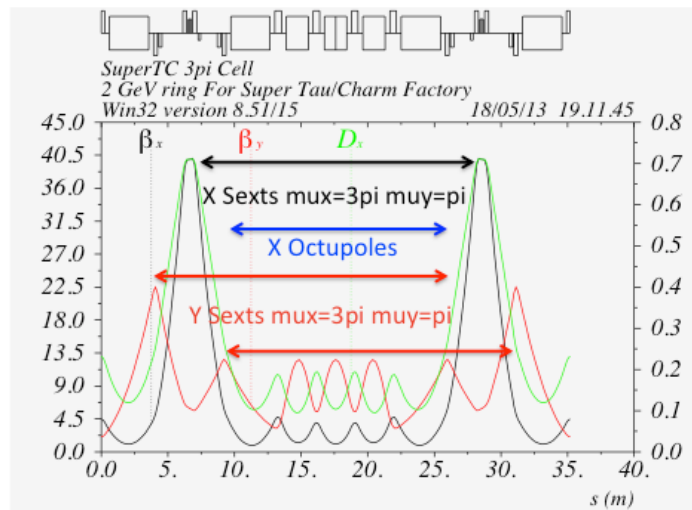

Figure 2.3.3 – Optical functions and position of sextupoles and octupoles in one Arc cell.

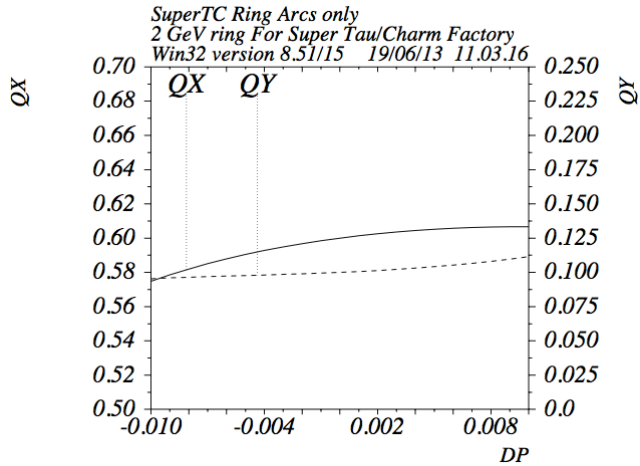

**Figure 2.3.4 – Tunes behavior as a function of the momentum deviation in the ARCs.**

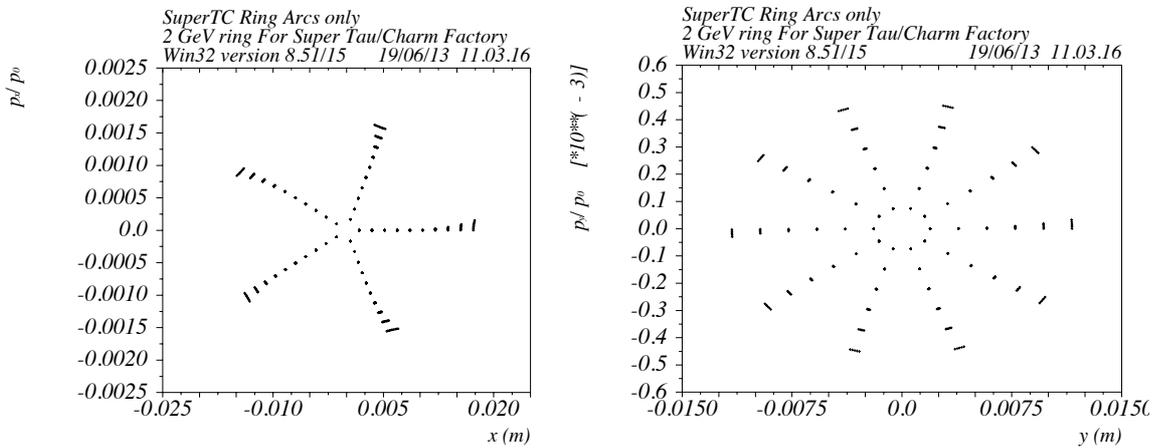

**Figure 2.3.5 – X and Y tracking for the ARCs only.**

The Final Focus (FF) optics is a scaled down version of the SuperB one. The geometric constraints for the polarization are no more necessary, thanks to the choice of a Siberian Snake scheme. The dipoles in the FF have fields very close (between 80-100%) to the ARC ones in order to maximize the polarization (exact match being not possible). The length of these dipoles is as long as possible in order to maximize the dispersion across the sextupoles. Further lengthening of the dipoles would increase the FF Curly-H and the FF contribution to the overall emittance. The ARCs emittance is about 2.4nm, including the FF the overall Ring emittance goes up to 2.8 nm. Also the intensity of the first doublet has been lowered, in order to minimize the effect of the fringing fields introducing unwanted nonlinearities and to reduce the synchrotron radiation coming from the dipoles. In Figure 2.3.6 the plot of the optical functions and Curly-H function in the FF is shown.

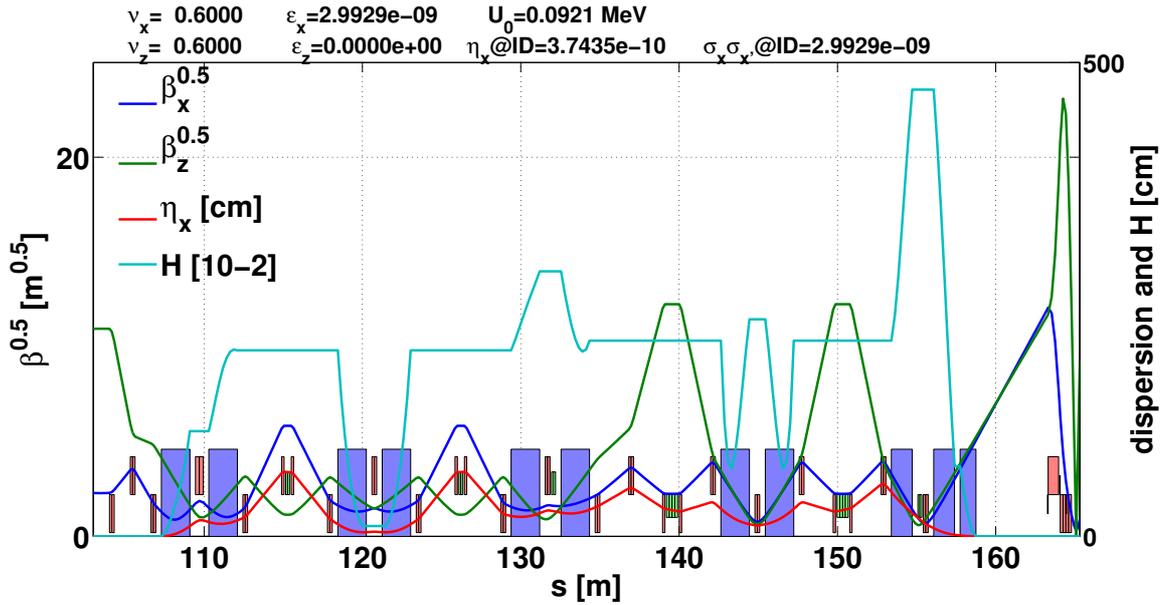

**Figure 2.3.6 – Optical functions and H function in the Final Focus.**

The optimization of nonlinearities in the FF has also required a special care. The main sextupoles, in phase with the Final Doublets, are paired (see Figure 2.3.7). Off Phase (in phase with the IP) sextupoles correct the third order chromaticity; their residual geometric aberrations are very small. A third sextupole further reduces them. Thanks to this arrangement the FF bandwidth becomes about 3 times larger (see Figures. 2.3 8 and 2.3.9 below).

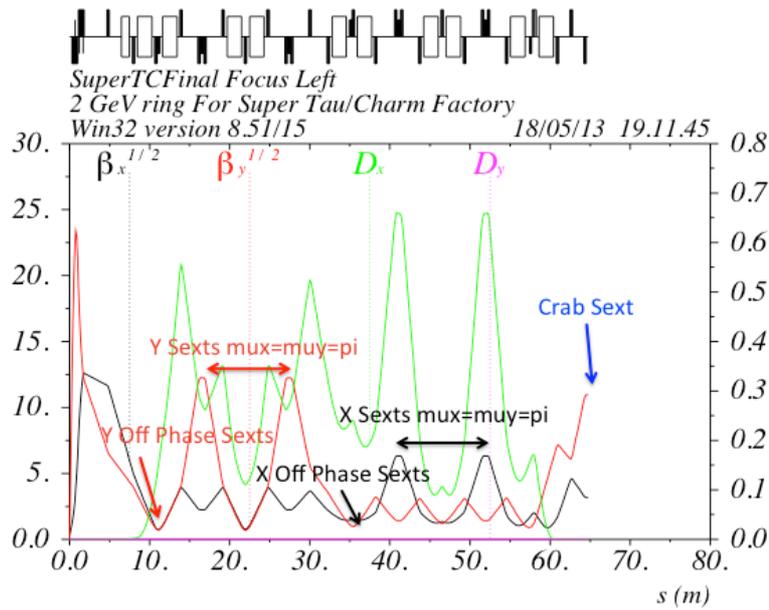

**Figure 2.3.7 – Sextupoles arrangement in thru Final Focus.**

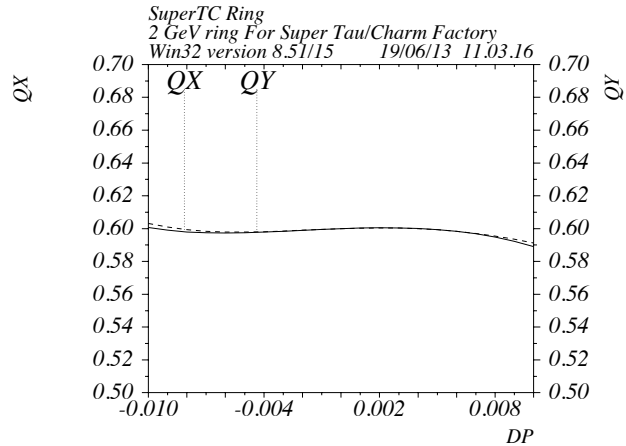

**Figure 2.3.8 – Tunes behavior as a function of the momentum deviation for the whole ring.**

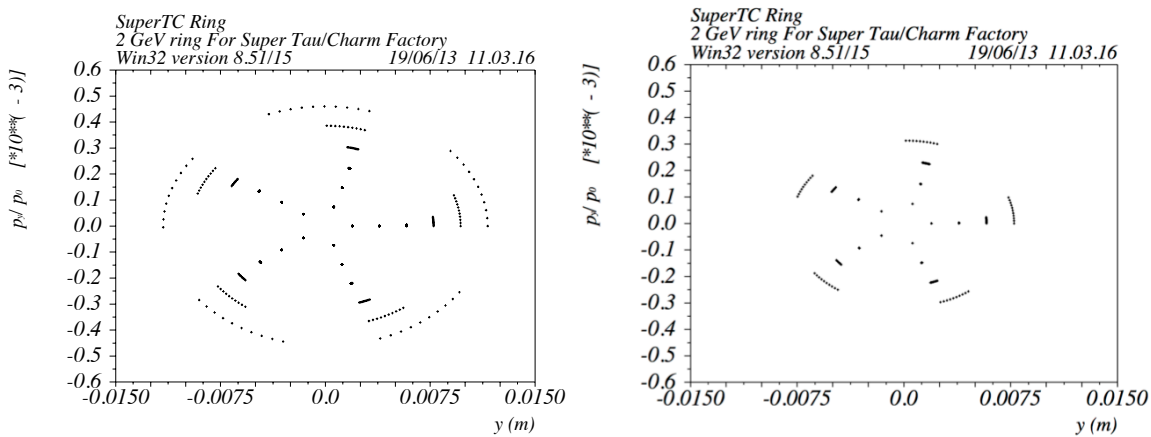

**Figure 2.3.9 – X and Y tracking for the whole ring.**

The Dynamic Aperture (DA) reduction due to the Final Doublets (FD) fringe fields and the Crab Sextupoles is a well-known issue of this kind of FF designs, and the compensation of such effects has required a dedicated study. Fringe fields are very weak third order non-linearities, ultimately related to the fact that the magnetic fields do satisfy the Maxwell equations. However in Super Flavor Factories their effect is very large and strongly reduces the DA, due to the strong FD quadrupoles and the high beta functions. The first possible cure is simply to make the FD quadrupoles as long and weak as possible, since all the terms do scale with the gradient. Since their main effect is to generate a strong detuning with particle amplitudes, a second cure is to add or modify some non-linear magnets. The most efficient solution is to have 3 octupoles to cancel the detuning due to fringes and the "kinematic" octupolar term (about ¼ of the fringes) introduced by the FD. The added complexity is very modest. The present solution is almost 100% effective in eliminating the DA reduction due to fringes and kinematic. In addition the octupoles can be optimized in order to reduce all the detuning due to lattice errors/imperfections on the real machine. The layout of the IR with the octupoles position is in Figure 2.3.10. The detuning with the amplitude is shown in Figure 2.3.11.

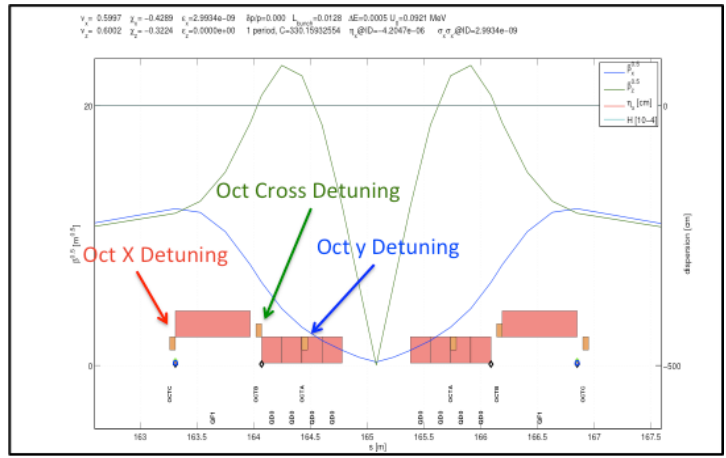

**Figure 2.3.10 – Layout of the IR with the octupoles.**

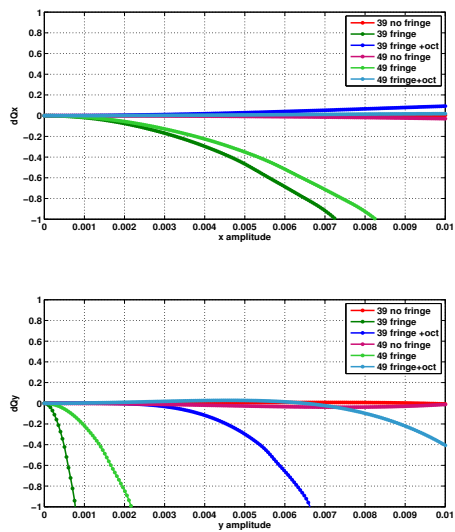

**Figure 2.3.11 – Detuning with amplitude: X-tune (top) and Y-tune (bottom) for different configurations.**

The FD quads must be very long and weak, and this has disadvantages and advantages. Main disadvantage is that the IR becomes very crowded. Advantages are: is easier to make lower gradients and larger aperture magnets, and there will be less Synchrotron Radiation in the Interaction Region (IR), with lower critical energy of the X-Rays. A first draft of the IR is described in the following section and more optimization seems possible and several solutions for the FD quadrupoles are under study (e.g. Permanent Magnets or Superconducting).

The crab sextupoles are very strong and the optics between them, the whole section that includes the left and right Final Focus Optics, has to be as much linear as possible. This has to be true for off energy particles too. The adopted solution for the Fringe-Fields greatly helps in this case. The tracking with fringe fields included shows a very good linear behavior. The FF optics has been re-optimized with the Crab sextupoles ON. A sextupole has been added to compensate for the aberrations induced by the off-phase sextupoles (see Figure 2.3.7). Beta functions and phase advance between the Y and X Chromatic Correction Sections have now optimal values. The FF linear and non-linear chromaticity has been readjusted (is not exactly zero anymore) to improve the off-energy behavior.

Figure 2.3.12 shows the effect of the introduction of crab sextupoles at 50% of their design value (bottom) compared with crab sextupoles OFF (top). The DA reduction due to the crab sextupoles is still large. However the transverse acceptance (at full crab sextupoles intensity) is still larger than the physical aperture and the energy acceptance is above +/-1%. From these preliminary results it seems that there are enough "tuning knobs" to trim the design and the nonlinearities to obtain a solution with very good DA at optimal crab sextupoles strength to about 90% of the design value.

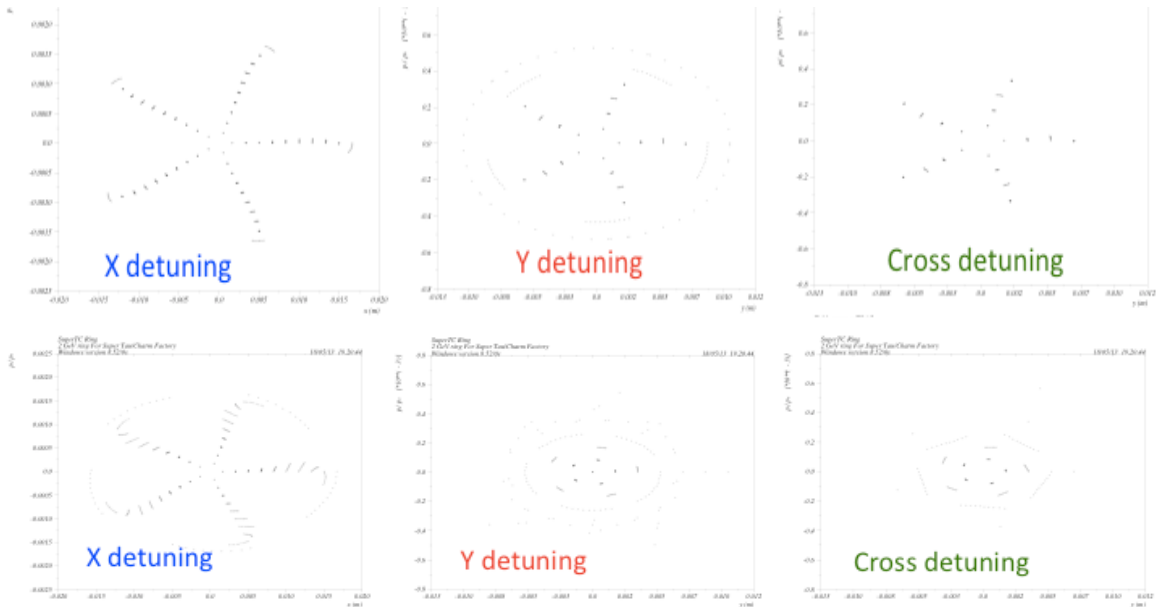

**Figure 2.3.12 – Detuning with amplitude: crab sextupoles OFF (top), crab sextupoles at 50% (bottom).**

Finally, Figure 2.3.13 shown the optical functions in the positron ring. A detailed description of the optimization of the DA also in presence of machine errors will be given in section 2.5.

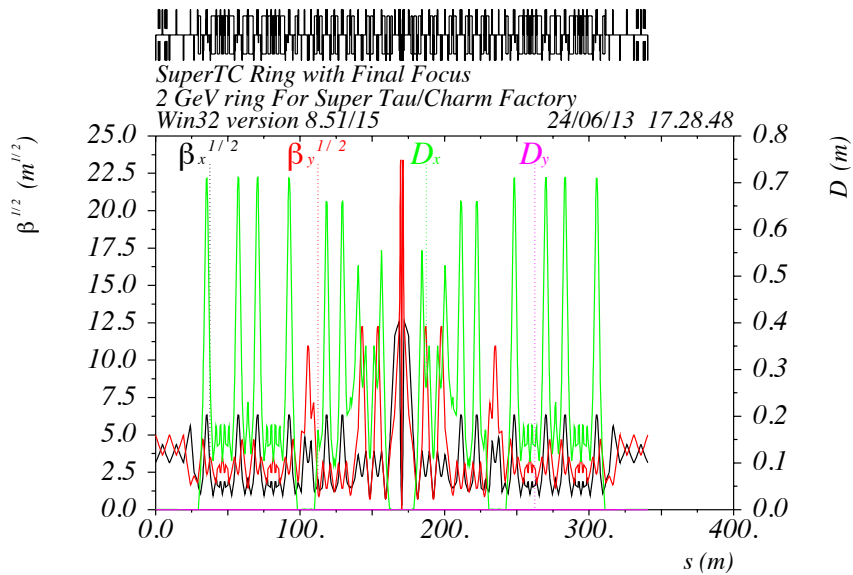

**Figure 2.3.13 – Positron ring optical functions.**

### Siberian Snake insertion

The longitudinal polarization of the electron beam at the IP will be assured by a special insertion called Siberian Snake [2.2]. This device will rotate the vertical spin injected into the electron ring to have a logitudinal orientation at the IP (see Figure 2.3.13).

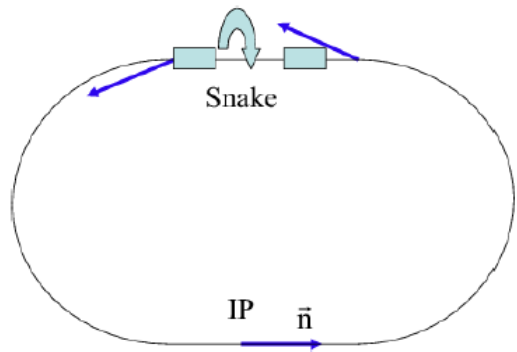

**Figure 2.3.13 - Two π/2 solenoids of the Siberian Snake installed at a half turn away from the interaction point rotate spin by π around the velocity direction. As a result, an equilibrium closed spin orbit n(θ) has a purely longitudinal spin direction at the IP. In the ARCs, the spin always lies in the horizontal plane.**

This is accomplished by means of 2 solenoids which rotate the spin by 180 deg, and quadrupoles arranged in a FODO lattice in between, to have a compensation of the x-y coupling induced. This arrangement results in the formation of a closed spin orbit **n(θ)** with a purely longitudinal equilibrium spin direction at the IP. Everywhere along the ARCs, the spin lies in the horizontal plane, rotating around the vertical axis, which is directed along the bending magnetic field of the ring. Perpendicular to **n**, spins make a half turn around **n** each turn, and thus the total spin tune equals **ν** = 0.5. The coupling induced by the two solenoids of a full Siberian Snake must somehow be compensated in the ring optics. The simplest, and at the same time very convenient way to do this, was suggested by Litvinenko and Zholents in 1980 [2.3]. If the matrices of the FODO lattice inserted between solenoids satisfies the requirement **T$_y$** = - **T$_x$**, then the horizontal and vertical betatron oscillations became fully decoupled. An additional requirement comes from the spin transparency condition [2.4]:

$$T_x = -T_y = \begin{pmatrix} 1 & 0 \\ 0 & 1 \end{pmatrix}$$

Figure 2.3.14 shows a sketch of such an insertion. In the Tau/Charm electron ring this can be easily accommodated in the straight section opposite to the IP.

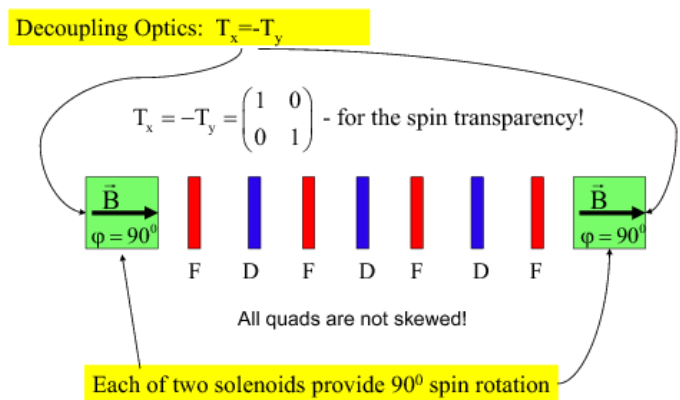

**Figure 2.3.14 - Siberian Snake with the FODO lattice to decouple horizontal and vertical motion.**

At 2 GeV the Siberian Snake will need 2 Superconducting solenoids about 2 m long, with a field of about 5 T. Seven quadrupoles, arranged in 4 families symmetric with respect to the center quadrupole, will provide the coupling correction.

## 2.4    Interaction Region design

The Interaction Region (IR) is usually the most difficult part of the collider to be laid out. The two rings are closely spaced, both the radial and vertical beta functions reach their maxima and the quadrupoles are the strongest of the lattice, moreover the constraints posed by the detector must be fulfilled too. Permanent magnets (PM) are an attractive technology to solve the final doublet design problems. Its main advantages are compactness, stability, field quality and simplicity.

The proposed layout (see Figure 2.4.1 and its caption for details) is composed by a set of Halbach [2.5] quadrupoles described in Table 2.4.1. The inner radius rin of the quadrupoles is such to provide a minimum beam stay clear of 40 σ, the outer radius rout is set by the Halbach formula for the quadrupole gradient:

$$\frac{\partial B_y}{\partial x} = 2 B_r \left( \frac{1}{r_{in}} - \frac{1}{r_{iout}} \right) K_2$$

in which it is assumed a remnant field Br = 1.2 T (which is a quite common figure for samarium cobalt), and K2, which is the figure of merit for a 16 sectors Halbach array, equals 0.94 [2.5]. The energy tunability of the final doublet can be implemented following the Halbach idea of combining two fixed strength PM quadrupoles to obtain a variable gradient: one quadrupole is located, tightly fitted, inside the aperture of a second one. If the gradients of the two are exactly matched to be the same G, then gradient in the bore of the inner one can be tuned from 0 to 2 G by rotating the two quadrupoles about the common axis by the equal amounts but in opposite directions [2.5].

The QD0s will be shared between the two rings. The magnetic axis of the QD0s lies on the centerline of the two beams. The main draw back of this configuration is the synchrotron radiation produced by the incoming beams, and the losses near the IP caused by the dispersion introduced in the spent beam by the offset of the QD0s axis.

It turns out that the first issue is not a severe problem. The low energy of the beam (<∼ 2 GeV) together with the moderate gradient of the QD0s and the small axis offset will set the critical energy of the synchrotron radiation at ∼665 eV. Albeit the total radiated power will in the range of 150 W per beam the 1mm thick beryllium beam pipe with a thin gold coating should provide enough shielding for the detectors close to the IP.

The losses downstream the IP by radiative Bhabha scattering together with the Touschek and beam gas losses will be contained and kept away from the detector by a high Z cylindrical shield surrounding the beam lines as in the last generation high intensity B-Factories.

Other options for the IR layout based on super conducting magnets are at present under study. This solution, albeit more challenging from several point of view and more expensive for the needed ancillary system, will relief the problems of energy tunability and backgrounds.

**Table 2.4.1 – IR quadrupoles parameters**

| Magnet name | Gradient (T/m) | $r_\text{in}$ (mm) | $r_\text{out}$ (mm) | Length (mm) | IP distance |
|---|---|---|---|---|---|
| QD0s | -24.912 | 28.0 | 41.1 | 300 | 200 |
| QD0 | -24.734 | 15.0 | 18.1 | 200 | 550 |
| | | 20.0 | 25.8 | 200 | 750 |
| | | 25.0 | 34.8 | 150 | 950 |
| QF1 | +12.606 | 36.0 | 45.4 | 100 | 1200 |
| | | 38.0 | 48.6 | 100 | 1300 |
| | | 41.0 | 53.6 | 100 | 1400 |
| | | 43.0 | 57.1 | 100 | 1500 |
| | | 45.0 | 60.6 | 100 | 1600 |
| | | 47.0 | 64.3 | 200 | 1700 |

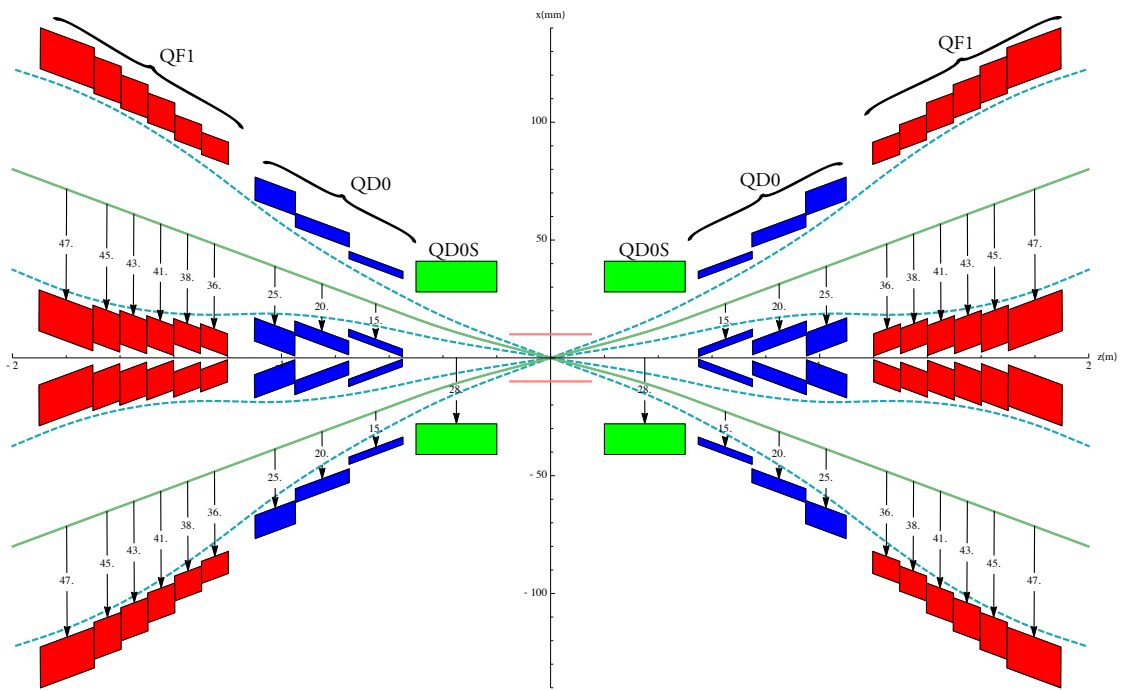

**Figure 2.4.1 - The IR layout horizontal cross section. The colored boxes represents the PM material (QD0s in green, QD0 in blue, QF1 in red). The two QD0s's are shared among the electron and positron machine. The 40σ beam stay clears are represented with dashed lines. The thin beam pipe facing the IP is represented with thin red lines. Note the different scale of the radial and longitudinal axes.**

## 2.5 Dynamic Aperture and Tolerances

The Dynamic Apertures for the previously described lattice have been evaluated with Accelerator Toolbox [2.6], by tracking particles for 512 turns. The tracking performed takes into account the effect of hard edge fringe fields in all quadrupoles [2.7] and the effects of the truncation of the Hamiltonian of drift spaces to higher orders. Figure 2.5.1 shows the reduction of the dynamic aperture due to the Final Focus. The plot is performed at the QD1 location in the center of the straight section where $\beta_x$ = 21m and $\beta_y$ = 9m. The 15 mm observed at this point must be rescaled at the point where the horizontal beta functions are larger in the lattice i.e. at the entrance of the QF1 with $\beta_x$ = 146m and $\beta_y$ = 137m. The physical aperture at this location is approximately 3.6 cm while the rescaled horizontal extent of the dynamic aperture is 4 cm. The effect of the fringe fields is shown in Figure 2.5.2, where is plotted the DA without fringe fields (black), with fringe fields in all quads but QD0-QF1 (cyan), with fringe fields in all quads including QD0-QF1 (blue) and with fringe fields in QD0-QF1 only (red). It is clear that the main contribution comes from the FF doublet (red curve is equal to the blue one), but fortunately this effect can be corrected with three octupoles lenses located in the doublet (shown in green).

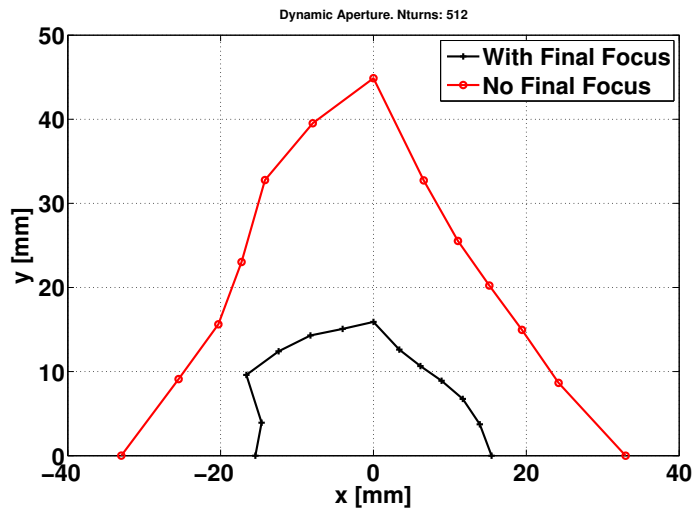

**Figure 2.5.1 – DA without (red) and with (black) the Final Focus.**

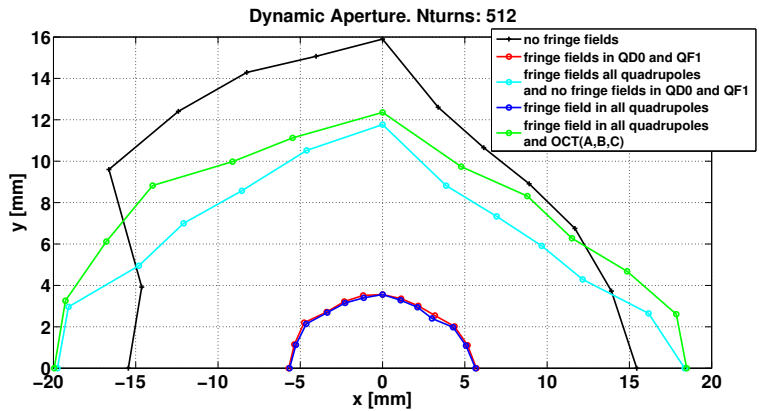

**Figure 2.5.2 – DA without (black) and with Fringe Fields in all quads (red, blue). In green is the correction by octupoles in the FF doublets.**

The DA behavior for off momentum particles is shown in Figure 2.5.3. The dynamic aperture is computed in the horizontal (full width from -x to +x) and vertical plane varying the energy deviation. A maximum deviation of [-2%; +2.5%] is achieved (no physical aperture set in the simulations). The reason of this limit is observable in Figure 2.5.4 where the non-linear chromaticity crosses the half integer and third order resonances.

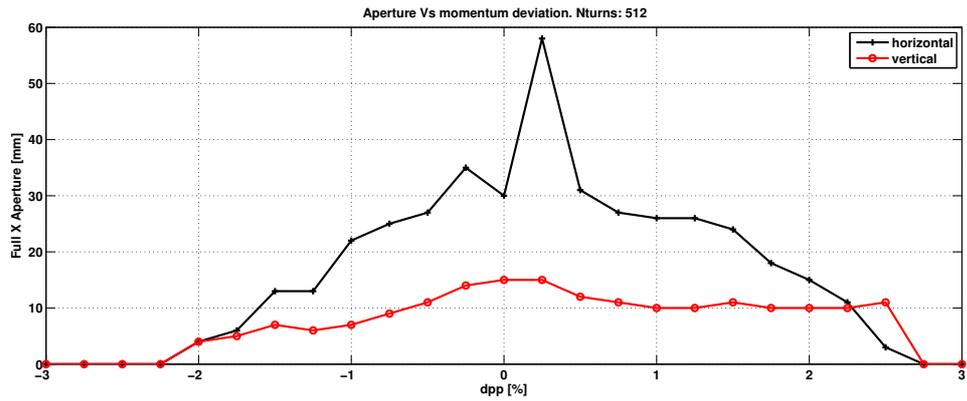

**Figure 2.5.3 – DA vs particle momentum deviation Δp/p.**

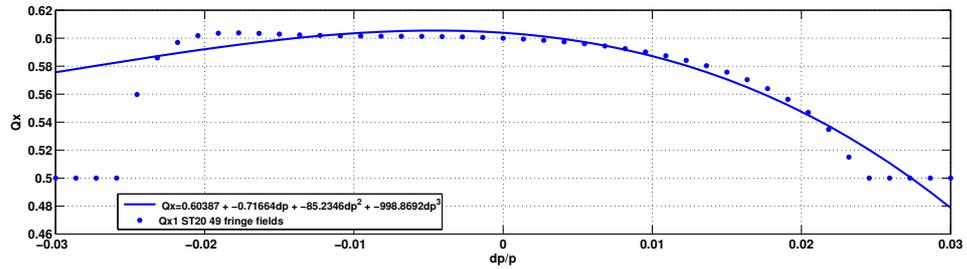

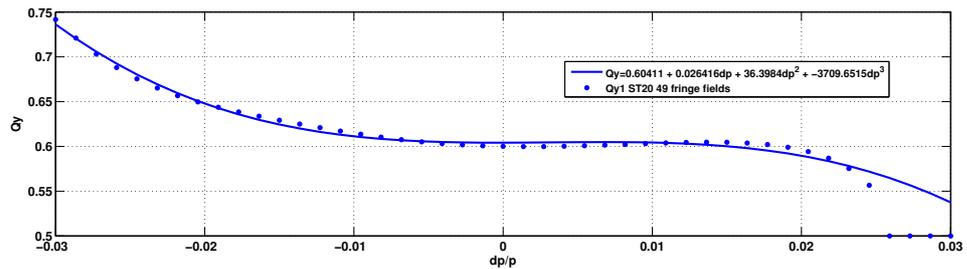

**Figure 2.5.4 – Non-linear chromaticity vs Δp/p.**

A scan for the dynamic apertures when changing the tunes is shown in Figure 2.5.5. Errors and correction are present in the lattice used for the scan. There is a large good region in the tune space analyzed. The choice of the working point is not done here since it needs to be performed considering also the beam-beam scan. However the Dynamic Apertures for some good working points are shown in Figure 2.5.6.

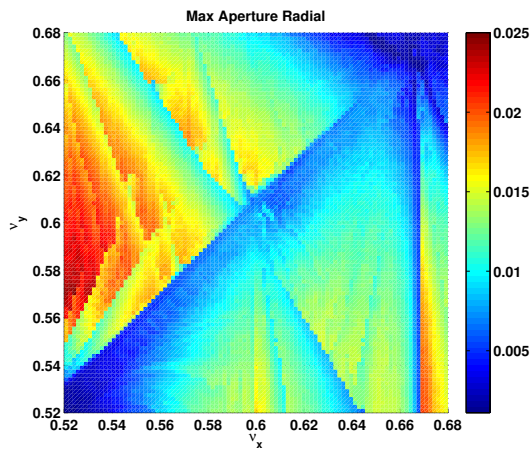

**Figure 2.5.5 – Dynamic Aperture vs tunes.**

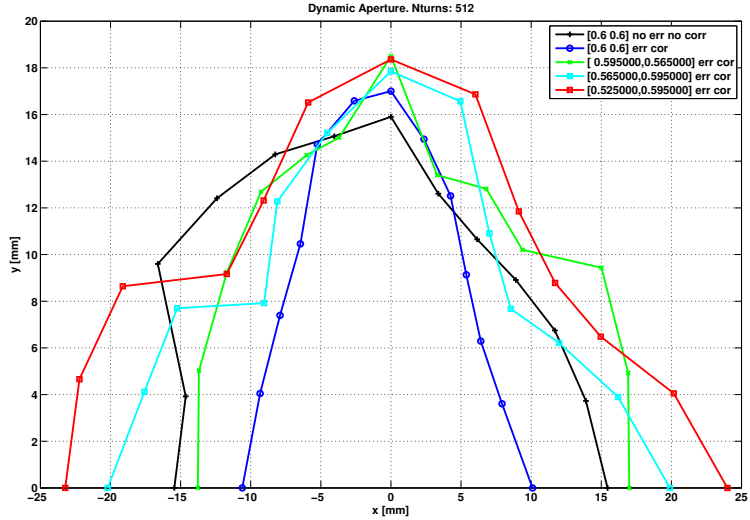

**Figure 2.5.6 – DA without (black) and with errors and correction for possible good working points.**

The frequency map for the lattice with small errors (and no correction) is shown in Figure 2.5.7 (performed using the code Elegant [2.8]). The diffusion is very limited as it is the detuning with amplitude (see Figure 2.5.8). Some points with wrong tune reconstruction (1-tune) may be seen at x=0. No main resonance crossing is recognized in the tune space.

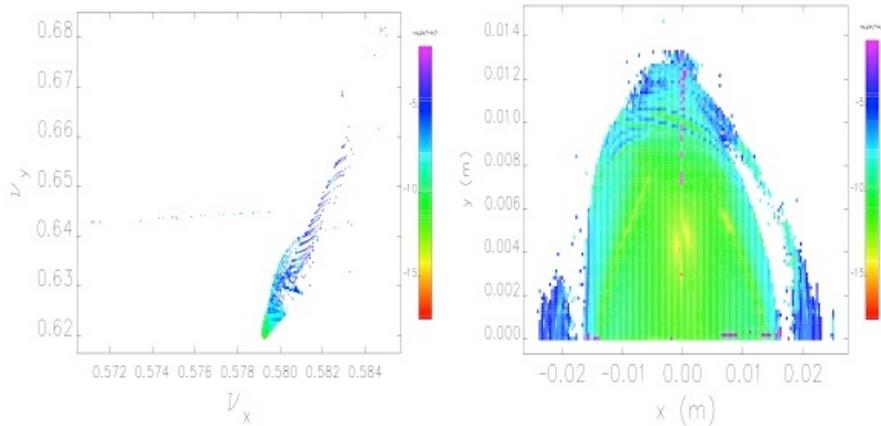

**Figure 2.5.7 - Frequency map. Diffusion rates are displayed on the left in the tune space and on the right in the configurations space (Elegant code).**

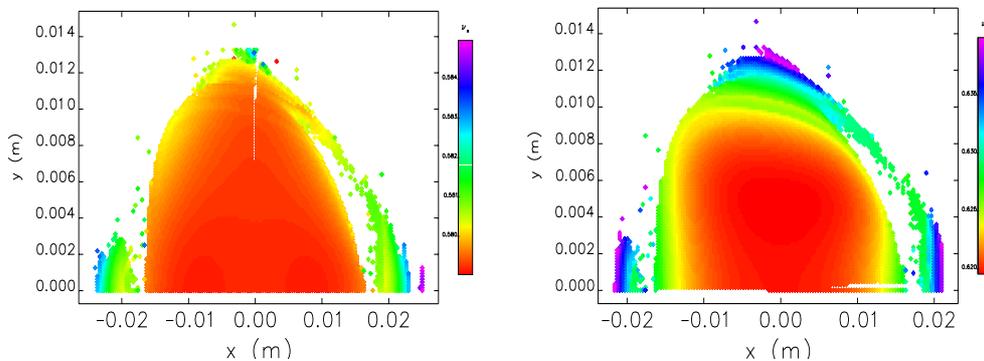

**Figure 2.5.8 - Horizontal (left) and vertical (right) tune computed with FFT in the configuration space (Elegant code).**

The correction scheme is done by placing Beam Position Monitors (BPM), Skew Correctors and H-V Steerers along the lattice. It is assumed that additional coils may be included in sextupoles to produce dipolar and quadrupolar fields (normal and skew) and that dipoles without gradient may act as horizontal correctors and defocusing quadrupoles may act as vertical correctors. Figure 2.5.9 shows the sampling of the beta functions and dispersion at the Beam Position monitor positions. The Skew Quadrupole correctors are added in every sextupole. Also a possible preliminary girder distribution is established. There are a total of 93 BPM, 56 Horizontal Steerers, 46 Vertical Steerers, 46 Skew Quadrupoles. All quadrupoles are assumed as independently powered and are used for the correction, while the dipole with gradient are not used for the correction. The tunes are set to (Qx = 16.58, Qy = 9.62) for the simulations.

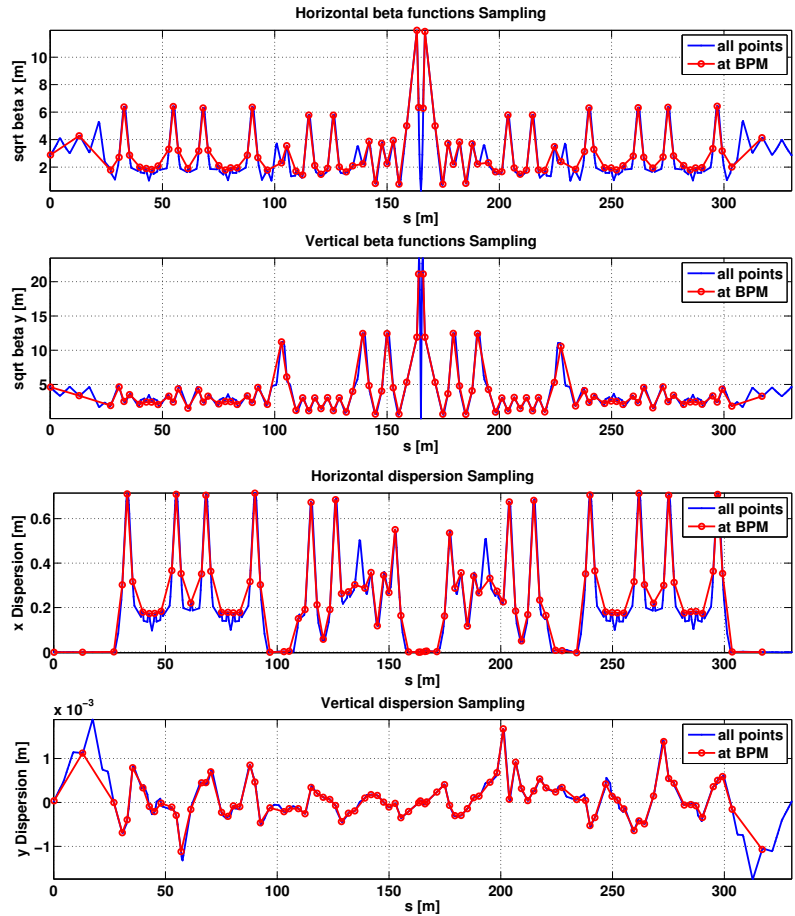

**Figure 2.5.9 – Beta-functions and dispersion at the BPM (red) and at all elements in the lattice.**

In order to apply appropriate misalignments, all magnets that are split in the lattice for matching and tracking purposes are merged into single magnets. Sliced Dipoles are also considered in the application of errors as a single magnet even if not explicitly merged into a single element. The errors considered are:

- $\Delta_{x,y}$ horizontal and vertical magnets misalignment of quadrupole, sextupoles and girders;
- $\Delta_{x,y}$ horizontal and vertical BPM offsets;
- $\Delta BL/BL$ dipole field integral error
- $\Delta_{\psi}$ roll: rotation about the longitudinal coordinate "s"
- $\Delta K_1 L/K_1 L$ quadrupole gradient error
- $\Delta K_2 L/K_2 L$ quadrupole gradient error

The dipole roll error is applied by adding transverse fields (PolynomA or B in AT [2.6]). All errors (except the girder errors) are extracted from a random Gaussian Distribution truncated at 2σ (the truncated gaussian is modified to achieve the correct rms [2.9]). Figure 2.5.10 shows an example of applied misalignment and roll errors. The final doublet quadrupoles position and gradients are unchanged in all simulations.

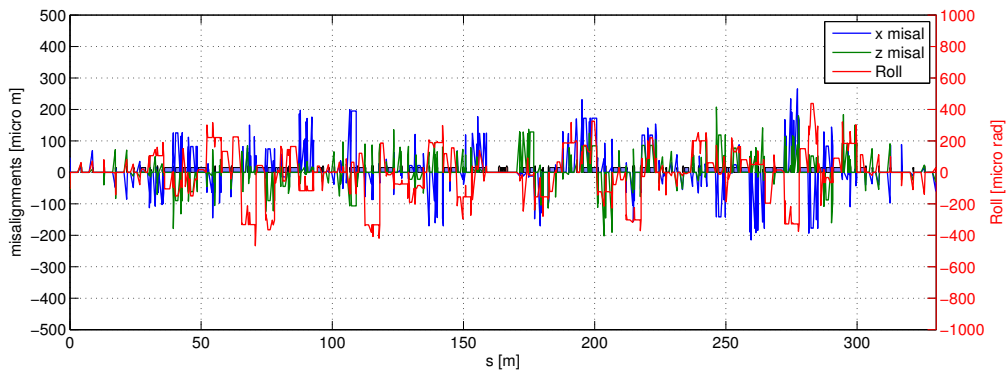

**Figure 2.5.10 – Example of applied horizontal and vertical alignment errors and roll errors.**

Errors are then corrected with the LET algorithm [2.10] using all the quadrupole gradients (excluding the dipole gradients). The correction is performed in four iterations:

- pre-correction of orbits increasing linearly the rms of errors applied up to the wished value;
- correction of horizontal Orbit and Dispersion with horizontal correctors;
- correction of vertical Orbit with vertical correctors;
- correction of vertical dispersion with skew quadrupoles;
- correction of β-beating with quadrupole magnets;
- reiteration of the above for 3 times;
- reset of the tune working point.

To describe the reduction in performance of the lattice and thus determine the tolerated error values we consider the evolution of DA and emittance with increasing rms of the random error distribution [2.11, 2.12]. Figures 2.5.11 and 2.5.12 show the variation of DA (Area) and Emittances, increasing linearly the rms of the error distribution. The maximum rms error distribution is different for the various errors and is taken from Table 2.5.1 (first column). The seeds are kept constant for each step when increasing the errors to avoid the large fluctuations due to the different error sets and emphasize the effect of the errors enlargement.

**Table 2.5.1 – Max studied errors and accepted error values**

| Error | max | accepted |
|---|---|---|
| Dipole $\Delta_\psi$ | $400\ \mu rad$ | $150\ \mu rad$ |
| Girder $\Delta_\psi$ | $400\ \mu rad$ | $175\ \mu rad$ |
| Girder $\Delta_x$ | $200\ \mu m$ | $100\ \mu m$ |
| Girder $\Delta_y$ | $200\ \mu m$ | $65\ \mu m$ |
| BPM $\Delta_x$ | $100\ \mu m$ | $100\ \mu m$ |
| BPM $\Delta_y$ | $100\ \mu m$ | $100\ \mu m$ |
| Quadrupole $\Delta_\psi$ | $400\ \mu rad$ | $75\ \mu rad$ |
| Quadrupole $\Delta_x$ | $100\ \mu m$ | $50\ \mu m$ |
| Quadrupole $\Delta_y$ | $100\ \mu m$ | $45\ \mu m$ |
| Sextupole $\Delta_x$ | $100\ \mu m$ | $50\ \mu m$ |
| Sextupole $\Delta_y$ | $100\ \mu m$ | $50\ \mu m$ |
| $\Delta BL/BL$ | $1\ 10^{-3}$ | $0.5\ 10^{-3}$ |
| $\Delta K_1 L/K_1 L$ | $2\ 10^{-3}$ | $1\ 10^{-3}$ |
| $\Delta K_2 L/K_2 L$ | $2\ 10^{-3}$ | $1\ 10^{-3}$ |

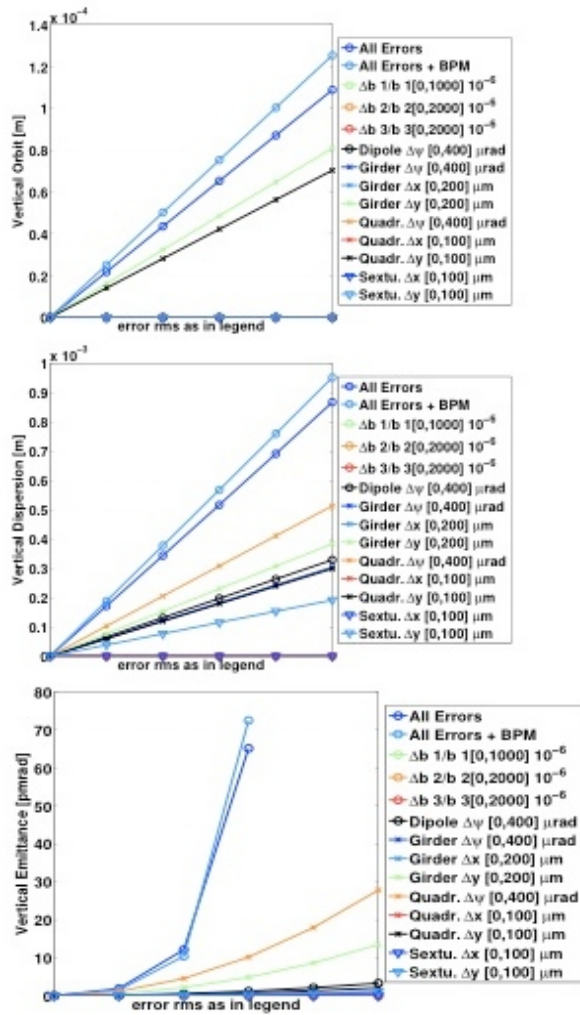

**Figure 2.5.11 – Orbit, Dispersion, Emittances for increasing error sources.**

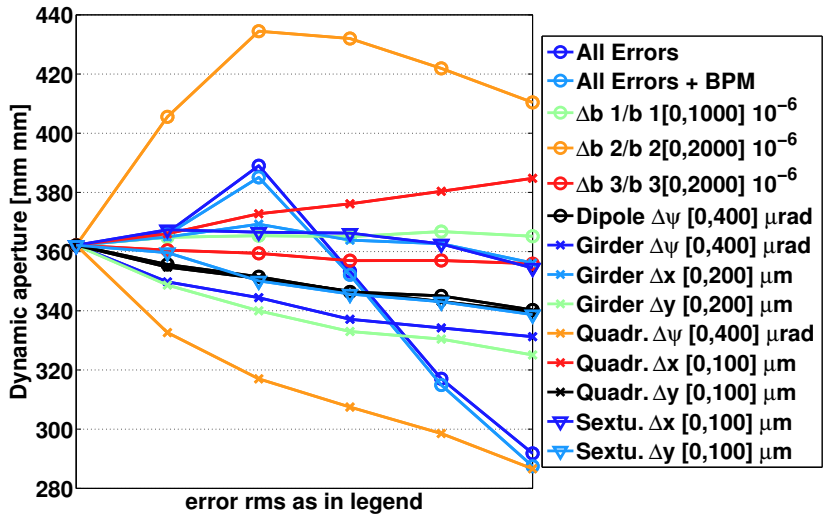

Figure 2.5.12 – Dynamic Aperture for increasing error sources.

The DA evaluated for every studied seed, including all errors, may be seen in Figure 2.5.13. The selected tolerated values are listed in Table 2.5.1 (second column, "accepted"). Only the vertical emittance suffers from the introduction of errors. In particular quadrupole roll errors, vertical girder displacements and dipole rotation are the dominant errors in enhancing the vertical emittance. These errors are reduced more compared to the global set (see Table 2.5.1).

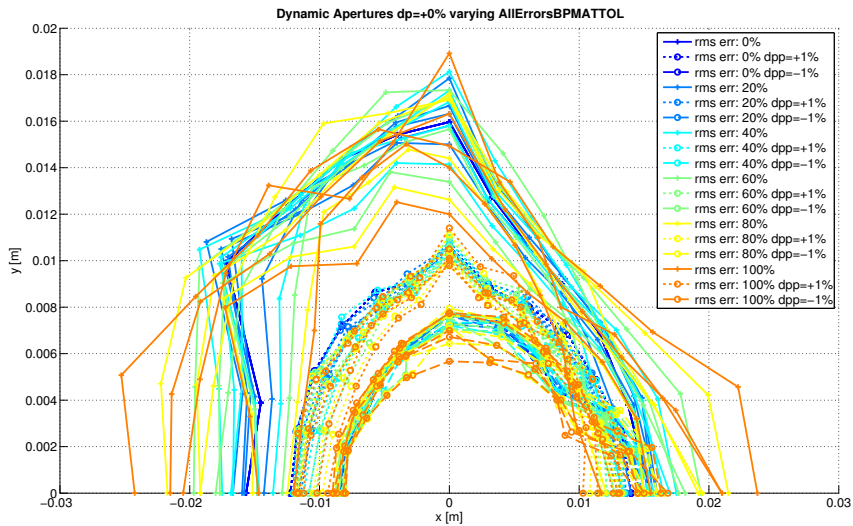

Figure 2.5.13 – DA with above set of accepted errors. The legend refers to the maximum value of error set in the simulation (DK1 here), all errors are applied together.

The rms orbit, dispersion, β-beating and corrector strengths introduced by each error source at the maximum of the studied range (column 1 of table 2.5.1) are shown in Table 2.5.2, while those realized for the accepted set of errors (column 2 of Table 2.5.1) are shown in Table 2.5.3, comparing to those obtained by the maximum range of studied errors. The dynamic apertures and emittances are also included for comparison.

**Table 2.5.2 - Orbit, dispersion, optics modulation and correctors strengths introduced by a given error source, with relative corrections. The table entries highlighted in red are the most influent. Table compiled using Figures. 2.5.9 and 10 and the like**

| $\sqrt{\langle()^2\rangle}$ | $x$ | $(\eta_x - \eta_x^0)$ | $\frac{\Delta\beta_x}{\beta_x^0}$ | $\theta_x$ | $y$ | $\eta_y$ | $\frac{\Delta\beta_y}{\beta_y^0}$ | $\theta_y$ | $a_2$ | $b_2$ |
|---|---|---|---|---|---|---|---|---|---|---|
| error | $\mu m$ | $\mu m$ | % | $\mu rad$ | $\mu m$ | $\mu m$ | % | $\mu rad$ | $T/m$ | $T/m$ |
| $AllErrors$ | 484.6 | 5166.9 | 2.7 | 424.0 | 108.8 | 867.6 | 4.5 | 150.1 | 0.022 | 0.010 |
| $AllErrors + BPM$ | 483.8 | 5147.4 | 2.7 | 428.3 | 125.4 | 952.0 | 4.7 | 163.7 | 0.024 | 0.011 |
| $\Delta b_1/b_1$ | 201.2 | 1583.5 | 1.1 | 389.6 | 0.0 | 0.0 | 2.4 | 0.0 | 0.000 | 0.005 |
| $\Delta b_2/b_2$ | 261.3 | 3331.7 | 2.1 | 31.1 | 0.0 | 0.0 | 2.3 | 0.0 | 0.000 | 0.007 |
| $\Delta b_3/b_3$ | 0.0 | 0.3 | 0.0 | 0.0 | 0.0 | 0.0 | 0.0 | 0.0 | 0.000 | 0.000 |
| $Dipole\,\Delta\psi$ | 0.1 | 1.1 | 0.0 | 0.0 | 0.0 | 328.9 | 0.0 | 0.0 | 0.007 | 0.000 |
| $Girder\,\Delta\psi$ | 0.1 | 1.5 | 0.0 | 0.0 | 0.0 | 303.7 | 0.0 | 0.0 | 0.005 | 0.000 |
| $Girder\,\Delta x$ | 155.7 | 1573.2 | 1.6 | 138.5 | 0.0 | 0.0 | 2.4 | 0.0 | 0.000 | 0.006 |
| $Girder\,\Delta y$ | 1.9 | 60.1 | 0.0 | 0.3 | 81.0 | 383.9 | 0.0 | 114.4 | 0.012 | 0.000 |
| $Quadrupole\,\Delta\psi$ | 0.2 | 4.3 | 0.0 | 0.0 | 0.0 | 513.6 | 0.0 | 0.0 | 0.012 | 0.000 |
| $Quadrupole\,\Delta x$ | 115.6 | 946.7 | 0.5 | 138.1 | 0.0 | 0.0 | 1.5 | 0.0 | 0.000 | 0.003 |
| $Quadrupole\,\Delta y$ | 0.1 | 4.2 | 0.0 | 0.0 | 70.4 | 299.0 | 0.0 | 111.3 | 0.005 | 0.000 |
| $Sextupole\,\Delta x$ | 77.2 | 1171.3 | 0.8 | 9.3 | 0.0 | 0.0 | 1.5 | 0.0 | 0.000 | 0.005 |
| $Sextupole\,\Delta y$ | 0.3 | 9.7 | 0.0 | 0.1 | 0.0 | 191.8 | 0.0 | 0.0 | 0.006 | 0.000 |

The realized emittances for 100 seeds with the tolerated set of errors are shown in Figure 2.5.14. The threshold is set to 7.5 pm rad for the vertical emittance (for a design 0.25% coupling).

**Table 2.5.3 - Rms value of orbit, dispersion, optics modulations and corrector values, at the maximum error range studied and at the accepted set of errors. Horizontal and vertical dynamic aperture and emittances are also reported**

| Parameter | max | accepted |
|---|---|---|
| y | $127\,\mu rad$ | $55\,\mu rad$ |
| $\theta_y$ | $163\,\mu rad$ | $70\,\mu rad$ |
| $\eta_y$ | $1\,mm$ | $0.3\,mm$ |
| $\beta - beat_y$ | $6\,\%$ | $1.6\,\%$ |
| x | $454\,\mu m$ | $232\,\mu m$ |
| $\theta_x$ | $476\,\mu rad$ | $237\,\mu rad$ |
| $\eta_x$ | $18\,mm$ | $9\,mm$ |
| $\beta - beat_x$ | $3\,\%$ | $1.5\,\%$ |
| $\epsilon_x$ | $3\,nm$ | $3\,nm$ |
| $\epsilon_y$ | $24\,pm$ | $5\,pm$ |
| H DA | $15\,mm$ | $15\,mm$ |
| V DA | $15\,mm$ | $15\,mm$ |
| $\delta$ DA | $10\,mm$ | $10\,mm$ |

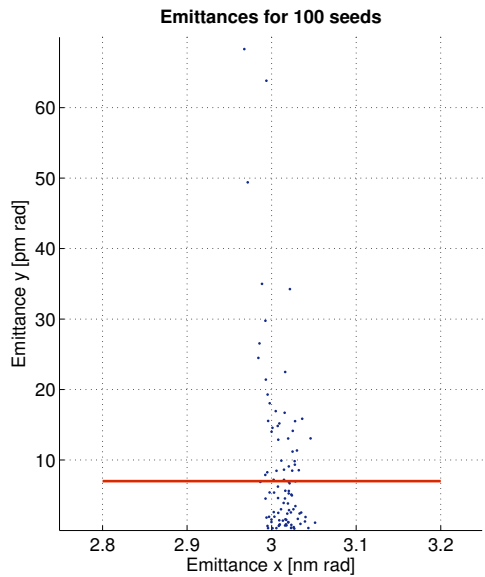

**Figure 2.5.14 – Realized emittances for tolerated set of errors for 100 seeds. In red the threshold V emittance.**

In summary: the Dynamic Aperture for on and off momentum particles has been studied and it is well above the size of the chamber in the region where the beam is larger. The influence of Fringe Fields in the final doublet is also under control thanks to the introduction of octupole lenses. A scan of the tune space has been performed and shows a large region where the optimal working point may be chosen. The frequency map analysis shows no main resonance limiting the dynamic aperture as little detuning is observed even in presence of errors. The influence of errors on the lattice has been studied for all sources independently. Very small influence is observed on the DA, while small quadrupole roll errors are required to achieve vertical emittances in the wished range (less than 7.5 pmrad).

## 2.6  Backgrounds and lifetimes

Backgrounds and lifetime are two issues strictly connected one to the other, even if they have different implications for the accelerator design and operation, being determined by the same physical process that may induce particle losses.

Namely, backgrounds can be cured with detectors shielding, masking and collimator systems while, on the other hand, a short lifetime can be handled with continuous top-up injections.

In particular, for the Tau/Charm factory the primary sources of backgrounds and lifetime are:

- Single and multiple Touschek effect;
- Synchrotron radiation photons produced in the machine magnetic elements;
- Beam-beam Bremsstrahlung (off-energy beam particles and photons are produced);
- Elastically (Coulomb) scattered electrons, produced in interactions with residual gas molecules;
- Beam-gas Bremsstrahlung.

The relatively low beams energy at 2 GeV determines the single Touschek effect as the primary source of particle losses, dominating both lifetime and backgrounds. In addition, due to a factor ten lower luminosity, radiative Bhabha scattering decreases accordingly and it is not the limiting effect for the lifetime anymore, as instead it was for SuperB.

Due to the high beam density also the multiple Touschek effect, usually addressed as intra-beam scattering (IBS), is a critical issue indirectly related to lifetime, as it deteriorates the emittance and the bunch length, as discussed in a dedicated section.

Table 2.6.1 summarizes the lifetimes for the different effects. At this design stage these evaluations can be used as a reference point, and there is space to increase these values with the appropriate knobs, dealing with the beam parameters, lattice design and physical aperture. More particularly, Touschek IR particle losses and lifetime can be optimized numerically by means of a trade-off between emittance, bunch current and bunch length.

**Table 2.6.1. - Summary of lifetimes from main processes as from tracking simulations**

| Physical process | Relevant Machine set | Lifetime (s) |
|---|---|---|
| Touschek | 30 $\sigma_x$ @QF1 with IBS | 376 |
| Touschek | 40 $\sigma_x$ @QF1 with IBS | 484 (8 minutes) |
| Touschek | 40 $\sigma_x$ @QF1  no  IBS | 208 |
| Radiative Bhabha (bb Brem) | $\Delta E/E$ = 1.3% | 680 (11.3 minutes) |
| Beam-gas | P=1nTorr; Z=8; Ry = 2cm | 5500 (1.5 hrs) |
| Gas Bremsstrahlung | P=1nTorr; Z=8; Ry = 2cm | 2.9 e+5 (80 hrs) |
| **TOTAL** | With $\tau_{Touschek}$ = 484 s | 268 |

Table 2.6.2 reports the most significant parameters for the Touschek effect. The studies presented here refer to *V49 lattice*, as a reference for further improvements. A circular beam pipe with a radius of 2.5 cm models the physical aperture everywhere but at the IR, where it is elliptical with the horizontal and vertical dimensions shown in the IR section. The beam-stay-clear at the low-β doublet is a critical parameter for the Touschek lifetime and machine induced backgrounds, being a hot spot (largest loss location) for particle losses. The IR physical aperture - for a fixed IR lattice- is a critical knob at this design stage of the project that can be used to handle this effect. Moreover, the Monte Carlo simulation is a powerful tool to determine the necessary beam-stay-clear to have acceptable Touschek loss rates.

**Table 2.6.2 - Relevant parameters for single and multiple Touschek effect**

| | |
|---|---|
| Beam energy | 2 GeV |
| Bunch current | 3.1 mA |
| Bunch particles | 2.1 $10^{10}$ |
| Beam coupling | 0.25 % |
| Emittance_x (no IBS/ with IBS) | 2.97 / 5.2 nm |
| Emittance_y (no IBS/ with IBS) | 7.425 /13.2 pm |
| Sigma z (no IBS/with IBS) | 4 / 5.6 mm |
| RF energy acceptance | 2.4 % |
| $V_{RF}$ | 2.4 MV |
| βx / βy @IP | 6 cm / 0.06 cm |

The problem of machine-related backgrounds is one of the leading challenges in the Factories. The detector must be sufficiently protected to prevent either excessive component occupancies or deterioration from radiation damage. These background sources can give rise to primary particles that can either enter the detector directly or generate secondary debris that ultimately reaches the detector. Consequently, the design of the interaction region (IR) is critical for the success of the project and important to reach this goal is both the control of the expected particle losses with full tracking simulations and the absorption of the radiated power. An efficient collimator system has been designed to counteract machine backgrounds in the Tau/Charm factory, namely horizontal and vertical collimators for Touschek and Coulomb scattering, respectively.

The Touschek effect and the elastic and inelastic beam-gas scattering have been studied for the Tau/Charm factory with the same approach. A Monte Carlo technique has been applied to track the macroparticles, allowing the evaluation of the particle losses for the scattering under investigation and the corresponding lifetime. This numerical code has been developed for DAΦNE and tested both with the KLOE data and with SIDDHARTA with the crab waist collision scheme showing good agreement, resulting very useful for understanding the critical beam parameters and optics knobs 312].

An accurate analysis of the critical positions where scattered particles are generated - mainly dispersive regions - can be performed, together with the optimization of collimators, both for finding the optimal longitudinal position along the ring and the optimal radial jaw position. IR losses can be studied in detail, like transverse phase space and energy deviation of these off-energy particle losses as a function of different beam parameters, of different optics and for different sets of movable collimators. The generation of the scattering events in the simulation code is done continuously all over the ring. In fact, by properly slicing all the elements in the lattice and evaluating there the transverse beam size, we obtain a good estimate of the scattering probability density function along the ring. We verified that a good accuracy is obtained also in regions where the optical functions change rapidly. We track about $10^6$ scattered macroparticles for a sufficient number of machine turns to have stable results.

A realistic tracking of the off-energy particles includes the main non-linear terms present in the magnetic lattice together with the kinematic term. The scattered particles are simulated from their generation to their loss point in the beam pipe. Lifetime $\tau$ is estimated from the ratio between the number of particles in a bunch N and the loss rate dN/dt, as $1/\tau = (dN/dt)/N$.

The primary losses at the beam pipe induced by this effect can be used for tracking secondaries, generated by feeding primary losses, using a Geant4 detailed model of the detector and IR.

Figure 2.6.1 shows the Touschek lifetime obtained for a given machine momentum acceptance by integrating the evaluation of the formula for each small lattice section, without tracking. This approach doesn't give any indication on the position of particle losses useful for backgrounds studies; however it can be useful to find lower and upper limits to be compared to dynamical aperture (DA). From simulations plotted in Figure 2.6.1 it results that Touschek lifetime is about 8 minutes if the machine momentum acceptance is of about 1%. However, as discussed in the dedicated section, the DA simulations indicate that it is larger than 2% - no physical aperture considered, as usual for these numerical studies. This implies that the physical aperture in the Final Focus (especially in the QF1) dominates on the machine momentum acceptance: Touschek particles are lost due to the small energy acceptance at the FF. This appears also from Figure 2.6.2 (a), where the longitudinal dependent momentum aperture is plotted together with the loss probability at each longitudinal position. This result is in good agreement with DA simulations including the physical aperture (Figure 2.6.2 (b)).

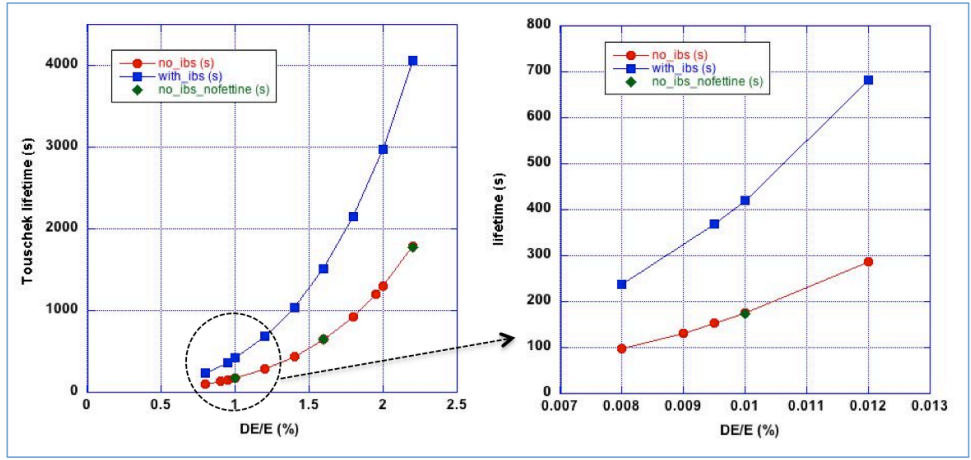

**Figure 2.6.1 - Touschek lifetime as a function of the machine momentum acceptance.**

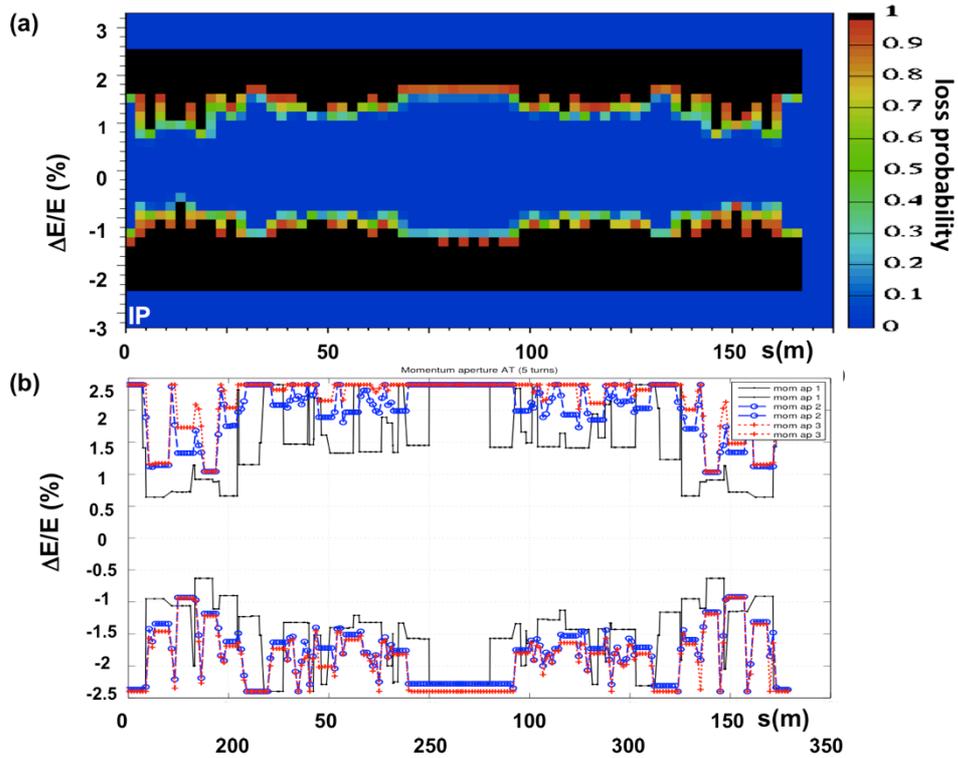

**Figure 2.6.2 - (a) Momentum acceptance with loss probability through the ring (with 40 $\sigma_x$ @QF1) from Touschek tracking simulation; (b) Momentum acceptance with Accelerator Toolbox (AT) [2.14], black curve is with the same physical aperture and parameters used in the (a) case, red and blue are for larger IR apertures.**

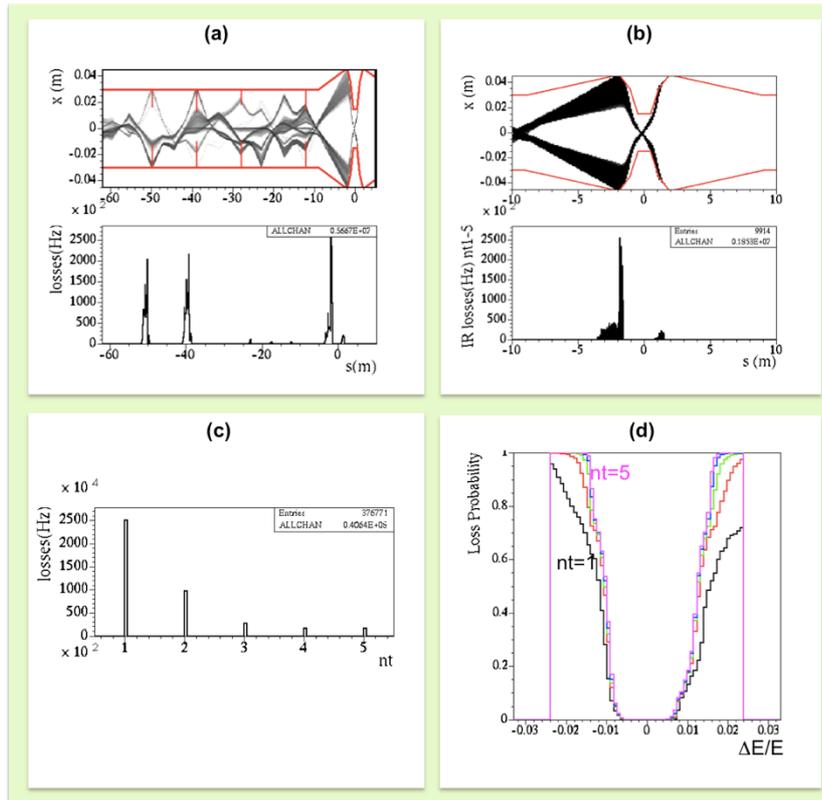

**Figure 2.6.3 - (a) Touschek trajectories and corresponding losses in the FF upstream the IP, the red lines represent the jaws of horizontal collimators; (b) IR trajectories and losses with a beam stay clear of 40 σ$_x$ at the QF1; (c) losses versus machine turns; (d) loss probability versus energy deviation for different machine turns.**

Touschek lifetime obtained with the tracking simulation for different machine sets is summarized in Table 2.6.1. Main indication is that a beam stay clear of at least 40 σ$_x$ at the QF1 is needed, as it gives a longer lifetime by a factor 1.3 wrt the one with a beam stay clear of 30 σ$_x$. (See Figure 2.6.3(b): Touschek particle trajectories and losses at the IR, the hottest spot corresponds to QF1 location). Touschek particles are lost in the first few machine turns, as found from the tracking simulations (see Figure 2.6.3(c)), consistently with the fact that particles are lost for the physical aperture limitations. The loss probability versus the ΔE/E for each machine turn is shown in Figure 2.6.3 (d).

The collimator system has been designed to reduce very efficiently IR particle losses to minimize Touschek and beam-gas backgrounds in the experiment. For Touschek particles only horizontal collimators are necessary, while vertical ones are needed to intercept Coulomb beam-gas scattered particles. Figure 2.6.4 shows the Final Focus (FF) optical functions and the proposed locations for collimators. The two primary horizontal ones have been placed upstream the FF, at about -50 m and -40 m, where the dispersion and the β$_x$ functions are maxima, they are close to two horizontal focusing sextupoles. They stop most of the energy deviated particles that otherwise would be lost at the IR. The secondary collimators are positioned at about -28 m and -12 m, closer to IP. As the vacuum chamber design will become more realistic, simulations including collimators will follow subsequently. In particular, we expect Coulomb scattering to increase its rates at the QD0, as smaller but more realistic vertical vacuum chamber will be designed. Collimators are modeled in the simulation as perfectly absorbing, and infinitely thin. This is a good approximation, for the precision of the simulations required up to this stage. More detailed studies are foreseen to optimize their design, especially the collimator closest to IP. This collimator system is expected to be efficient, as it appears also from Figure 2.6.3 (a) where the horizontal jaws are superimposed to the Touschek trajectories.

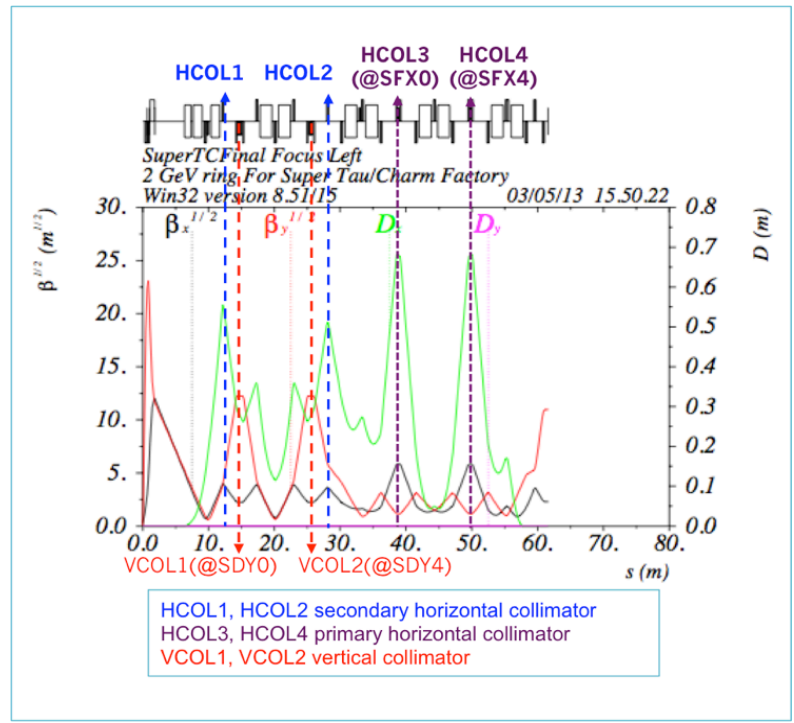

Figure 2.6.4 - Collimators system in the Final Focus upstream the IP.

Radiative Bhabha scattering is an important effect in high luminosity colliders. The cross-section for the Tau/Charm factory is similar to that of SuperB but with a factor 10 smaller luminosity. The luminosity lifetime results 11.3 minutes at $10^{35}$ cm$^{-2}$ s$^{-1}$, where we have considered a momentum aperture at IP of 1.3%, as from radiative Bhabha tracking simulations performed with the same Monte Carlo technique used for Touschek and beam-gas scattering.

Regarding backgrounds studies, large scattered angles are generated with the BBBrem code [2.15] and then tracked into the detector. These particles are lost at the first turn, very close downstream the IP. Low radiative Bhabha scattering angle beam particles (with energy deviation lower than RF acceptance) are studied with the same Monte Carlo technique used for Touschek and beam-gas scattering to check for multi-turn losses. These trajectories are shown in Figure 2.6.5(a) and their dependence on the machine turns in shown in Figure 2.6.5(b). From this plot is appears that, fortunately, radiative Bhabha multi-turn losses are not dangerous.

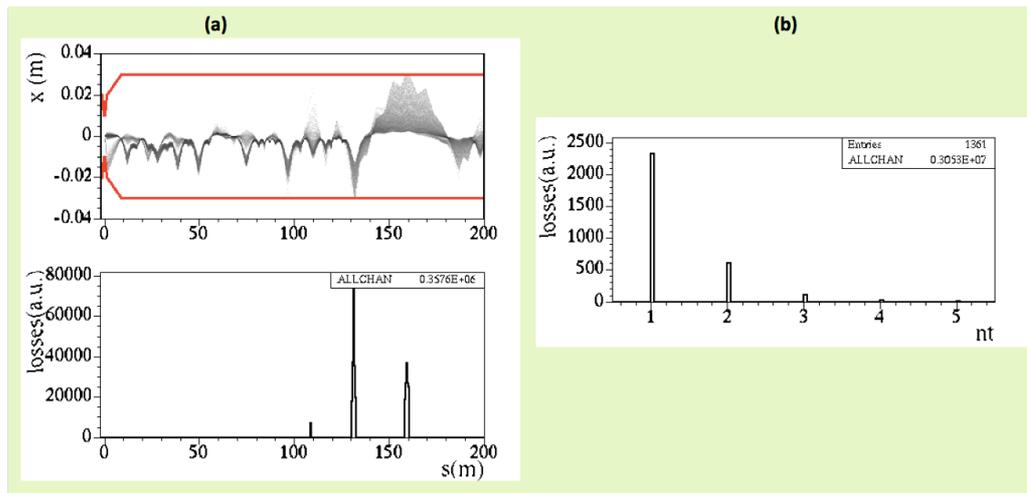

**Figure 2.6.5 - (a) Radiative Bhabhas trajectories and losses; (b) losses versus machine turns.**

## 2.7    Intra-Beam Scattering

Multiple small angles Touschek scatterings, usually addressed as Intra-Beam Scattering (IBS), deteriorate the emittance and the bunch length, if the bunch is dense enough and beam energy relatively low. At the Tau/Charm factory due to the high beams density and to the relatively low energy this effect is significant – more than for SuperB - and has to be carefully taken into account.

In most electron storage rings the growth rates arising from IBS are usually much longer than synchrotron radiation damping times so that the effect is not observable. However, as bunch charge density increases, the IBS growth rates become large enough to induce significant emittance increase. IBS growth rates depend on the bunch sizes, which vary with the lattice functions through the ring; several formalisms have been developed for calculating them [2.16, 2.17]. Accurate growth rates should be calculated at each point in the lattice, and then averaged over the circumference. Furthermore, since IBS results in an increase in emittance, which dilutes the bunch charge density and affects the IBS growth rates, it is necessary to iterate the calculation to find the equilibrium, including radiation damping, quantum excitation and IBS emittance growth. The full IBS formulae include complicated integrals that must be evaluated numerically, and can take significant computation time; however, methods have been developed [2.18] to allow reasonably rapid computation of the equilibrium emittances, including averaging through the circumference and iterating.

For calculation of the IBS emittance growth in the Tau/Charm rings, we use the K. Bane model, in the high energy approximation for Gaussian beams, discussed in [2.18]. In our calculations, the average growth rates are found from the growth rates at each point in the lattice, by integrating over the circumference; we assume lattice natural emittances as equilibrium values at low bunch current and use iteration to find the equilibrium emittances in the presence of radiation and IBS.

Figure 2.7.1 shows the horizontal, vertical and longitudinal emittances ratios together with the bunch length as a function of the bunch charge in presence of IBS. At the nominal bunch charge of $2.1 \times 10^{10}$, corresponding to a bunch current of 3.1 mA, the nominal horizontal emittance of 2.97 nm is increased by a factor 1.69 with $r_e = 0$, i.e. assuming that vertical emittance comes only from coupling. Horizontal emittance due to IBS becomes 5.02 nm. The increase in the vertical emittance is by a factor 1.58, so that the nominal $\varepsilon_y$ goes from 7.425 pm to 11.7 pm (see upper plots in Figure 2.7.2). The longitudinal emittance is enlarged by a factor 2.20. Consequently, IBS induces a significant bunch lengthening: from 4 mm at zero current its final value at nominal current is to 5.9 mm, together with an increase of the rms energy spread: from $4.9 \times 10^{-4}$ to $7.25 \times 10^{-4}$. Some of these parameters, significant for IBS, are listed in Table 2.6.2 of the **Backgrounds and Lifetimes** section.

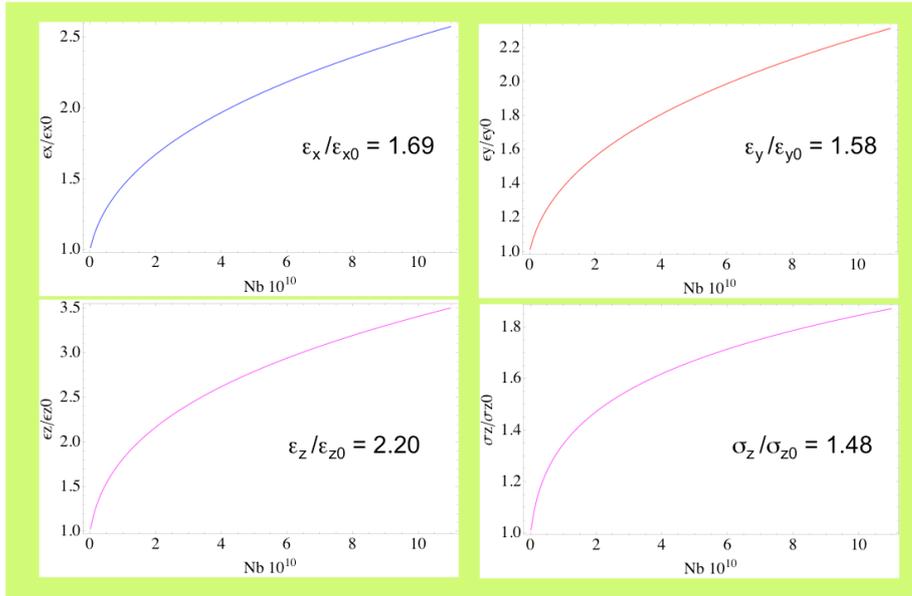

**Figure 2.7.1** - Horizontal, vertical and longitudinal emittance growth ratio, and bunch length growth ratio, as a function of the bunch charge, assuming r=0. The ratio reported on plots refers to the nominal bunch charge (2.1x10^10).

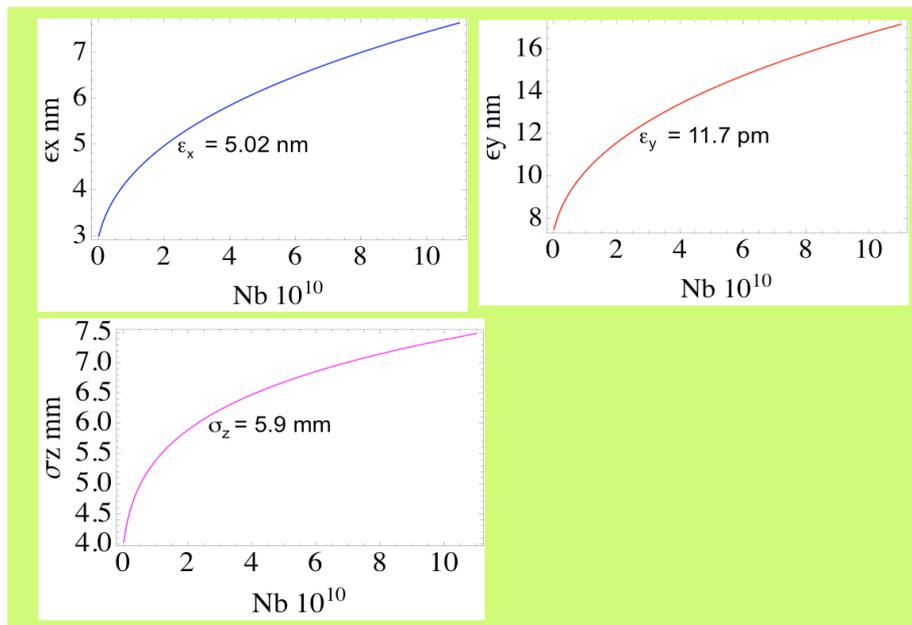

**Figure 2.7.2** - Horizontal, vertical emittance and bunch length as a function of the bunch charge in presence of IBS, assuming r=0. The values reported on plots refer to the final values with IBS at the nominal bunch charge (2.1x10^10).

## 2.8   E-cloud instability

In the beam pipe of the positron storage ring of the Tau/Charm collider, an electron cloud may be first produced by photoelectrons and ionization of residual gases and then increased by the secondary emission process.   Primary electrons are generated by the interaction of beam synchrotron radiation with the chamber walls.   The primary electrons are accelerated by the beam potential with sufficient energy to impact the vacuum chamber walls and produce secondary electrons.   Under certain conditions, electrons can accumulate in the vacuum chamber and grow to a high density levels.

Subsequently, coupling between electrons in the cloud and the circulating beam can cause coupled-bunch instabilities, coherent single-bunch instabilities or incoherent tune spreads that may lead to increased emittance, beam blow-up and ultimately to beam losses.   All these effects would directly affect the collider luminosity, and therefore it is important to plan for the suppression of the electron cloud in the positron ring.   During the last decade, several machines such as the LHC, the B-Factories, DAΦNE, CesrTA at Cornell University and others have confirmed the presence of the electron cloud and measured its effects. In these colliders, the electron cloud correlates with the machine performances, being a limiting factor to increase the beam intensity, decrease the bunch spacing and ultimately increase the collider luminosity.

The electron cloud effect is expected to be an issue in future colliders such as the linear colliders ILC and CLIC, Super-KEKB and the Tau/Charm positron (HER) ring.

### Electron cloud assessment

During the last years, much work has been done for the prediction of the electron cloud effect in the SuperB Factory.  Simulations indicate that a peak surface secondary electron yield (SEY) as low as 1.1 and a challenging 99% antechamber protection result in a cloud density below the instability threshold [2.19, 2.20, 2.21].

A parallel experimental effort is ongoing, and needs further burst at project approval, to study and identify material properties able to reduce SEY below 1.1. Experiments at LNF have already individuated the chemical processes occurring during "scrubbing" (that is during electron bombardment of the accelerator wall surfaces), which is the mitigation baseline adopted for LHC [2.22]. Such studies shines light onto the profound nature of surface properties causing the desired SEY reduction, and do suggest further work to individuate innovative low SEY materials which may be implemented in Tau/Charm positron (HER) ring.

The build-up of the electron cloud is strongly dependent on the bunch separation, which decreases with the storage ring circumference. Reduction of the circumference in the Tau/Charm with respect to the SuperB could make the electron cloud effects more severe.  On the other hand, in a shorter ring, the integrated cloud density along the ring is smaller and an equal synchrotron tune would result in a stronger damping.

The electron cloud assessment in the Tau/Charm positron storage ring has started.  The simulation plan consists in three phases: the evaluation of photoelectron production and their distribution, the evaluation of the electron cloud build-up in magnetic and non-magnetic regions and the estimate of beam instabilities. The photoelectrons production is not only necessary as input for the build-up calculation. Ohmi and Zimmermann [2.23], in 2000, when attempting to explain the observed vertical beam-size blow up for KEKB, introduced  the concept of "single beam instability threshold" suggesting that the mere existence of a certain electron density in the accelerator (for the KEKB case around $7 \times 10^{11}$ e-/m$^3$) is able to detrimentally affect the beam quality. Hence, even in the absence of resonant phenomena, such electron density has to be carefully simulated, controlled and careful material choice is needed for its mitigation. This call for a more careful prediction of such electron density for Tau/charm accelerator, and a complete experimental campaign to study Photoemission Yield and photo reflectivity for all candidate materials to be used in the accelerator vacuum system.

In the following we present preliminary estimates, based on numerical simulations performed with the CMAD code [2.24], of the cloud density at which single-bunch instability is expected to set in with Tau-Charm beam parameters for a luminosity of $10^{35}$ cm$^{-2}$ s$^{-1}$. The evolution of the single-bunch vertical emittance is shown in Figure 2.8.1 for different average cloud densities. These preliminary simulations are performed with the so-called *continuous focusing* approach. In this approach, the ring is modeled with 100 elements and thus there are 100 beam-cloud interactions per turn, and with constant average beta functions. The Left side of Figure shows the threshold for single-bunch instability between 2-3 × 10$^{12}$ electrons/m$^3$. The right side of Figure 2.8.1, shows the typical linear (incoherent) emittance growth which is persistent even at low electron cloud densities, lower than the instability threshold. Considering the average cloud density shown in the of Figure of 6 × 10$^{11}$ e/m$^3$, for which the emittance growth is about 0.25% in 1000 turns, and linearly extrapolating to 10 vertical damping times (440,000 turns), the emittance would increase by 100%. Similarly for the case of an average cloud density of 1 × 10$^{11}$ e/m$^3$, the emittance would increase by 8.5%. Radiation damping though should partially compensate the linear emittance increase, and remains to be evaluated.

These considerations should be taken into account together with estimates of the electron cloud build-up to characterize the maximum allowed chamber's surface secondary electron yield. Furthermore, technical mitigations should aim at reducing the cloud density at the lowest possible value.

Single-bunch simulations with a realistic lattice should be performed next. Furthermore, studies are needed to fully characterize the effect taking into account the photoelectron distribution and electron cloud build-up.

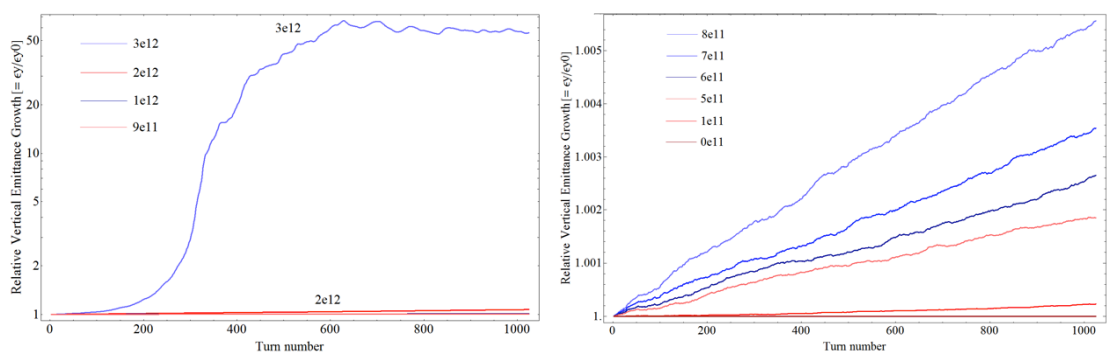

**Figure 2.8.1 - Relative vertical emittance growth as a function of electron cloud densities in units of e/m³. (Left) Threshold for single-bunch instability between 2 - 3 × 10¹² e/m³. (Right) Linear (incoherent) emittance growth below the instability threshold.**

### Mitigations Plan

Several high energy physics laboratories around the world joined a ten-year long effort to develop mitigation techniques to overcome the electron cloud effect.

In the B-Factories, the installation of solenoids and coating of the chamber walls was required to suppress the build-up of the cloud. In the DAΦNE positron ring, the installation of clearing electrodes has been proven to be very beneficial to suppress the build-up and increase beam intensities [2.25]. Grooves have been proven to work efficiently in dedicated tests in PEP-II at SLAC, in KEK-B and in CesrTA. Amorphous Carbon coating has been proven to efficiently lower the surface secondary electron yield at CERN and in CesrTA. Studies of the secondary electron yield and surface morphology have been crucial to identify proper materials and coatings to counteract the electron cloud effect.

Some of the potential remedies developed during the last few years are shown in Figure 2.8.2. In particular, the copper electrodes inserted in all dipole and wiggler chambers of the DAΦNE collider is shown in the center Figure.

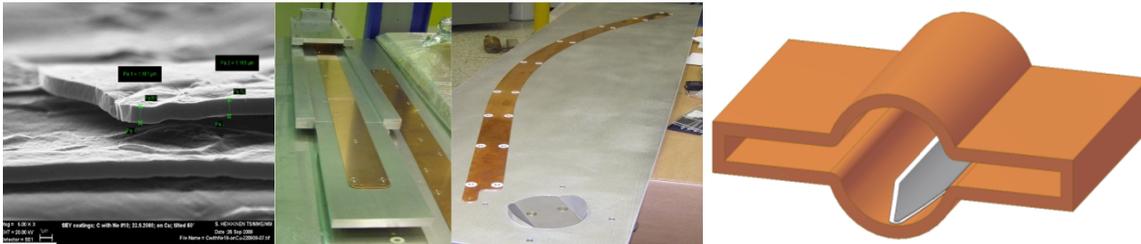

**Figure 2.8.2 - Amorphous carbon coating (left) at CERN, clearing electrodes about 1.5 m long before installation in the DAΦNE collider INFN LNF (center) and 3D model of thermal spray tungsten on alumina insulator clearing electrodes (right) for the ILC DR wiggler chambers.**

The mitigation plan adopted by the International Linear Collider (ILC) [2.26] and similarly by SuperKEKB is shown in Table 2.8.1. Baseline Mitigations I and II, presented in the Table, have been extensively tested and will be applied to the various regions of the positron storage and damping rings.

**Table 2.8.1 - Electron cloud mitigations plan to be adopted for the Tau/Charm collider, similarly to the plan for ILC**

|  | Drift | Dipole | Wiggler | Quadrupole |
|---|---|---|---|---|
| Baseline Mitigation I | TiN Coating | Grooves with TiN coating | Clearing Electrodes | TiN Coating |
| Baseline Mitigation II | Solenoid Windings | Antechamber | Antechamber |  |
| Alternate Mitigation | Amorphous Carbon or NEG coatings | TiN Coating | Grooves with TiN or Amorphous Carbon coating | Clearing Electrodes or Grooves |

To reduce the expected electron cloud effects in the Tau/Charm positron storage ring, we plan to adopt the mitigation techniques shown in Table 2.5.1 as adopted by future colliders. Future work will be addressed to further develop potential remediation techniques and to integrate mitigation techniques specifically into the Tau/Charm collider vacuum chambers.

### Conclusions

The electron cloud is a severe effect that might affect the luminosity reach of future colliders and it is expected to be an issue for the Tau/Charm. The electron cloud assessment by simulations for the Tau/Charm positron storage ring has started.

Mitigations developed during the last several years to counteract the electron cloud effect have been subject of extensive studies by several laboratories. In view of the expected electron cloud effect in the Tau/Charm, we plan to adopt the mitigation techniques that have been proven to be the most efficient protection against the electron cloud effect and that will be adopted by the next generation of particle accelerators. Future work should be addressed to advance the evaluation of the electron cloud effect and to further develop potential mitigations for the Tau/Charm.

## 3   Injection Complex

The Tau/Charm injection system delivers full energy, low emittance beams to the main rings. The two rings hereafter called Electron Ring (ER) and Positron Ring (PR) have the same energy E = 2 GeV. The present evaluation assumes that the maximum luminosity is achieved at 2 GeV and the maximum ring Energy is 2.3 GeV. The injection has been designed to be continuous in order to keep nearly constant beam current and luminosity.

### 3.1   General Layout

The preliminary layout of the injection system is shown in Figure 3.1.1. This layout is based on the design of the SuperB injection system, which is described in the following references [1.2, 3.1, 3.2]. A layout very similar to the SuperB one has been adopted in order to use the same design for the linac and damping ring lattice and very similar transfer lines. The main difference with respect to the SuperB design is the fact that only the positrons are stored in the Damping Ring (DR).

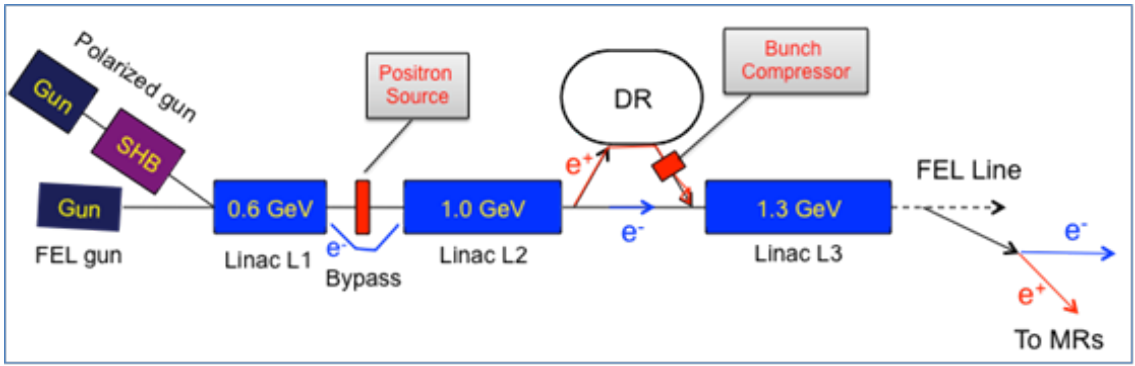

Figure 3.1.1 - Preliminary layout of the injection system.

The system consists of a polarized electron gun, a positron production system, electron and positron linac sections, a damping ring and the transfer lines connecting these systems to the collider main rings. The injection parameters are listed in Table 3.1.1. The charge per bunch per pulse required to replace the lost particles at the maximum luminosity is ~100 pC but the system is designed to provide 200 pC in order to have a good safety margin.

**Table 3.1.1- Injection Parameters**

|  | $e^-$ | $e^-$ |
|---|---|---|
| Max beam energy (GeV) | 2.3 | 2.3 |
| N particles/bunch @ L=2x10$^{35}$ cm$^{-2}$s$^{-1}$ | 3.2x10$^{10}$ | 3.2x10$^{10}$ |
| Number of bunches | 530 | 530 |
| Total beam lifetime (s) | 300 | 300 |
| Particles lost/beam/sec | 6x10$^{10}$ | 6x10$^{10}$ |
| N bunches per injection pulse | 4 | 4 |
| Injection repetition frequency (Hz) | 25 | 25 |
| $\Delta$N injected/bunch/pulse | 6.0x10$^{8}$ | 6.0x10$^{8}$ |
| Required injected charge/bunch/pulse (pC) | 96 | 96 |
| MAX injected charge/bunch/pulse (pC) | 200 | 200 |
| $\Delta L/L_{peak}$ (%) | 1.9 | |

The gun, similar to the one used by the SLC collider at SLAC [3.3], produces a short train of up to 4 bunches with up to 10nC charge and 85% polarization [3.4]. The small train of 4 bunches will be injected in each of the Main Rings with a repetition cycle of 40 ms for each beam.

For electron injection the electrons are accelerated through linacs L1, L2, L3 up to 2.3 GeV and then transported to the Electron Ring (ER) with a transfer line. A pulsed magnet is used to bypass the positron source.

For Positron Ring (PR) injection the electrons are accelerated up to 0.6 GeV in linac L1 and focused on a tungsten target to produce pairs $e^+e^-$ by bremsstrahlung with a conversion efficiency of about 3%. After the converter, the positrons are collected and accelerated up to 1.0 GeV in linac L2 and injected in the damping ring. Positron beams are stored in the damping ring for 40 ms, corresponding to 4.5 damping times, in order to reduce the beam emittance. At the damping ring exit a bunch compressor reduces the length of the positron bunches in order to accelerate them in the linac L3 without producing a large energy spread that would not be acceptable for main ring injection.

The three Linacs are based on S-band, SLAC type, accelerating sections operating at a repetition frequency of 100 Hz. The injection repetition cycle is 40 ms for each beam, corresponding to 25 Hz. The timing scheme allows to accelerate two beam pulses for a SASE FEL facility, during the store time of the positrons in the DR, without affecting the injection rate for the Tau/Charm. A possible design for an FEL facility [3.5] was proposed for the SuperB project, a similar proposal will be studied for the Tau/Charm. A sketch of the possible timing scheme is shown in Figure 3.1.2.

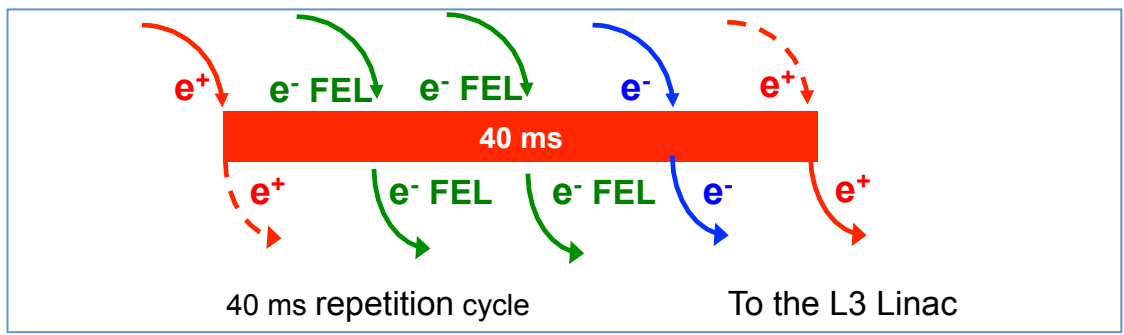

**Figure 3.1.2 Timing scheme of the beams accelerated in the Linacs.**

## 3.2 Positron Source

The positron production has been studied for SuperB [3.2, 3.6]. The positrons are created through a target downstream an electron drive beam, are then captured in an Adiabatic Matching Device (AMD) and accelerated by a capture section made of 4 accelerating cavities encapsulated in a solenoidal field.

The accelerating capture section takes the beam up to the energy of ~ 300 MeV. Then 4 quadrupoles are used to match the beam transverse phase space to the periodic focusing structure used for the following sections, which accelerate the beam up to 1 GeV. Two different lattices have been considered: a FODO cell, and a FDOFDO (doublets) cell. The Phase advance per cell is $\pi/2$ in both cases resulting in roughly the same period length (~ 4 m).

The positron yield at the end of the linac is reported in Figure 3.2.1 as a function of the energy of the drive beam. The yield is calculated for the positrons within the longitudinal and transverse DR acceptance. Both cells present roughly the same behaviour. The doublet solution is preferred since it allows using the same 3 m long accelerating sections used in the other linacs L1 and L3. The yield within the DR acceptance varies between 3% and 7% for electron energies between 0.6 and 1.5 GeV. We assume to have 10 nC/bunch from the SLAC type gun and a bunching efficiency of ~90%. We need to inject 96 pC/bunch to restore the particle losses and keep constant the average current but we require 200 pC from the injection system to shorten the injection time when injecting from zero current and to have some safety margin. Therefore a positron yield of ~2% satisfies the injection requirements. The choice of 0.6 GeV conversion energy gives a yield of 3% with further safety margin.

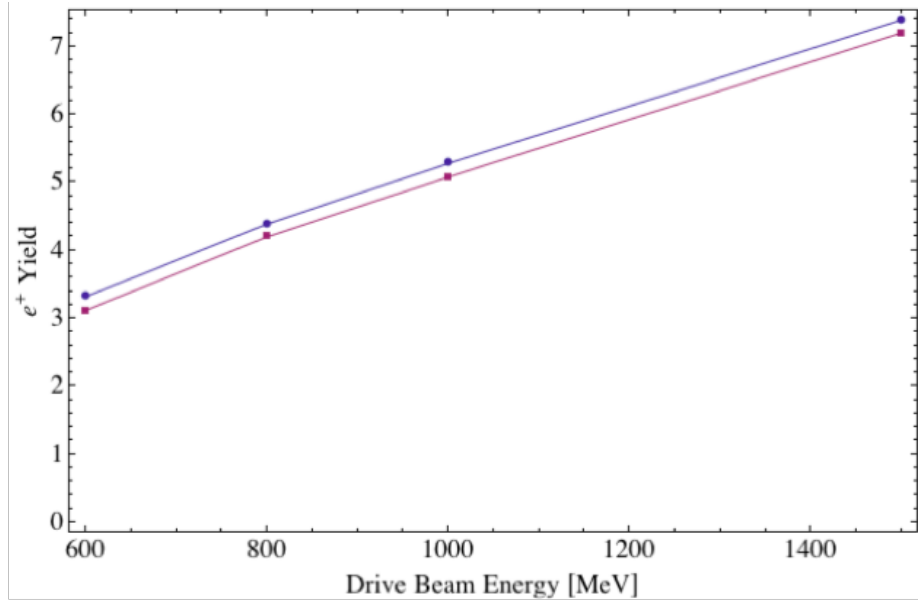

**Figure 3.2.1 - Yield of positrons within the longitudinal and transverse DR acceptance as a function of the drive beam energy for a FODO cell (red) and FDOFDO cell linac.**

### 3.3 Damping Ring

The parameters for the injected and extracted positron beams are listed in Table 3.3.1. The injection acceptance $A_x$ ($A_y$) is defined as the maximum betatron amplitude:

$$A_x = \gamma_x x^2 + 2\alpha_x x x' + \beta_x x'^2$$

where $\alpha_x$, $\beta_x$ and $\gamma_x$ are the Twiss parameters.

The emittance of the beams extracted from the DR is given by:

$$\varepsilon_{out} = \left[ \left( \varepsilon_{in} - \varepsilon_0 \right) e^{-2t/\tau} + \varepsilon_0 \right]$$

where $\varepsilon_{in}$ is the injected emittance, $\varepsilon_0$ the equilibrium emittance, $\tau_{x,y}$ the damping time and "t" is the storing time.

The DR design is the same as the SuperB one [3.7, 3.8] with a main modification: the energy has been lowered to 1.0 GeV and the dipole field has been reduced from 1.9 T down to 1.7 T to reduce the power consumption. As a result the transverse damping time is increased from 6.6 ms to 8.9 ms. The positron storing time in the DR has been increased to 40 ms, which corresponds to 4.5 damping times, in order to reduce the large emittance produced by the source down to the value required for injection into the MR. The equilibrium emittance is reduced from 30 nm down to 25 nm. The DR layout is shown in Figure 3.3.1 and the optical functions are shown in Figure 3.3.2. The DR parameters are listed in Table 3.3.2.

**Table 3.3.1 - Parameters of injected and extracted positron beams**

|  | Injection | Extraction |
|---|---|---|
| Energy (GeV) | 1.0 | |
| Number of bunches | 4 | |
| Charge/bunch (pC) | 200 | |
| Repetition frequency (Hz) | 25 | |
| Max betatron amplitude $A_x=A_y$ (m rad) | $1.0\times10^{-5}$ | - |
| Max energy error $\delta_{max}$ | ±1.5% | - |
| Horizontal emittance | $1.1\times10^{6}$ | $25\times10^{-9}$ |
| Vertical emittance | $1.1\times10^{6}$ | $0.67\times10^{-9}$ |
| Relative energy spread $\sigma_p$ | - | $5.8\times10^{4}$ |
| Bunch length $\sigma_z$ (mm) | - | 4.5 |

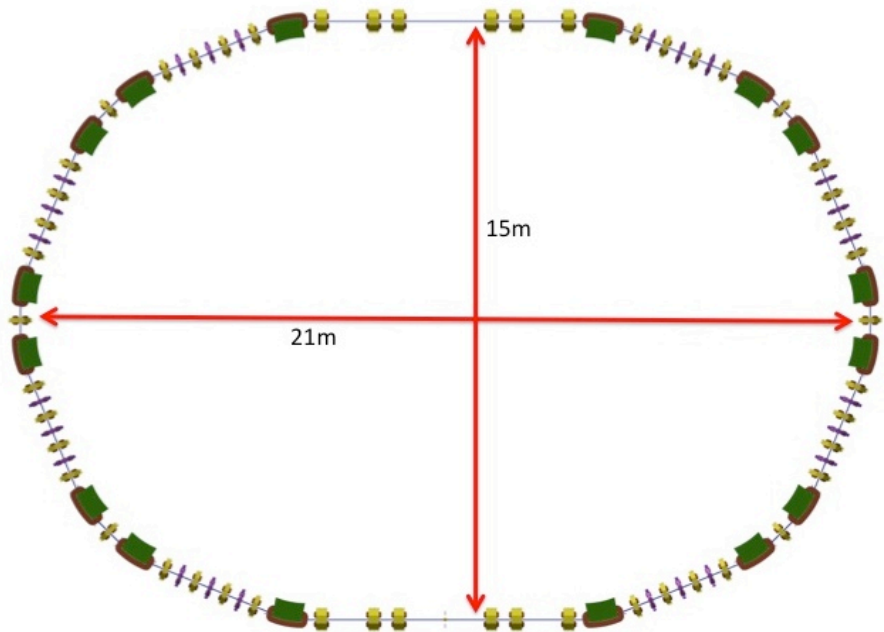

**Figure 3.3.1 – Damping ring layout.**

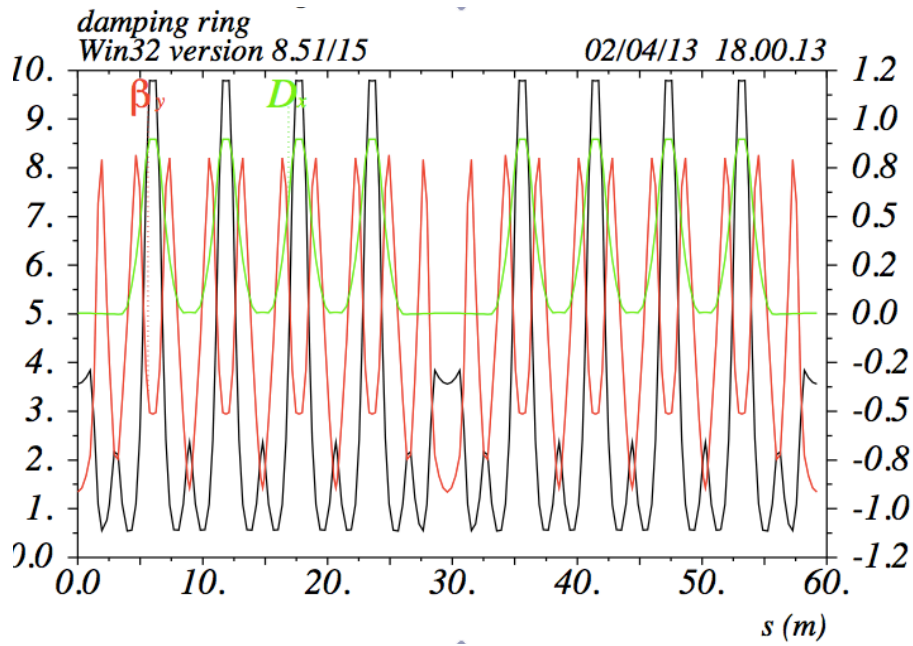

**Figure 3.3.2 – DR optical functions.**

**Table 3.3.2 – Damping Ring Parameters**

| | |
|---|---|
| **Energy (GeV)** | **1.0** |
| **Circumference (m)** | 59.2 |
| **Hor. Betatron tune** | 7.41 |
| **Ver. Betatron tune** | 2.71 |
| **Horizontal chromaticity** | -11.6 |
| **Vertical chromaticity** | -9.3 |
| **Horizontal emittance (nm rad)** | 24 |
| **Momentum compaction** | 0.0052 |
| **H/V damping time (ms)** | 8.9 |
| **Syn. Damping time (ms)** | 4.4 |
| **Energy loss/turn (MeV)** | 0.045 |
| **RF frequency (MHz)** | 476 |
| **Harmonic number** | 476 |
| **RF peak voltage (MV)** | 0.56 |
| **Bunch length (mm)** | 4.5 |

### 3.4 Linac Specifications

Linacs L1, L2 and L3 have same parameters as in SuperB, only the number of sections and klystrons is changed, since they scale with the respective energies. The Linac L3 provides, both to electron and positron beams, the final energy for the injection in the ER and PR of the collider. The maximum energy of the rings is 2.3 GeV. The linacs will also be used to accelerate high peak current, low emittance electron bunches for FEL experiments. The three operating modes are alternated every 10 msec with a repetition cycle of 40 msec. The parameters allow reaching the nominal positron beam energy of 2.3 GeV also in case of a klystron failure.

The 3 linac sections L1, L2, L3 are based on S-band technology. The accelerating structures are the 3 m, constant gradient, 2856 MHz units, known as SLAC-type sections, operating at 100 Hz. They are equipped with SLED systems.

The single RF modules consist of one klystron each feeding 3 sections as shown in Figure 3.4.1. This choice is a good compromise between the need to have a rather high accelerating gradient and to keep the number of the klystrons as low as possible.

The drive linac L1 is a high current 0.6 GeV machine to produce the electron beam for positron generation through the Tungsten target. In the electron mode, the positron converter is by-passed with a magnetic chicane. The successive L2 accelerator is a 1.0 GeV linac to inject the positrons in the damping ring. In Linac L3 focusing is done by a FODO cell with one quadrupole each two accelerating sections. One beam position monitor and one corrector each 4 sections are used to correct the orbit in the linac L3 [3.9, 3.10].

We identified the characteristics of the RF sources and network. The high power RF sources are 60 MW klystrons, supplied by solid state pulsed modulators, both commercially available from the industry. The RF power is transmitted to the accelerating structures with a network of rectangular, under-vacuum WR284 copper waveguides. The linac parameters are listed in Table 3.4.1.

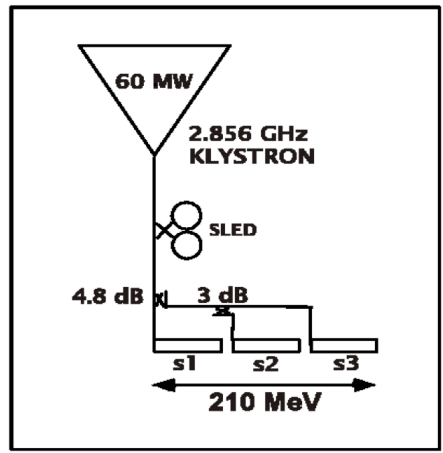

**Figure 3.4.1 – Layout of the RF Power Station.**

**Table 3.4.1 - S-band Linac Parameters**

| Section | L1 | L2 | L3 |
|---|---|---|---|
| Energy (GeV) | 0.63 | 1.26 | 1.47 |
| Repetition rate (pps) | 100 | 100 | 100 |
| Length (m) | 31.5 | 63 | 73.5 |
| Number of klystrons | 3 | 6 | 7 |
| Klystron peak power (MW) | 60 | 60 | 60 |
| Number of sections | 9 | 18 | 21 |
| Gradient (MV/m) | 23 | 23 | 23 |

### 3.5   Transfer Lines

The design of the transfer lines and bunch compressor will be based on the same criteria used for the SuperB project [1.2, 3.9, 3.10]. The TL layout for Tau/Charm is simpler and the lower energy reduces costs and power consumption of the magnetic elements (2.3 GeV instead of 4.2 GeV and 6.7 GeV). In particular the DR injection and extraction lines are simpler since they transport only the positron beam, there is no need for pulsed magnets and kickers to separate the beams. The electron beam goes straight from linac L2 to L3 and only a few quadrupoles are needed.

### 3.6   Injection into the Main Rings

Injection in the MR needs to have a very high efficiency, more than 99%. The injection efficiency strongly depends on the parameters of the injected beam and on the acceptance of the rings. At present the injection parameters have been evaluated assuming that the ring transverse acceptance is 20σx (in fact it is larger) and the injected beam energy spread is well within the ring energy acceptance. The required injected beam energy spread will be obtained by optimizing the bunch compressor parameters. The injection parameters for the positron ring are listed in Table 3.6.1.

**Table 3.6.1 – Injection into Positron Ring Parameters**

|  | PR |
|---|---|
| Septum thickness $\Delta$s (mm) | 3 |
| $\beta_{xs}$ (m) @ septum | 90 |
| $\beta_{xi}$ (m) @ injection line | 40 |
| $\beta_{xk}$ (m) @ kicker | 30 |
| Injected emittance $\varepsilon_{xi}$ (m rad) | 1.3E-08 |
| Stored emittance $\varepsilon_{xs}$ (m rad) | 6.0E-09 |
| Stored $\sigma_x$ (m) | 7.3E-04 |
| Injected $\sigma_x$ (m) | 7.1E-04 |
| $xmax_{inj}/\sigma_x$ | 13.9 |
| $xmax_{inj}$ (m) | 1.02E-02 |
| bsc/sx | 20 |
| $\theta$ kick (mrad) | 0.23 |

The kickers strength $\theta_{kick}$ is nearly the same as the DAΦNE kickers and therefore it is possible to use the same type of fast pulsers, allowing single bunch injection with a small, even negligible, perturbation of the neighbouring bunches.

The betatron oscillations of the injected beams (14 $\sigma_x$) are well within the ring acceptance. A simulation tracking the distribution of injected particles through the ring, taking into account the effect of the beam-beam kick and the machine errors and nonlinearities, will be performed to set the tolerances on the injection parameters.

PART 2 *Accelerator Systems*

# CONTENTS

# 4. Accelerator Systems

In this Part a description of the technical systems of the Accelerator complex will be given. Some of these systems are in a preliminary definition stage, are in a more mature stage, having profited of the work done for the SuperB Factory project in the past years.

## 4.1 Diagnostics

The beam diagnostics play a crucial role for the achievement of the nominal performances and establishing repeatable operation. Comprehensive characterization of relevant beam properties (see summary Table 4.1.1 below) must be provided in terms of:

- Beam position, by means of strip-line BPMs in the Linacs and transfer lines and button BPMs in the damping ring and main rings;
- Beam size/emittance, by means of fluorescent or OTR screens and synchrotron light monitors in the rings;
- Energy/Energy spread, exploiting the dispersive properties of the transfer lines and/or by special spectrometer magnets;
- Charge/current, by means of BPMs (sum mode), Toroidal Current Monitors, Faraday cups and photodiodes;
- Bunch length, by means of streak cameras on light from Cherenkov radiators or SR from bending magnets;
- Polarization, by means of Mott and Compton polarimeters;
- Luminosity;
- Beam losses.

**Table 4.1.1 - Summary of diagnostics**

|  | LINAC | Transfer Lines | Damping Ring | Main Rings |
|---|---|---|---|---|
| Beam Passage / Presence | Screen |  |  |  |
| Position / Closed Orbit | Striplines |  | Button BPM |  |
| Emittance | Screen |  | SR Monitor (Visible / Xray PinHole) |  |
| Energy / Energy Spread | Magnet + SEM Hodoscope |  |  |  |
| Charge / Current | Faraday Cup / WCM / BCT/ FCT |  | DCCT |  |
| Bunch by Bunch current |  |  | WCM / FCT / Fast Photodiode / BbB Fbk |  |
| Bunch Length |  |  | Streak Camera |  |
| Beam Size | Screen |  | SR Monitor |  |
| Coherent Beam Response |  |  | Tune Monitor / BbB Feedback |  |
| Incoherent Response |  |  | SR Monitor |  |
| Fast Loss | Long Ionization Chamber / Cherenkov Fiber |  |  |  |
| Slow Losses |  |  | Coincidence PIN Diode + Counter |  |
| Polarization |  | Mott | Compton |  |

We distinguish different phases of operation, namely:

a) commissioning
b) recovery after a major shut-down or an important hardware modification
c) machine studies
d) routine operation.

Phases a) and b) mostly affect the injection system, but not only. The diagnostic devices employed must allow single pass measurement. The beam charge is likely to be much smaller than the nominal one, and the beam behaviour is mostly dissimilar from what expected from a more or less precise machine model, for various flaws involving cabling, auxiliary systems, interlocks, computer control system etc. The diagnostic devices in first place must be able to record the beam presence or passage *anyhow*, even with invasive devices such as screens, radiators etc. General purpose laboratory instruments such as TV cameras, digitizers and signal analyzers are used to look at signals from various devices and pickups. This is quite demanding on the control system and network, which must be fully operational and flexible.

The absolute calibration and ultimate resolution of current and position measurement is not crucial at this stage; on the other hand, the sensitivity to very small beam current, even in the presence of electrical noise, is considered more important.

Even not being a beam diagnostic system in a strict sense, a valuable complement to the conventional beam instrumentation can be given by suitable beam loss monitors distributed in various places of the facility.

During phases c) and d), most value of the diagnostic systems is assigned to precise absolute calibration and finer resolution of beam position, size, current and frequency (tune) measurement, for which a deep integration of the various devices in the computer control system is necessary.

All relevant beam properties at the interface between different components of the injection chain must be fully characterized and compared. The following features are distinctive for the ultimate performances and affect the diagnostics specification:

- Short bunch length;
- Low-$\beta^*$ and *crab* optics;
- Very low x-y coupling;
- High colliding currents;
- Polarization;
- Continuous injection.

The bunch length is within the measurement capability of state of art commercial streak cameras.

Low $\beta^*$, crab optics and low coupling imply fine-tuning of the machine optics and tight control of the IP both in the transverse and longitudinal planes. The importance of having an accurate working model cannot be overstressed. The start point of modeling the optics is the beam-based alignment of BPM's and quadrupoles (it is implied that the BPM blocks are preferably integral with the quadrupole supports and that each quadrupole can be programmed separately). The model is iteratively derived mainly from (difference) orbit and tune measurements.

In first place the beam position monitor (BPM) system must have the smallest intrinsic absolute and relative accuracy, same for the tune monitor. The BPM system must allow the direct measurement of betatron phase advance between monitors in both planes, implying turn-by-turn time resolution. Turn-by-turn capability allows applying modern analysis methods for further refinement of the working model and measuring on-line the energy dispersion function. Furthermore, single turn capability can considerably shorten the run-in during phases a) and b).

The beam position measurement system is certainly the major system in terms of number of units, complexity and cost. The requirements are very challenging in terms of resolution, dynamic range, throughput of data and low latency. Commercial units suitable to our requirements are available and have become a de-facto standard in many low-emittance storage rings of synchrotron light facilities worldwide. If there is the possibility to form an internal team covering adequately all the skills of digital, RF and microwave electronics needed for such system, a measuring board can be developed in-house, but the risks connected are to be considered.

Regarding the small (vertical) emittance, experience at Diamond shows that a measurement station based on a X-ray pinhole camera is adequate if installed in a region with sufficiently high value of the vertical beta function. Another method with visible light, used at SLS and under further development, appears adequate and has to be considered [TIARA WP6]. The synchrotron radiation from a bending magnet is brought to an outside laboratory, possibly accessible during operation. It is highly desirable the addition of a beam line from a region with substantial value of the dispersion function to allow the observation of energy spread dependent beam size.

High currents imply thorough comprehension of the beam transverse and longitudinal dynamics, reliable operation and adequate margin of the feedback systems and RF servos. High current has implications in prompt protection of machine hardware from synchrotron radiation, HOM losses, and beam losses. Moreover, the risk to exceed safe radiation levels and to damage the hardware of the accelerator and of the detector is great: uncontrolled beam losses must be prevented and avoided as much as possible.

A beam abort system, based on a fast kicker must be provided to dump the beam in a controlled way at the occurrence of anomalies or beam misbehaviour. The kicker field must rise to a flat top in a time shorter than the ion-gap. In the beam abort line a spectrometer magnet can be added, allowing precise measurement of the incoherent energy spread.

The post mortem analysis can give important information, thus it is very important that all major accelerator systems provide adequate buffer memory of relevant waveforms, which can be stop-triggered by the abort trigger.

We will pursue bunch-by-bunch capability in the measurement of charge, lifetime, transverse and longitudinal beam size, luminosity, transverse and longitudinal displacement and tunes. All-digital approach is feasible using the Dimtel fast feedback systems IGp. A relevant amount of software for time-domain and frequency-domain analysis has already been developed. The use of a fast-gated camera for along-the-train transverse beam size observation can be very useful to characterize electron cloud effects.

Although "button" pickups are at all suitable for the orbit system, we propose to employ 50-Ohm back-terminated stripline monitors as pickups for the feedback systems. Such devices can provide strong signal, hence high S/N ratio, without reflections that can be a possible cause of unwanted bunch to bunch coupling and must be minimized.

Continuous injection at high efficiency is necessary to keep the average luminosity high. For this scope, the injection process must be continuously monitored and corrected for possible drifts. The bunch-by-bunch current is monitored by means of a fast photodiode with DC response, to get rid of the baseline shift associated with EM pickups. The transfer lines accommodate non-multiplexed single-pass BPM's after each bending magnet and quadrupole. Beam charge monitors of the toroidal type (position insensitive) will be used along the transfer lines to localize beam losses and to initiate all the corrective actions.

## 4.2  Feedbacks

Since 1992, PEP-II and DAΦNE bunch-by-bunch feedback systems have been developed within a large collaboration frame, sharing common approach and technologies and involving in common design and implementation also other circular lepton accelerators like ALS-Berkeley (USA), KEK (Japan), Bessy (Germany). The SuperB synchrotron and betatron bunch-by-bunch feedback systems have been designed keeping in mind the previous experience but also making necessary upgrades in terms of better signal resolution, larger dynamic range, most modern and powerful components and the necessary update of the software, firmware and gateware releases.

Looking to the Tau/Charm versus SuperB parameter table from feedback system point of view, it is possible to note a strong reduction of main ring lengths and of the harmonic number and, as consequence, the amount of stored bunches is also smaller. This consideration brings to a reduced need of separated real time processing channels to be implemented but, considering the present status of the digital technology, the advantage is basically negligible. Another important difference between Tau/Charm and SuperB colliders is the minimal distance between bunches: in the Tau/Charm it is 1/RF = 2.1ns while, previously, in the SuperB parameter list, it was the double (4.2ns) because not all the buckets were to be filled. This parameter, for the feedback systems, has no influence on the front end design and the signal processing units while it has a direct impact on the back end power section, mainly on the longitudinal and transverse kicker lengths. Indeed, to avoid cross-talk between adjacent bunches, it should be necessary to plan a complete redesign because both PEP-II and DAΦNE kickers seem to be too long for the Tau/Charm rings specification. From the same consideration, also the transverse and longitudinal power amplifier bandwidths should be carefully evaluated and probably the reusing of the PEP-II power amplifiers asked to SLAC could become more problematic or not useful: this point has to be checked. Apart from these considerations, the upgrade of the feedback systems designed for the SuperB Factory continues to be absolutely valid also for the Tau/Charm.

Going more in depth, the last version of the bunch-by-bunch feedback systems has been installed in the DAΦNE main rings to make real tests with beams and collisions. This choice have been done also considering that the design beam currents for DAΦNE and Tau/Charm main rings is about of the same order of magnitude, even if the emittance and the transverse beam sizes in DAΦNE are much bigger than for the Tau/Charm Factory. Presently DAΦNE is running with feedback versions that are exactly as foreseen for the Tau/Charm, both for synchrotron and for betatron motion, as shown in Figure 4.2.1. Furthermore with the current technology, the longitudinal and the transverse digital processing units are now identical while, in the past, they were strongly different. This fact will simplify strongly the system maintenance and the future software and firmware upgrades as well as the debugging phase.

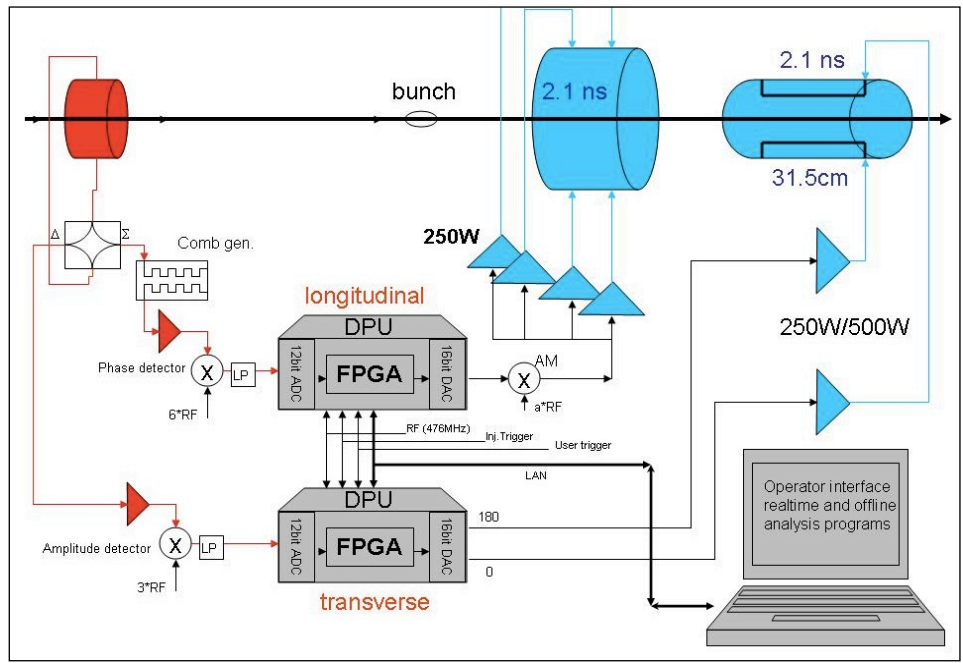

**Figure 4.2.1 – Tau/Charm bunch-by-bunch feedbacks are based on identical DPU (digital processing unit) for both longitudinal and transverse systems. The DPU core is implemented by a single powerful FPGA (field programmable gate array) containing >2000 DSP (digital signal processor).**

In particular four topics are currently under investigation for the feedback systems: the new digital processing units including the operator interface; an experimental low noise analog front-end implemented in vertical system; the synchrotron feedback analog back-end and a new horizontal kicker with enhanced performance. In synthesis the status of the tests is the following one.

Up to now, all the digital and software parts are running very well. In the analog to digital conversion (now 12 bits, in the past version, 8 bits), the better resolution achievable is a powerful feature. In particular with lower emittance rings and beam currents up to ~1A (in the current DAΦNE runs), the beams don't show any vertical enlargement due to the feedback gain and power. Nevertheless it will be necessary to continue the tests until the stored beam currents will increase to ~2A to confirm or not that the feedback resolution is good enough; otherwise we need to consider 14 or 16 bits conversion systems. In the Figure 4.2.2 it is possible to see a plot of the e-/e+ real time feedback-OFF/feedback-ON operation. These plots are automatically generated, after an operator request, by the new 12-bit feedbacks installed in DAΦNE main rings. This is a clear demonstration of the good behavior and performance of the new systems as well as of the perfect compatibility with the previous versions.

A low noise analog front-end has been assembled at LNF to be compatible with ultra-low emittance beam and it is now under test at DAΦNE. This system has been developed in collaboration with SuperKEKB feedback team. The main goal of the new design is to bring far from magnet fringe fields and from RF klystrons some parts of the system (mainly the "hybrids" making fast pulse difference and sum) and in the same time to detect bunch signals at higher frequency (4*RF for DAΦNE, 3*RF for Tau/Charm).

The third subsystem under investigation in DAΦNE during the current runs, is the analog back-end of the synchrotron feedback that has been simplified respect to the previous release. Indeed the present version implements only the signal amplitude modulation while, in the past, the system was based on two modulation techniques: amplitude and QPSK. This modification makes a design that is much easier to setup and to make in time, but that could also waste a small percentage of the power. Up to now, with the achieved beam currents during last DAΦNE runs, this approach has worked adequately, but it is necessary to increase the beam currents at least up to 2A for a complete evaluation and commissioning. It should be noted that in the past, beam currents up to 2.6 A were stored in the DAΦNE electron ring and 2-4 A in the PEP-II main rings.

A new feedback kicker having larger shunt impedance has just installed (April 2013) in the DAΦNE electron main ring and it is going to be tested in the next DAΦNE run (July/September 2013). The kicker is very similar to the horizontal kicker installed few years ago in the DAΦNE positron ring with excellent results, but it features also a new vacuum feed-through recently designed at LNF. The experience both done and in progress on this kicker can be very useful for the Tau/Charm kicker design.

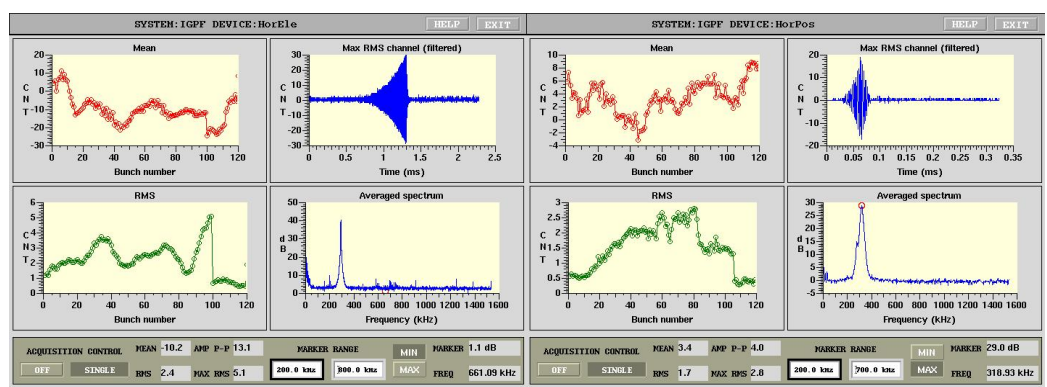

**Figure 4.2.2 – DAΦNE: e- (left) and e+ (right) real time feedback-off followed by feedback-on plots, that are automatically generated by the 12-bit feedback systems installed.**

## 4.3 Controls

An innovative accelerator like Tau/Charm requires a huge amount of effort devoted to the design, development and maintenance of the Computing Infrastructure and its Controls System (CS).

The Computing Infrastructure has to guarantee powerful tools for simulations and computations (beam dynamics, backgrounds, optics, magnet design, …) and services needed for the machine operation (storage, documentation and project management, security and access rights). In the same time the Controls System have to meet the innovative requirements coming from more, and more performing diagnostic devices and software trends, permitting intrinsic scalability, reliability, versatility and future portability.

Moreover, controls, computing infrastructure, as well as experiment, Synchrotron Light, and extracted beam information need to be integrated by common hardware & software tools, in order to guarantee the sharing of the information.

Here below are summarized the activities needed to build a computing infrastructure and control system for a new accelerator:

1. **Computing infrastructure:** design, develop and maintenance of:
   a. Electronics Management Data System (EMDS) dedicated to the storing and presentation of all (accelerators & experiment) project documents, CADs;

b. Project Management Data System (PDS) for the accelerator and experiment, in order to efficiently allocate and monitor efforts and costs;

c. common infrastructure and tools for the experiments in order to share and correlate data;

d. accelerator simulation code FARM/TIER2 share;

e. servers and services needed for the accelerator controls;

f. software tools (Mathlab, Matematica, LabView..).

2. **Software infrastructure, Control Systems:** design, develop and maintain of:

a. control system for the accelerator devices providing the possibility to integrate very fast data acquisition, interface with experiment data, electronic logbook, trouble ticket, high level software, simulation code interface;

b. controls system libraries, drivers, and interfaces of the accelerator devices;

c. user interfaces and high level accelerator software;

d. infrastructure to monitor accelerator subsystems device like PLC, field bus, etc..;

e. simulation code interface and controls systems in order to permit an easy and standardized data flow;

f. logbook and trouble ticketing system in order to monitor, store and perform statistics on accelerator devices and subsystems;

g. web services for public and private data presentation and correlation, online analysis, and monitoring.

3. **Users infrastructure, remote Control Room:** the infrastructure previously introduced (hardware and software) requires developing identification and **security tools** and the implementation of **collaborating tools** for the community participating to the project. In the mean time, the international community interested in the development of the accelerator, push also to foreseen a Remote Control Room in order to permit and guarantee participation in the operation and high efficiency in diagnostics and fault solution.

### !CHAOS introduction

The !CHAOS (Control System based on Highly Abstracted Operating Structure) is the proposed software infrastructure to realize the Control System (CS) of TauCharm. !CHAOS is an INFN project, the main goal is to create the framework and the services needed to build an efficient and scalable control system, mainly addressed to large experimental apparatus and particle accelerators. !CHAOS is under test at DAFNE and SPARC accelerators and has been developed to overcome the strong requirements throughput of new accelerators, like SuperB and TauCharm.

The !CHAOS general architecture (see Figure 4.3.1) consists in three development **frameworks** (a group of API – Application Program Interface) and two **services**. The schema below represents the whole !CHAOS structure. The dotted line represents the boundary between the core of the system and the specific implementation needed to integrate control (EU), User Interface (UI) and driver (CU) for a specific system/device/sensor.

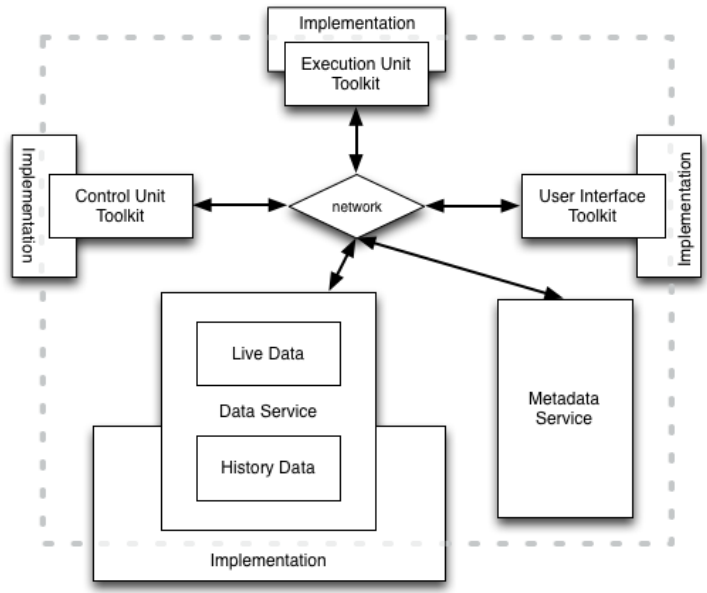

**Figure 4.3.1 – !CHAOS flowchart.**

The !CHAOS architecture relies on five **nodes**: Control Unit (CU), Execution Unit (EU), User Interface (UI), Metadata Services and Live (MS) and History Data Services (HDS).

The !CHAOS framework is composed by the following three nodes:

- The **Control Unit** (CU) toolkit abstracts the instrument's drivers. It consists in a set of API and C++ class that help the developer to realize a device driver and the hardware integration in the !CHAOS system.

- The **Execution Unit** (EU) toolkit abstracts the control's algorithms. It consists in a set of API and C++ class that help the developer to realize a general-purpose algorithm (math library, feedbacks, etc) specialized by setting algorithms parameters and input/output data. It can be used in two mode:
  o Collect data from device and push it on the data services (HDS);
  o Collect data from a device to control another device.

- The **User Interface** (UI) toolkit abstracts the user interface and connection with specialized Graphical User Interface (GUI). It consist in a set of API and C++ class that help the developer to realize the user interface for monitoring and for control the devices/systems/subsystems and/or general-purpose algorithms

The !CHAOS services rely on the following two nodes:

- The **Metadata Services** (MS) is the service that maintains the information about the state, the type and the structure of all nodes, it answers to search queries and to the management of the control algorithm and global management tasks.

- **Live Data and History Data Services** (HDS) services components are:
  o The Data Proxy that manages the insertion and the query for either live and history data;
  o The Indexer that takes care to apply the index rule (specified for the node that has generate the data) to the new archived data by the proxy;

o  The Storage management that takes care to remove the archived data that is no more needed according to ageing information (specified for the node that has generate the data).

The communication between nodes is performed in three different ways:

- **Event** is a lightweight data protocol in multicast UDP that is used to bring information about internal node event (heartbeat, fault detection, etc.) or to handle other general purpose data (locking, discover, load balance information, etc. etc.);
- **RPC** is used to call node API. This method permits to be sure that a called API can be executed by the node and permits to asynchronously receive an answer; this method is used for commands.
- **Direct Stream I/O** permits the fast transfer of data (packet or raw data) between two nodes; this methods is used for high throughput data transfer.

The Cabibbo Lab is fully involved in the development of the !CHAOS framework in prevision of the TauCharm Control System implementation. In the mean time the main part of this Control System architecture is under test by using existing accelerator facilities at LNF.

### 4.4  Vacuum System

The vacuum system is one of the key components in the Tau/Charm Factory. The performances of the injector depend strongly on the vacuum pressure. Extreme care must be then adopted during each step of design, construction and assembling of all the vacuum system. An accurate ultra high vacuum technology practice must be adopted during each step of the design of each part of the vacuum chamber, only all metal components and devices are permitted as well as only oil free vacuum pumping systems. Special care must be adopted for the design of the RF Gun vacuum system, because of the very high pollution sensitivity of the photo cathode.

The mean vacuum working pressure for each subsystem is described in Table 4.4.1 and this distinction reflects the different vacuum levels and performances that are requested for each part of the injector.

**Table 3.4.1 - Injector Vacuum Requirements**

| SUBSYTEM | WORKING PRESSURE |
|---|---|
| LINAC | $1\times10^{-9}$ mbar (RF Gun $1\times10^{-10}$ mbar) |
| DAMPING RING | $1\times10^{-9}$ mbar |
| TRANSFER LINES | $1\times10^{-8}$ mbar |
| RADIO FREQUENCY WAVE GUIDES | $1\times10^{-8}$ mbar |

In particular the vacuum system of the LINAC can be divided mainly in three parts as shown in Table 4.4.2 below:

**Table 3.4.2 - LINAC Vacuum Requirements**

| Zone | Pressure |
|---|---|
| RF Gun | $1\times10^{-10}$ mbar |
| LINAC | $1\times10^{-9}$ mbar |
| RF WAVE GUIDES | $1\times10^{-8}$ mbar |

For each module of LINAC was created a scheme of pumping system which consists of:

- 5 pumps on WG (50 l/s)
- 3 pumps on RF cavity (100 l/s)

- 2 Service manual valves
- 1 Vacuum gauge
- 1 Gate valve each 2 modules

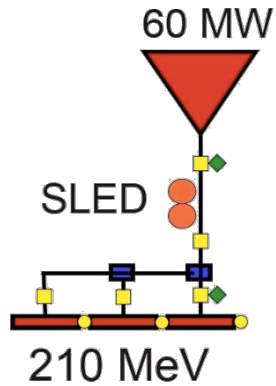

**Figure 4.4.3 – Scheme of Pumping System in the LINAC (yellow squares are the pumps).**

For what concern the Damping Ring, the vacuum specifications are set by the need to limit build up of electron cloud (e-cloud) and to avoid pressure instability in the ring. In addition, the vacuum system should deal with power and gas desorption due to synchrotron radiation. Modern vacuum technology can provide a wide variety of means for reaching the required gas density along the beam trajectory. In this case, many of the vacuum issues can be addressed using conventional technology. For the vacuum dimensioning we can start by dividing the ring in four similar quadrants by means of four gate valves. Each quadrant comprises four dipoles and some other magnets. The damping ring vacuum requirements can be satisfied putting a couple of ion pumps on each bending magnet, one on the RF cavity and two more pumps on injection and extraction straights, 9 gauges, 2 for each quadrant plus one on the RF cavity and 4 valves, as shown in the Figure 4.4.2 below.

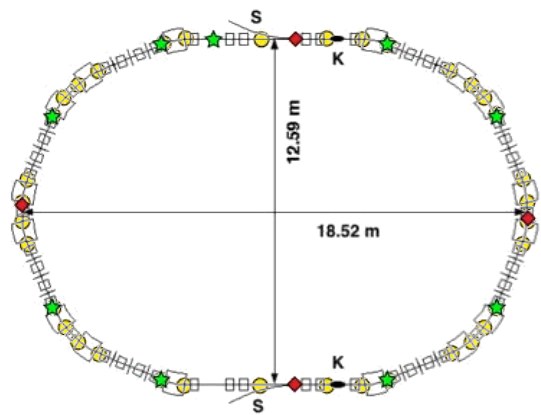

**Figure 4.4.2 – DR Vacuum Pumps Gate Valves and Gauges positioning.**

For the Transfer Lines the vacuum requirements are more relaxed and can be achieved in a simple way, as shown in Table 4.4.3 below.

**Table 4.4.3 – Transfer Lines requirements**

| TL Pressure [mbar] | $1\times10^{-8}$ mbar |
|---|---|
| Vacuum Chamber Cross Section- Straight [mm] | 60 |
| Pumping Speed [l/s] every 8 m | 120 |

The Main Rings are the most critical part of the Tau/Charm accelerator complex. Their vacuum system requirement is very stringent, and several actions are foreseen to avoid ion effects (ion trapping, or fast ion instability) and to limit build up of electron cloud (e-cloud) and to avoid ion-induced pressure instability in the rings. Anyway, the very crucial point is the synchrotron radiation induced gas load (see Figure 4.4.3). Table 4.4.4 below shows which are the main parameters and requirements.

**Table 4.4.4 – Values for Main Rings vacuum system**

| Parameters | |
|---|---|
| LUMINOSITY [ $cm^{-2}s^{-1}$ ] | $1\times10^{35}$ |
| Beam Energy [GeV] | 2.00 |
| Beam Current [A] | 1.7 |
| Circumference [m] | 340.70 |
| Total Photon Flux [photons/s] | $3\times10^{21}$ |
| Desorption Coefficient | $1\times10^{-6}$ |
| Total Gas Load [mbarl/s] | $1.2\times10^{-4}$ |
| Mean working pressure [mbar] | $1\times10^{-9}$ |
| Total Pumping Speed [l/s] | $1.2\times10^{5}$ |
| Net Linear Pumping Speed [l/sm] | 350 |

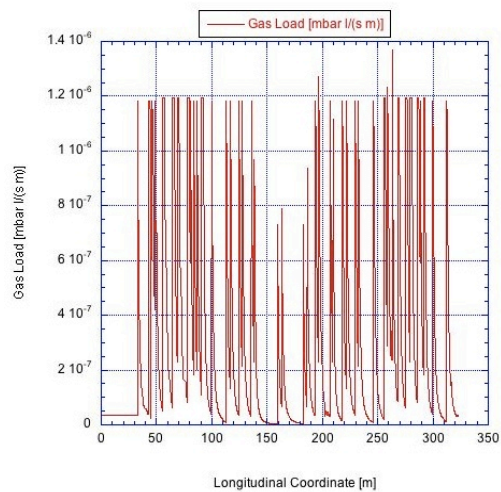

**Figure 4.4.3 - Gas Load Distribution Along the Machine.**

The main process of gas desorption in beam operation is photodesorption, i.e. emission of gas from the vacuum chamber surface caused by synchrotron radiation. In order to reach the required vacuum level in the beam chamber several approaches can be used, all together if possible. A possible and promising solution is a combination of distributed pumping system made up by strip-type non evaporable getter (NEG), that can be used irrespective of the presence of magnetic fields, lumped NEG pumps, sputter ion pumps and titanium sublimation pumps. Besides the active elements, the pumps, passive elements must be used too. Passive elements are: coatings (NEG, TiN or Graphite), special vacuum chamber geometry (antechamber design, synchrotron radiation aborbers) and special vacuum chamber surface machining (grooves).

Let us consider now the net Linear Pumping Speed needed for the Baseline Luminosity. The value shown in red in Table 4.4.4 is the net amount actually needed on the beam chamber, that is, - considering the conductance of RF screen, pump connection, slots and so on, - the installed gross pumping speed should be more or less twice that value. In other words, the gross installed pumping speed will be about 600 l/s/m. These values of pumping speed are very high if compared to that obtainable only with Distributed Ion Pumps (max 160 l/s/m gross pumping speed) and Distributed NEG Pumps (max 300 l/s/m gross pumping speed). The final design, that surely will make use of a suitable system of distributed pumps, must include a solution capable to manage very high gas loads. For this reason a deep investigation on a new and special custom-made vacuum system must be undertaken.

According to current information about the accelerator, a possible solution to fulfill the requirements of the vacuum system relies on what has been done in DAΦNE (considering the large amount of synchrotron radiation that is produced in the arcs of both machines), where synchrotron radiation absorbers, titanium sublimators (about 2000 l/s each) and sputter ion pumps were used. In Figure 4.4.4 is represented an example of bending magnet vacuum chamber where it should be possible to allocate the synchrotron radiation absorbers, the sputter ion pumps and the titanium sublimation pumps. The effectiveness of this solution should be confirmed by a more accurate work of research and development, strongly supported by prototype studying.

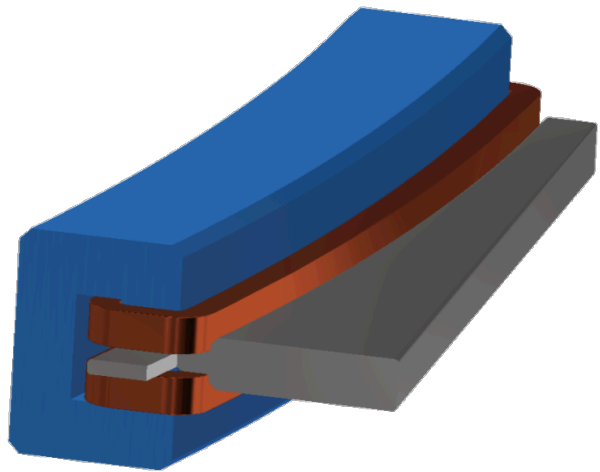

**Figure 4.4.4 – Example of bending magnet chamber.**

Regarding the choice of the vacuum chamber material there are two parameters to take in count for the selection: the synchrotron radiation power density and the secondary electron yield coefficient. In particular, Aluminum is a good candidate to satisfy the SR radiation power density, provided that a suitable water-cooling will be adopted. For the secondary electron yield coefficient, special surface coating, like NEG, TiN or graphite, will be adopted; in addition, the vacuum chamber surface will be machined with a special grooved shape, which acts as a trap for the secondary electrons.

The NEG coating, in some cases, could be used to reduce secondary electron yield, not as a pumping aid. Indeed, because of its limited absorption capacity, in a high gas load machine, this kind of NEG should need a reactivation process every few hours of operation. This seems not practical, considering that the reactivation process requires the heating of the vacuum chamber to a temperature of about 150°C.

In light of the information currently available, it is possible to define a first hypothesis for a vacuum system that could fulfill the requirements for the Baseline Luminosity case. However, this assumption must be confirmed by a careful investigation on mechanical constraint between magnets and pumping system. An intense activity of R&D must be started as soon as possible in order to test materials and solutions and to go ahead with the design of the vacuum system. The deep interaction with other systems (Beam Diagnostics, RF Feedback, Mechanics, etc.) is of crucial importance, as well as the necessary and careful evaluation of the Vacuum Chamber Impedance, to define the final design of the vacuum system.

### 4.5  Radio Frequency

For the Tau/Charm RF system we propose to re-use the main elements of the PEP-II RF system as klystrons, modulators and cavities [1-8]. The RF frequency is 476 MHz allowing a bunch distance of 2.1 ns with all the buckets filled. SLAC PEP-II RF operational experience shows that the power limit for each cavity window is 500 kW. Stable operational voltage in one cavity should be limited to 750-800 kV to avoid cavity arcs [9-12]. Parameters of a PEP-II cavity are shown in Table 4.5.1. A sketch of one PEP-II RF cavity is in Figure 4.5.1. Detailed information about calculated and measured parameters of the longitudinal and transverse modes of the PEP-II cavity is given in reference [1.1].

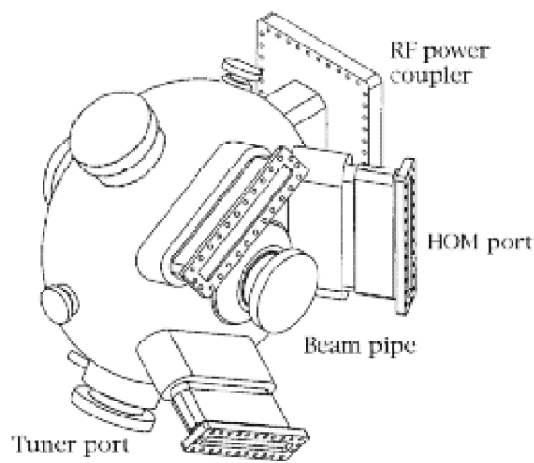

**Figure 4.5.1 – Sketch of PEP-II RF cavity.**

**Table 4.5.1 - PEP-II RF cavity parameters**

| Parameter | value | units |
|---|---|---|
| RF frequency | 476 | MHz |
| Shunt impedance | 3.8 | MOhm |
| Unloaded Q | 32000 | |
| R/Q | 118 | Ohm |
| Maximum incident power | 500 | kW |
| Maximum cavity voltage | 750-900 | kV |

The main parameters of the machine relevant for the RF system are shown in Table 4.5.2. For the nominal baseline configuration a beam current of 1.76 A is stored in 530 bunches. The value of the RF voltage is chosen in order to get a 5 mm bunch length at low current. The beam power reported below is the synchrotron radiation plus a very preliminary estimate of other possible sources of power losses (HOM). The power losses have been estimated by scaling the values calculated for SuperB.

For each ring, the beam power and the total RF voltage is shared among 3 cavities. These are located in the straight section opposite to the IP, roughly 40 m long, which can accommodate up to 8 cavities. Each klystron can feed 2 cavities, and therefore the minimum number of cavities needed to feed all the 6 cavities in both rings is 3. If we assume to have 4 klystrons (2 per ring) there is enough safety margin in case of failure.

For the luminosity upgrade (L = $2 \times 10^{35}$ cm$^{-2}$s$^{-1}$) we assume that the beam current can be increased up to 2.8 A and the bunch length (at low current) can be shortened down to 4 mm.

The parameters listed in Table 4.5.2 are evaluated at the nominal energy of 2.0 GeV per beam but the maximum ring energy is 2.3 GeV. The parameters of the RF system, see Table 4.5.3, have been optimized at 2.0 GeV; for the operation at the maximum energy it is assumed that the beam power is kept constant at 260 kW, and the voltage is below 2.4 MV (these are the values for the luminosity upgrade).

This is only a preliminary configuration of the RF system, for the technical design a full evaluation including the HOM power losses, the optimum coupling of the cavities and the effect of the transient due to the ion clearing gap and related frequency detuning is needed.

**Table 4.5.2 – Main Rings parameters relevant for the RF system**

| Parameter | Units | value | value |
|---|---|---|---|
| | | *Baseline* | *Luminosity upgrade* |
| Beam Energy | GeV | 2.0 | |
| Beam Current | A | 1.76 | 2.78 |
| Revolution frequency | kHz | 808 | |
| Bunch spacing | ns | 2.1 | |
| Harmonic number | | 541 | |
| Number of bunches | | 530 | |
| S.R. Energy loss per turn | MeV | 0.09 | |
| Momentum compaction | | $2.34 \times 10^{-3}$ | |
| Relative Energy spread | | $4.90 \times 10^{-4}$ | |
| Longitudinal damping time | ms | 30.9 | |

**Table 4.5.3 - RF System Parameters**

| Parameter | Unit | Baseline | Luminosity upgrade |
|---|---|---|---|
| Frequency | MHz | 476 | 476 |
| Total RF voltage | MV | 1.6 | 2.4 |
| Beam Current | A | 1.76 | 2.78 |
| S.R. Energy loss per turn $U_{SR}$ | kV | 90 | 90 |
| Bunch length (@low current) | mm | 5.0 | 4.0 |
| Ring loss factor $k$ | V/pC | 8 | 10 |
| Bunch charge | nC | 3.8 | 5.7 |
| Parasitic loss per turn $U_{par}$ | kV | 30.4 | 57 |
| Overvoltage factor | | 13.3 | 16.3 |
| Synchrotron frequency | KHz | 10.988 | 10.988 |
| Synchrotron tune | | 0.012 | 0.012 |
| Number of cavities/ring | | 3 | 3 |
| Cavity RF voltage | MV | 0.53 | 0.8 |
| Cavity RF power dissipation | kW | 37.0 | 84.2 |
| Total SR Beam power | kW | 160 | 260 |
| Total Parasitic Beam power | kW | 53.5 | 158.5 |
| Total RF power | kW | 324.5 | 671.1 |
| RF power/cavity | kW | 108.2 | 223.7 |
| Cavity input coupling factor | | 2.92 | 2.66 |
| Number of klystrons (2 rings) | | 2 | 2 |

## 4.6 Magnets

### 4.6.1 Damping Ring Magnets

A first preliminary design of the dipole, quadrupole and sextupole magnets of the DR has been made to verify their feasibility so that the layout of the ring and its main dimensions could be set up. The main parameter that guided the dimensioning of the magnets has been the current density that has been chosen at values that minimize the overall costs of the magnets (basically copper, iron and electric power). Particular attention has been paid to the dipole due to its high magnetic field (1.7 Tesla), even if the optimization of the field quality in the good field region will subject of future refinements. The simulations have been made in 2D, using the PoissonSuperfish code, version 7.17, from LANL. Table 4.6.1 lists the basic parameters of the dipole magnets, where also the electric parameters in the hypothesis that all the dipoles are series connected are reported. Figure 4.6.1 shows ½ of the dipole cross section and the magnetic field distribution (Poisson output).

**Table 4.6.1 - Damping Ring Dipole magnet main parameters**
**(Type of magnet: Curved, C shape, parallel ends, laminated (1-1.5 mm)-massive (t.b.c.))**

| Parameter | Units | |
|---|---|---|
| Nominal Energy | GeV | 1.0 |
| Nominal Mag. Field (@ pole center) | T | 1.7 |
| Bending Radius | m | 1.96 |
| Dipole number | | 16 |
| Gap (@ pole center) | m | 0.027 |
| Magnetic Length | m | 0.77 |
| Deflection angle | rad | 0.3927 |
| Ideal orbit sagitta | m | 0.03766 |
| Max. Iron Induction (Pole shoe) | T | 2.17 |
| Max. Iron Induction (Back Leg) | T | 1.7 |
| Pole/Gap ratio | | 4.74/4.15 |
| Pole width | m | 0.112/0.128 |
| Back leg width | m | 0.172 |
| Nominal Amper*turns/pole (@ 1.0 GeV) | A | 35010 |
| Conductor (Copper) | mm*mm | 8*8 |
| Conductor coolant hole | mm | Ø 5 |
| Number of turns | | 16(h)*18(w) |
| Nominal Current Density | A/mm$^2$ | 2.8 |
| Nominal Current (@ 1.0 GeV) | A | 121.6 |
| Magnet Resistance | Ω | 0.573 |
| Nominal Voltage per magnet | V | 69.6 |
| Nominal Power per magnet | kW | 8.47 |
| Total series voltage (no cable voltage drop)) | V | 1115 |
| Estimated cable voltage drop = 10% | V | 55.7 |
| Power Supply dc output voltage | V | 1170 |
| Power Supply dc output power | kW | 142.2 |
| Number of hydraulic circuit in parallel per coil | | 9 |
| Number of hydraulic circuit in parallel per magnet | | 18 |
| Temperature increase (max) | °C | 6 |
| Total Water Flow Rate | m$^3$/s | 0.000338 |

| Water speed | m/s | 0.96 |
|---|---|---|
| Pressure drop | Pa | 283250 |
| Yoke Weight per Magnet | kg | 1603 |
| Coil Weight per Magnet | kg | 538 |
| Total Weight of 1 Magnet (inc. ancillary) | kg | 2385 |
| Iron Longitudinal Mechanical Length | m | 0.738 |
| Overall Magnet Length | m | 1.078 |
| Overall Magnet Width | m | 0.613 |
| Overall Magnet Height | m | 0.743 |

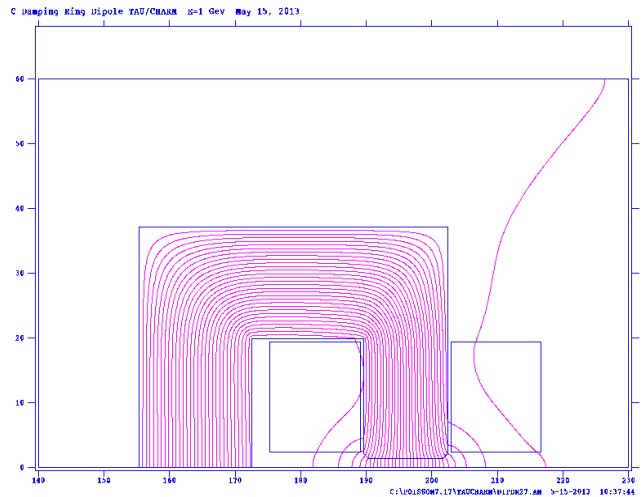

**Figure 4.6.1 - Dipole: ½ cross section and magnetic field distribution (Poisson output).**

The same considerations made for the dipole apply also for the quadrupole magnet and the sextupole magnet. Tables 4.6.2 and 4.6.3 list the quadrupole and sextupole basic parameters respectively and Figures 4.6.2 and 4.6.3 show the ¼ of the quadrupole cross section and 1/12 of the sextupole one.

**Table 4.6.2 - Damping Ring Quadrupole magnet main parameters**
**Type of magnet: Four Fold Symmetry - laminated (1-1.5 mm)**

| Parameter | Units | |
|---|---|---|
| Nominal Energy | GeV | 1.0 |
| Nominal Gradient | T/m | 20 |
| Quadrupole number | | 12/38[*] |
| Bore Radius | m | 0.035 |
| Magnetic Length | m | 0.30/0.15 |
| Max. Iron Induction | T | 1.6 |
| Pole width | m | 0.06/0.08 |
| Nominal Ampere*turns/pole (@ 20 T/m) | A | 10120 |
| Conductor (Copper) | mm*mm | 10*10 |
| Conductor coolant hole | mm | Ø 4 |
| Number of turns | | 30 |
| Nominal Current Density | A/mm$^2$ | 3.9 |
| Nominal Current (@ 20 T/m) | A | 337.4 |
| Magnet Resistance | mΩ | 23.13/15.66 |
| Nominal Voltage per magnet | V | 7.8/5.29 |
| Nominal Power per magnet | kW | 2.631/1.782 |
| Number of hydraulic circuit in parallel per coil | | 1 |
| Number of hydraulic circuit in parallel per magnet | | 4 |
| Temperature increase (max) | °C | 9/5 |
| Total Water Flow Rate | m³/s | 7*10⁻⁵/8.5*10⁻⁵ |
| Water speed | m/s | 1,4/1,7 |
| Pressure drop | Pa | 260700/250000 |
| Yoke Weight per Magnet | kg | 265/118 |
| Coil Weight per Magnet | kg | 90.4/60.9 |
| Total Weight of 1 Magnet (inc. ancillary) | kg | 373/188 |
| Iron Longitudinal Mechanical Length | m | 0.27/0.12 |
| Overall Magnet Length | m | 0.374/0.224 |
| Overall Magnet Width | m | 0.52/0.52 |
| Overall Magnet Height | m | 0.52/0.52 |

(*)(Low Gradient Quads have been assumed to have same gradient but half magnetic length)

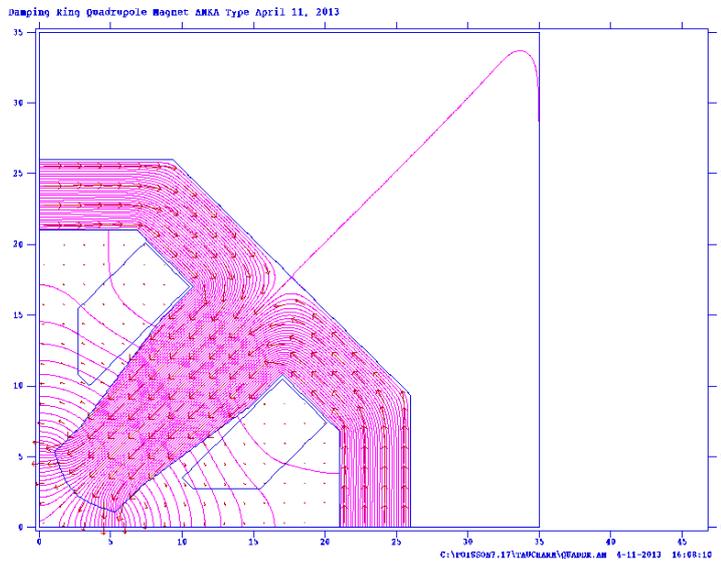

**Figure 4.6.2 - Quadrupole: 1/4 cross section and magnetic field distribution (Poisson output).**

**Table 4.6.3 - Damping Ring Sextupole magnet main parameters**
**Type of magnet: Six Fold Symmetry - laminated (1-1.5 mm)**

| Parameter | Units | |
|---|---|---|
| Nominal Energy | GeV | 1.0 |
| Nominal Gradient | T/m² | 154 |
| Sextupole number | | 24 |
| Bore Radius | m | 0.035 |
| Magnetic Length | m | 0.1 |
| Max. Iron Induction | T | 0.45 |
| Pole width | m | 0.08 |
| Nominal Ampere*turns/pole (@ 154 T/m²) | A | 1763.5 |
| Conductor (Copper) | mm*mm | 7*7 |
| Conductor coolant hole | mm | Ø 3 |
| Number of turns | | 20 |
| Nominal Current Density | A/mm² | 2.15 |
| Nominal Current (@ 154 T/m²) | A | 88.2 |
| Magnet Resistance | mΩ | 14.2 |
| Nominal Voltage per magnet | V | 1.25 |
| Nominal Power per magnet | kW | 0.111 |
| Number of hydraulic circuit in parallel per coil | | 1 |
| Number of hydraulic circuit in parallel per magnet | | 1 |
| Temperature increase (max) | °C | 4 |
| Total Water Flow Rate | m$^3$/s | 0.66*10$^{-5}$ |
| Water speed | m/s | 0.94 |
| Pressure drop | Pa | 218000 |
| Yoke Weight per Magnet | kg | 87.7 |
| Coil Weight per Magnet | kg | 12.6 |
| Total Weight of 1 Magnet (inc. ancillary) | kg | 121 |
| Iron Longitudinal Mechanical Length | m | 0.075 |
| Overall Magnet Length | m | 0.125 |
| Overall Magnet Width | m | 0.58 |
| Overall Magnet Height | m | 0.58 |
| Number of families | | 2 |
| Number of magnets per family | | 12 |
| Nominal Voltage per family | V | 15 |
| Cable Voltage Drop (80%) | V | 12 |
| P.S. Output Voltage | V | 27 |
| P.S. Output Power | kW | 2.4 |

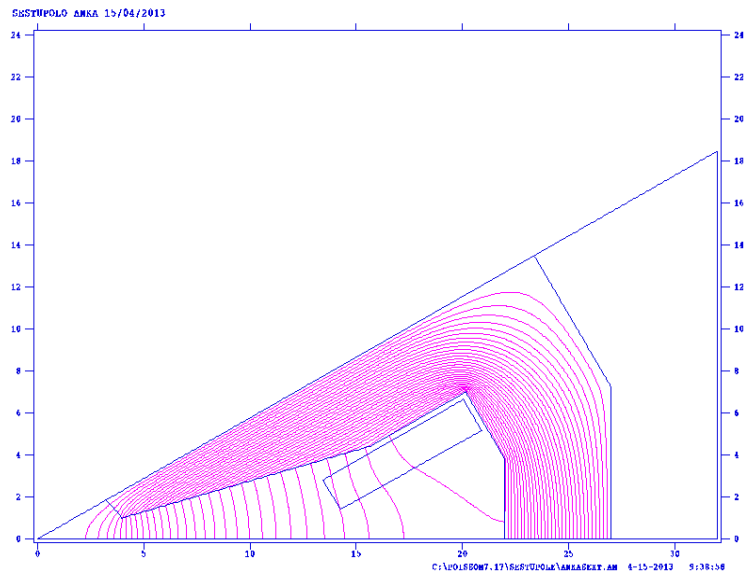

**Figure 4.6.3 - Damping Ring Sextupole: 1/12 cross section and magnetic field distribution (Poisson output).**

Figures 4.6.4 to 4.6.6 show the mechanical drawings for DR the Dipole, long and short Quadrupoles and Sextupole.

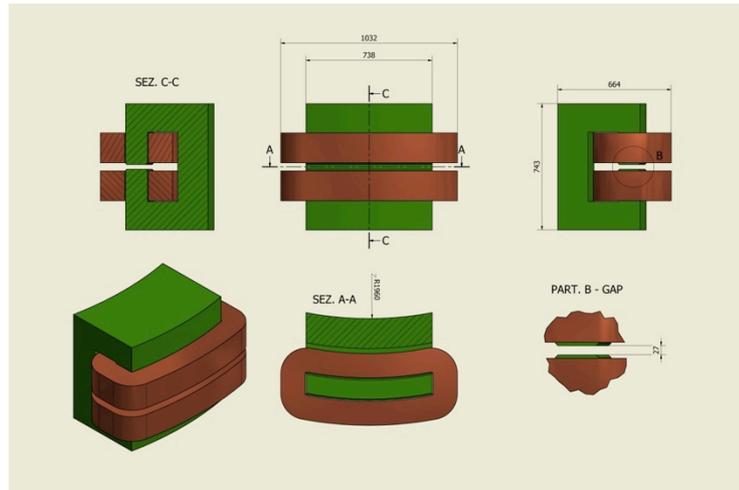

**Figure 4.6.4 - Dipole main dimensions.**

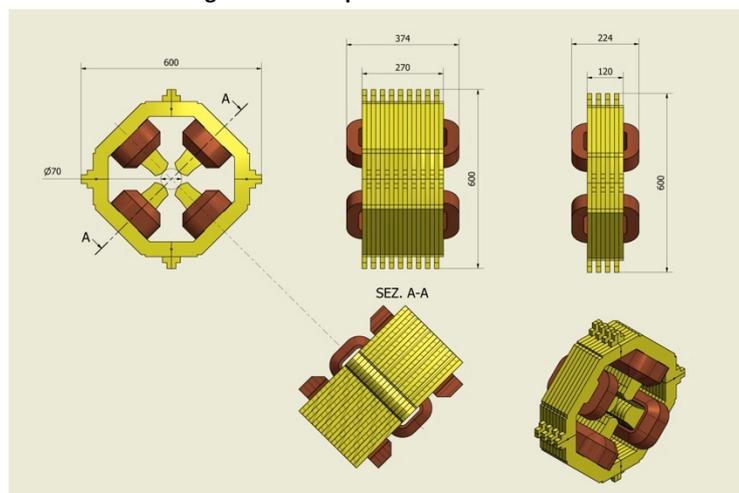

**Figure 4.6.5 – Long and short quadrupole main dimensions.**

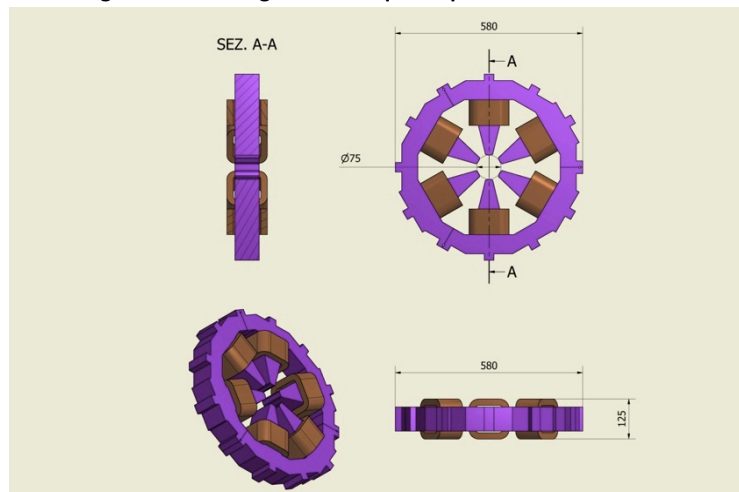

**Figure 4.6.6 - Sextupole main dimensions.**

### 4.6.2    Main Ring Magnets

The main topologies of the collider magnets can be summarized as follow:

❖ Bending Magnet – Field Index = 0
❖ Bending Magnet – High Field Index = 108-127

❖ Quadrupole Magnet
❖ Sextupole Magnet
❖ Octupole Magnet
❖ Horizontal and Vertical Steering
❖ Wiggler Magnets

The guideline in the design of these magnets has been the current density to be chosen in the dimensioning of the excitation coil. This parameter can be optimized according to three fundamental prices: 1) the copper cost; 2) the iron cost; 3) the energy cost. The last one is particularly important considering the lifetime of the accelerator, and related magnets, and is strongly dependent from country to country. The cost of the energy, including any tax and duty, for a research infrastructure like the Frascati National Laboratory of INFN, is near 0,2 €/kWh in the middle of the 2013. This price has been adopted in the calculations considering a functioning time of 50.000 hours. For the bending magnets the "optimum" current density value is around 2-2.5 A/mm$^2$.

Another consideration made in the magnet design has been the standardization, to avoid too much different types of magnets and, consequently, to reduce the construction costs reusing, where possible, the same stamping die, assembling tools, coil winding machine, etc.. The result has been that all the zero field index magnets will have the same cross-section, even if the lengths will be different according to the needs, the bending magnets with not-zero field index will have the same cross-section (and coil cross-section) but different pole profile. Finally, the quadrupole and sextupole magnets will have the same cross-section of the ones designed and described for the Damping Ring. Since the octupole and steering magnets have not been specified, their design will be done in a second phase of the project.

The data that follow come from simulations made with POISSON (Los Alamos) and FEMM42 (Aladdin Enterprise free license). These codes are bi-dimensional then the simulations are 2D. Common experienced rules have been adopted to determine the third dimension and calculate the related parameters but, obviously, simulations with 3D codes have to be done to get the appropriate design of the magnets. This part will be subject of future developments and when the lack of dedicated personnel will be solved.

### Zero field index bending magnets

A zero field index bending magnet is a bending magnet expected to have a flat magnetic field profile in the good field region. A total of 38 bending magnets, having different magnetic field, curvature radius and magnetic length are needed for each storage ring. Table 4.6.4 lists the relevant data concerning such kind of magnets. Table 4.6.5 shows the magnetic field harmonic content for each type of magnet. These figures can be ameliorated with an appropriate shimming of the pole profile, not done at the moment. Note that for BSB1 (2 units) no cooling is foreseen since the calculated power is very low and it seem reasonable to have a complete different magnet, probably air cooled.

**Table 4.6.4 - Zero field index bending magnet main parameters**
**Type of magnet: Curved, C shape, Parallel Ends, Laminated (1-1,5 mm), Straight**

| Parameter | Units | BSUP | BARC | B5/B4/B2 | B3 | B1 | BSB1 |
|---|---|---|---|---|---|---|---|
| Quantity | | 4 | 12 | 12 | 4 | 4 | 2 |
| Magnetic Field | T | 0.4456 | 0.4456 | 0.444 | 0.368 | 0.2772 | 0.013 |
| Bending Radius | m | 14.96 | 14.96 | 15 | 18.141 | 24.05 | 500 |
| Gap (@ pole center) | m | 0.063 | 0.063 | 0.063 | 0.063 | 0.063 | 0.063 |
| Field index | | 0 | 0 | 0 | 0 | 0 | 0 |
| Magnetic Length | m | 1.570 | 2.96 | 1.8 | 1.8 | 1.8 | 1 |
| Deflection angle | rad | 0.1050 | 0.1979 | 0.12 | 0.0992 | 0.07484 | 0.002 |
| Ideal orbit sagitta | m | 0.0206 | 0.0732 | 0.027 | 0.022 | 0.01684 | 0.00025 |
| Max. Iron Induction (back leg) | T | 0.64 | 0.64 | 0.64 | 0.55 | 0.4 | 0.02 |
| Pole/Gap ratio | | 2.54 | 2.54 | 2.54 | 2.54 | 2.54 | 2.54 |
| Pole width | m | 0.16 | 0.16 | 0.16 | 0.16 | 0.16 | 0.16 |
| Back leg width | m | 0.16 | 0.16 | 0.16 | 0.16 | 0.16 | 0.16 |
| Nominal Ampere*turns/pole (@ 2 GeV) | A | 11281.7 | 11281.7 | 11238.7 | 9303 | 7019 | 339,5 |
| Conductor (Copper) | mm$^2$ | 11.2*11.2 | 11.2*11.2 | 11.2*11.2 | 11.2*11.2 | 11.2*11.2 | 11.2*11.2 |
| Conductor coolant hole | mm | Ø 6 | Ø 6 | Ø 6 | Ø 6 | Ø 6 | Ø 6 |
| Number of turns | | 6(h)*8(w) | | | | | |
| Nominal Current Density | A/mm$^2$ | 2.44 | 2.44 | 2.43 | 2.01 | 1.52 | 0.07 |
| Nominal Current (@ 2 GeV) | A | 235 | 235 | 234.1 | 193.8 | 146.2 | 7.1 |
| Magnet Resistance | Ω | 0.067 | 0.117 | 0.075 | 0.075 | 0.075 | 0.046 |
| Nominal Voltage per magnet | V | 15.7 | 27.4 | 17.53 | 14.51 | 10.95 | 0.33 |
| Nominal Power per magnet | kW | 3.68 | 6.44 | 4.1 | 2.8 | 1.6 | 0.0023 |
| Number of hydraulic circuit in parallel per coil | | 2 | 2 | 2 | 2 | 2 | 2 |
| Number of hydraulic circuit in parallel per magnet | | 4 | 4 | 4 | 4 | 4 | 4 |
| Temperature increase (max) | °C | 8 | 18 | 10 | 6 | 4 | No cooling |
| Total Water Flow Rate | m$^3$/s | 0.00011 | 0.00009 | 0.00098 | 0.00011 | 0.0001 | No cooling |
| Water speed | m/s | 0.97 | 0.756 | 0.87 | 0.992 | 0.847 | No cooling |
| Pressure drop | Pa | 269750 | 303115 | 248120 | 313250 | 237850 | No cooling |
| Yoke Weight per Magnet | kg | 2518.63 | 4841.73 | 2903.03 | 2903.03 | 2903.03 | 1566 |
| Coil Weight per Magnet | kg | 306.6 | 536.1 | 344.6 | 344.6 | 344.6 | 212.5 |
| Total Weight of 1 Magnet (inc. ancillary) | kg | 3108 | 5916 | 3572 | 3572 | 3572 | 1956 |
| Overall Magnet Length | m | 1.7232 | 3.1132 | 1.9532 | 1.9532 | 1.9532 | 1.1532 |
| Overall Magnet Width | m | 0.51 | 0.51 | 0.51 | 0.51 | 0.51 | 0.51 |
| Overall Magnet Height | m | 0.602 | 0.602 | 0.602 | 0.602 | 0.602 | 0.602 |

**Table 4.6.5 - Magnetic field harmonic content**

---

**Magnet Type: BSUP**

**Magnetic Field Harmonic Analysis** @ = O (0.0;0.0) – Code: POISSON   Field coefficients

Normalization radius =  1.0 cm

Interpolation radius = 1.0 cm

$(Bx - iBy) = i[sum\ n*(An + iBn)/r * (z/r)**(n-1)]$

| n | n(An)/r | n(Bn)/r | Abs(n(Cn)/r) | Units |
|---|---------|---------|--------------|-------|
| 1 | 4.4570E+03 | 0.0000E+00 | 4.4570E+03 | Gauss |
| 2 | 1.1812E+00 | 0.0000E+00 | 1.1812E+00 | Gauss/cm |
| 3 | -2.2442E+00 | 0.0000E+00 | 2.2442E+00 | Gauss/cm$^2$ |
| 4 | 2.2361E+00 | 0.0000E+00 | 2.2361E+00 | Gauss/cm$^3$ |
| 5 | -2.6285E+00 | 0.0000E+00 | 2.6285E+00 | Gauss/cm$^4$ |

**Magnet Type: BARC**

**Magnetic Field Harmonic Analysis** @ = O (0.0;0.0) – Code: POISSON   Field coefficients

Normalization radius =  1.0 cm

Interpolation radius = 1.0 cm

$(Bx - iBy) = i[sum\ n*(An + iBn)/r * (z/r)**(n-1)]$

| n | n(An)/r | n(Bn)/r | Abs(n(Cn)/r) | Units |
|---|---------|---------|--------------|-------|
| 1 | 4.4570E+03 | 0.0000E+00 | 4.4570E+03 | Gauss |
| 2 | 1.1812E+00 | 0.0000E+00 | 1.1812E+00 | Gauss/cm |
| 3 | -2.2442E+00 | 0.0000E+00 | 2.2442E+00 | Gauss/cm$^2$ |
| 4 | 2.2361E+00 | 0.0000E+00 | 2.2361E+00 | Gauss/cm$^3$ |
| 5 | -2.6285E+00 | 0.0000E+00 | 2.6285E+00 | Gauss/cm$^4$ |

**Magnet Type: B2, B4, B5**

**Magnetic Field Harmonic Analysis** @ = O (0.0;0.0) – Code: POISSON   Field coefficients

Normalization radius =  1.0 cm

Interpolation radius = 1.0 cm

$(Bx - iBy) = i[sum\ n*(An + iBn)/r * (z/r)**(n-1)]$

| n | n(An)/r | n(Bn)/r | Abs(n(Cn)/r) | Units |
|---|---------|---------|--------------|-------|
| 1 | 4.4400E+03 | 0.0000E+00 | 4.4400E+03 | Gauss |
| 2 | 1.1763E+00 | 0.0000E+00 | 1.1763E+00 | Gauss/cm |
| 3 | -2.2357E+00 | 0.0000E+00 | 2.2357E+00 | Gauss/cm$^2$ |
| 4 | 2.2276E+00 | 0.0000E+00 | 2.2276E+00 | Gauss/cm$^3$ |
| 5 | -2.6185E+00 | 0.0000E+00 | 2.6185E+00 | Gauss/cm$^4$ |

**Magnet Type: B3**

**Magnetic Field Harmonic Analysis** @ = O (0.0;0.0) – Code: POISSON   Field coefficients

Normalization radius =  1.0 cm

Interpolation radius = 1.0 cm

$(Bx - iBy) = i[sum\ n*(An + iBn)/r * (z/r)**(n-1)]$

| n | n(An)/r | n(Bn)/r | Abs(n(Cn)/r) | Units |
|---|---------|---------|--------------|-------|
| 1 | 3.6755E+03 | 0.0000E+00 | 3.6755E+03 | Gauss |
| 2 | 9.5765E-01 | 0.0000E+00 | 9.5765E-01 | Gauss/cm |
| 3 | -1.8496E+00 | 0.0000E+00 | 1.8496E+00 | Gauss/cm$^2$ |
| 4 | 1.8441E+00 | 0.0000E+00 | 1.8441E+00 | Gauss/cm$^3$ |
| 5 | -2.1676E+00 | 0.0000E+00 | 2.1676E+00 | Gauss/cm$^4$ |

**Magnet Type: B1**

**Magnetic Field Harmonic Analysis** @ = O (0.0;0.0) – Code: POISSON   Field coefficients

Normalization radius =  1.0 cm

Interpolation radius = 1.0 cm

$(Bx - iBy) = i[sum\ n*(An + iBn)/r * (z/r)**(n-1)]$

| n | n(An)/r | n(Bn)/r | Abs(n(Cn)/r) | Units |
|---|---------|---------|--------------|-------|
| 1 | 2.7721E+03 | 0.0000E+00 | 2.7721E+03 | Gauss |
| 2 | 7.0118E-01 | 0.0000E+00 | 7.0118E-01 | Gauss/cm |
| 3 | -1.3945E+00 | 0.0000E+00 | 1.3945E+00 | Gauss/cm$^2$ |
| 4 | 1.3908E+00 | 0.0000E+00 | 1.3908E+00 | Gauss/cm$^3$ |
| 5 | -1.6347E+00 | 0.0000E+00 | 1.6347E+00 | Gauss/cm$^4$ |

| **Magnet Type: BSB1** | | | |
|---|---|---|---|
| **Magnetic Field Harmonic Analysis @ = O (0.0;0.0) – Code: POISSON   Field coefficients** | | | |
| Normalization radius =   1.0 cm | | | |
| Interpolation radius = 1.0 cm | | | |
| (Bx - iBy) = i[sum n*(An + iBn)/r * (z/r)**(n-1)] | | | |
| n    n(An)/r | n(Bn)/r | Abs(n(Cn)/r) | Units |
| 1    1.3343E+02 | 0.0000E+00 | 1.3343E+02 | Gauss |
| 2    3.3456E-02 | 0.0000E+00 | 3.3456E-02 | Gauss/cm |
| 3   -6.7523E-02 | 0.0000E+00 | 6.7523E-02 | Gauss/cm$^2$ |
| 4    6.6872E-02 | 0.0000E+00 | 6.6872E-02 | Gauss/cm$^3$ |
| 5   -7.8597E-02 | 0.0000E+00 | 7.8597E-02 | Gauss/cm$^4$ |

At this stage of the project all the magnets have the same iron and coil cross sections. Figure 4.6.7 shows the magnetic field profile on the mid-plane around the beam orbit of BSUP bending magnet. Figure 4.6.8 shows the magnetic flux line distribution in the cross-section of BSUP bending magnet. Similar figures, not reported here, apply to the other type of magnets.

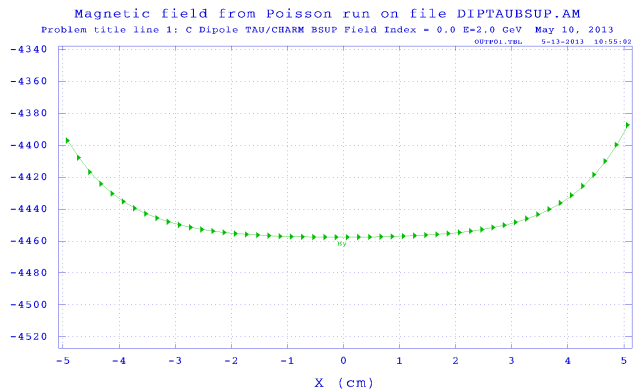

**Figure 4.6.7 - Magnetic field profile on the mid-plane, around the beam orbit, of BSUP bending magnet (POISSON).**

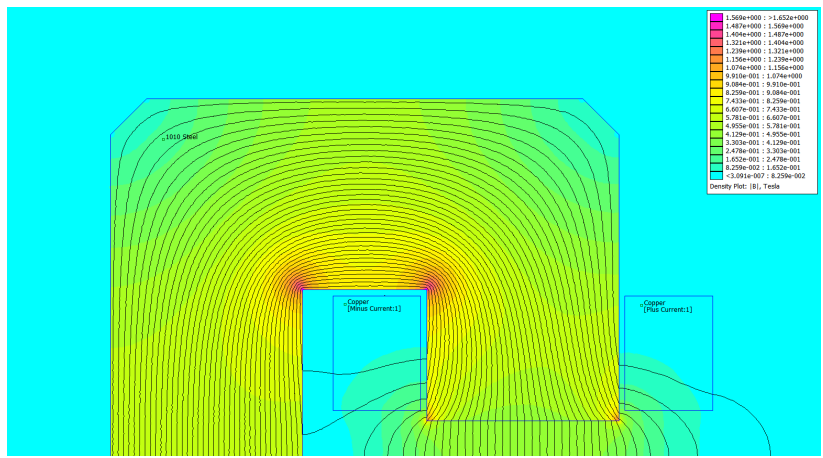

**Figure 4.6.8 - Flux line distribution in the BSUP bending magnet (FEMM42).**

## Bending magnets with non zero field index

A not zero field index bending magnet is a combined magnet where the bending function and the focusing (or defocusing) function are concentrated in one magnet. Unfortunately, the two magnetic functions are not separately settable. Two type of bending magnets, BQDM and BQDMA are requested with a strong magnetic field index. To fulfill this requirement, the pole profile has been designed following a suitable hyperbola, so that the required bending magnetic field and gradient can be achieved. Tables 4.6.6 and 4.6.7 list the basic parameters of BQDM and the magnetic field harmonic content around the beam orbit. Figures 4.6.9 and 4.6.10 show the geometry and the magnetic flux distribution and the magnetic field profile around the beam orbit on the magnet mid-plane respectively.

**Table 4.6.6 - BQDM bending magnet parameter list**
**Type of magnet: Curved magnet, C shape, parallel ends, laminated (1-1.5 mm)**

| Parameter | Units | |
|---|---|---|
| Nominal Energy | GeV | 2.0 |
| Nominal Mag. Field (@ pole center) | T | 0.4457 |
| Bending Radius | m | 14.955 |
| Dipole number | | 8 |
| Min/Max Gap | m | 0.063/0.473 |
| Gap (@ pole center) | m | 0.1112 |
| Magnetic Length | m | 1.68 |
| Deflection angle | rad | 0.112334 |
| Ideal orbit sagitta | m | 0.02359 |
| Field Index | | 126.74 |
| Gradient (@ pole center) | T/m | 3.77 |
| Max. Iron Induction (Back Leg) | T | 0.9 |
| Pole/Gap ratio (@ pole center) | | 1.44 |
| Pole width | m | 0.16 |
| Back leg width | m | 0.16 |
| Nominal Amper*turns/pole (@ 2.0 GeV) | A | 19663 |
| Conductor (Copper) | mm*mm | 11.2*11.2 |
| Conductor coolant hole | mm | Ø 6 |
| Number of turns | | 8(h)*12(w) |
| Nominal Current Density | A/mm$^2$ | 2.127 |
| Nominal Current (@ 2.0 GeV) | A | 204.8 |
| Magnet Resistance | Ω | 0.145 |
| Nominal Voltage per magnet | V | 29.7 |
| Nominal Power per magnet | kW | 6.09 |
| Total set voltage (no cable voltage drop)) | V | 237.8 |
| Estimated cable voltage drop = 10% | V | 23.8 |
| Power Supply dc output voltage | V | 261.6 |
| Power Supply dc output power | kW | 53.6 |
| Number of hydraulic circuit in parallel per coil | | 3 |
| Number of hydraulic circuit in parallel per magnet | | 6 |
| Temperature increase (max) | °C | 11 |
| Total Water Flow Rate | m$^3$/s | 0.000132 |
| Water speed | m/s | 0.78 |
| Pressure drop | Pa | 265900 |
| Yoke Weight per Magnet | kg | 3148 |
| Coil Weight per Magnet | kg | 702 |
| Total Weight of 1 Magnet (inc. ancillary) | kg | 4234 |
| Overall Magnet Length | m | 1.827 |
| Overall Magnet Width | m | 0.564 |
| Overall Magnet Height | m | 0.74 |

**Table 4.6.7 - Magnetic field harmonic content of BQDM bending magnet**

| **Magnetic Field Harmonic Analysis** @ = O (0.0;0.0) – Code: POISSON | | | Field coefficients |
|---|---|---|---|
| Normalization radius = 1.0 cm | | | |
| Interpolation radius = 1.0 cm | | | |
| (Bx - iBy) = i[sum n*(An + iBn)/r * (z/r)**(n-1)] | | | |
| n | n(An)/r | n(Bn)/r | Abs(n(Cn)/r) | Units |
| 1 | 4.4577E+03 | 0.0000E+00 | 4.4577E+03 | Gauss |
| 2 | -3.7760E+02 | 0.0000E+00 | 3.7760E+02 | Gauss/cm |
| 3 | 5.5058E+00 | 0.0000E+00 | 5.5058E+00 | Gauss/cm$^2$ |
| 4 | 2.7150E+00 | 0.0000E+00 | 2.7150E+00 | Gauss/cm$^3$ |
| 5 | -2.1722E+00 | 0.0000E+00 | 2.1722E+00 | Gauss/cm$^4$ |

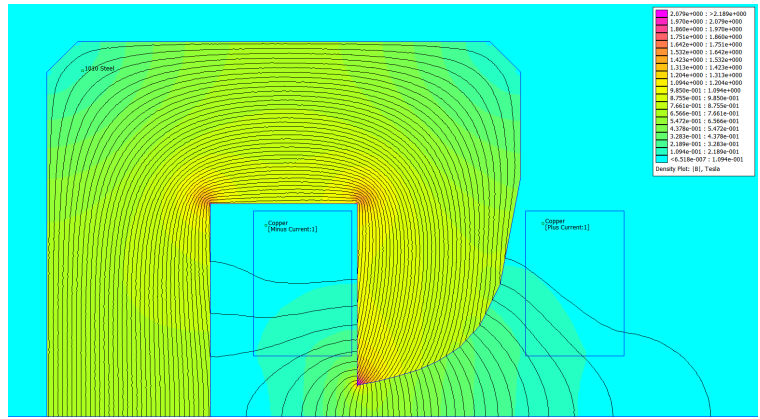

**Figure 4.6.9 - Magnetic flux distribution and geometry of BQDM (FEMM42).**

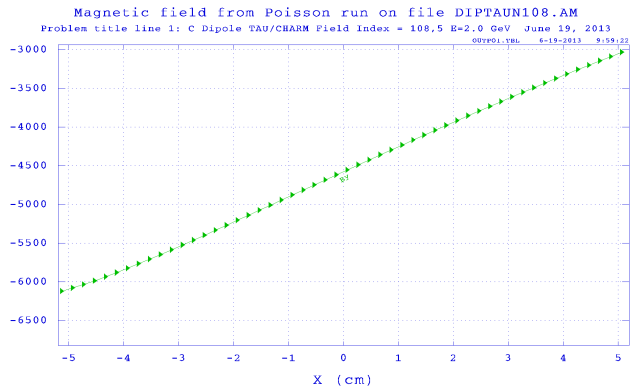

**Figure 4.6.10 - Magnetic field profile around the beam orbit on the magnet mid-plane of BQDMA (POISSON).**

## Quadrupole magnet

As said before, the same geometry and cross-section of the Damping Ring quadrupoles have been adopted also for the quadrupoles of the collider for standardization reasons. However, since the magnetic lengths are different and also the maximum gradient is different, 16 T/m for the collider against 20 T/m for the Damping Ring, a new parameter lists, shown in Table 4.6.8, has been calculated. Double data refer to the two magnetic lengths with the exception of the pole width, that is always the same, but that refers to the minimum and maximum pole width (see Figure 4.6.11). Table 4.6.9 reports the harmonic content of the magnetic field in the four-fold symmetry assumption. The IP quadrupoles QD0 and QF1 are not considered here.

**Table 4.6.8 - Quadrupole magnet parameter list**
**Type of magnet: Four Fold Symmetry - laminated (1-1.5 mm)**

| Parameter | Units | |
|---|---|---|
| Nominal Energy | GeV | 2.0 |
| Nominal Gradient | T/m | 16 |
| Quadrupole number | | 101/22 |
| Bore Radius | m | 0.035 |
| Magnetic Length | m | 0.3/0.5 |
| Max. Iron Induction | T | 1.1 |
| Pole width | m | 0.06/0.08 |
| Nominal Ampere*turns/pole (@ 16 T/m) | A | 7870 |
| Conductor (Copper) | mm*mm | 10*10 |
| Conductor coolant hole | mm | Ø 4 |
| Number of turns | | 30 |
| Nominal Current Density | A/mm$^2$ | 3.03 |
| Nominal Current (@ 16 T/m) | A | 262.4 |
| Magnet Resistance | mΩ | 23.13/33.1 |
| Nominal Voltage per magnet | V | 6.1/8.7 |
| Nominal Power per magnet | kW | 1.591/2.278 |
| Number of hydraulic circuit in parallel per coil | | 1 |
| Number of hydraulic circuit in parallel per magnet | | 4 |
| Temperature increase (max) | °C | 6/10 |
| Total Water Flow Rate | m$^3$/s | $6.4*10^{-5}/5.4*10^{-5}$ |
| Water speed | m/s | 1,3/1,1 |
| Pressure drop | Pa | 219900/241400 |
| Yoke Weight per Magnet | kg | 265/461 |
| Coil Weight per Magnet | kg | 90.4/125 |
| Total Weight of 1 Magnet (inc. ancillary) | kg | 373/616 |
| Iron Longitudinal Mechanical Length | m | 0.27/0.47 |
| Overall Magnet Length | m | 0.374/0.574 |
| Overall Magnet Width | m | 0.52/0.52 |
| Overall Magnet Height | m | 0.52/0.52 |

**Table 4.6.9 - Magnetic field harmonic content of the quadrupole magnet**

**Magnetic Field Harmonic Analysis** @ = O (0.0;0.0) – Code: POISSON   Field coefficients
Normalization radius =   1.0 cm
Interpolation radius = 1.0 cm
(Bx - iBy) = i[sum n*(An + iBn)/r * (z/r)**(n-1)]

| n | n(An)/r | n(Bn)/r | Abs(n(Cn)/r) | Units |
|---|---|---|---|---|
| 2 | 1.6008E+03 | 0.0000E+00 | 1.6008E+03 | Gauss/cm |
| 6 | -1.2947E-04 | 0.0000E+00 | 1.2947E-04 | Gauss/cm$^5$ |
| 10 | -1.4274E-05 | 0.0000E+00 | 1.4274E-05 | Gauss/cm$^9$ |
| 14 | -6.8930E-08 | 0.0000E+00 | 6.8930E-08 | Gauss/cm$^{13}$ |
| 18 | -3.7770E-10 | 0.0000E+00 | 3.7770E-10 | Gauss/cm$^{17}$ |

Figure 4.6.11 shows the gradient quality of the quadrupole on the mid-plane and Figure 4.6.12 shows the magnet geometry and the related magnetic flux distribution.

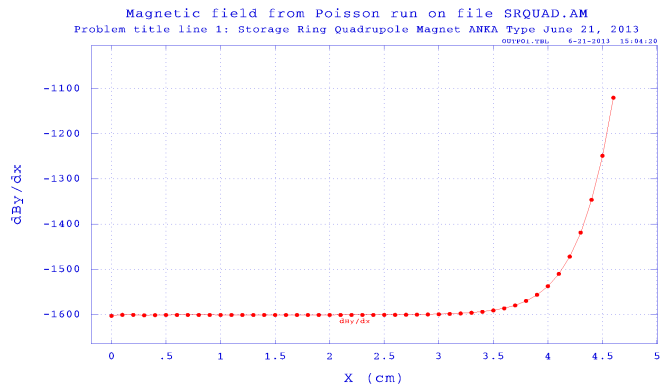

**Figure 4.6.11 - Gradient quality of the qudrupole magnet on the magnet mid-plane (POISSON).**

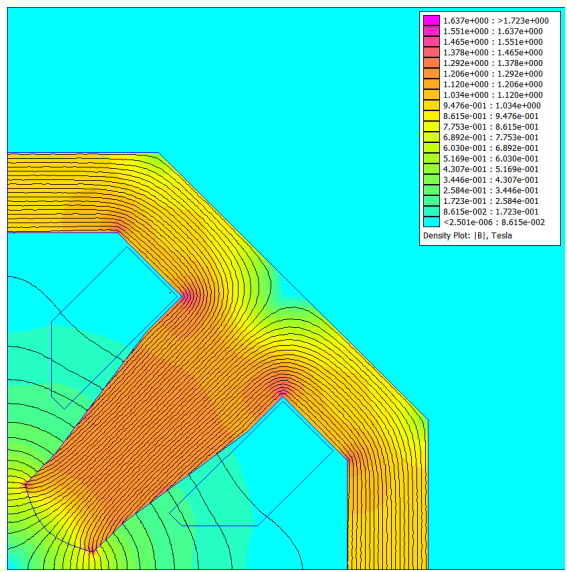

**Figure 4.6.12 - Geometry and magnetic flux distribution of the quadrupole magnet (FEMM42).**

## Sextupole magnet

Same considerations made for the quadrupole apply also to the sextupole magnet. Table 4.6.10 lists the sextupole parameter list and Table 4.6.11 the related magnetic field harmonic content. Figures 4.6.13 and 4.6.14 show the magnetic field profile on the mid-plane and 1/12 of the full geometry of the magnet.

**Table 4.6.10 - Sextupole magnet parameter list**
**Type of magnet: Six Fold Symmetry - laminated (1-1.5 mm)**

| Parameter | Units | |
|---|---|---|
| Nominal Energy | GeV | 2.0 |
| Nominal Gradient | T/m$^2$ | 195 |
| Sextupole number | | 38 |
| Bore Radius | m | 0.035 |
| Magnetic Length | m | 0.25 |
| Max. Iron Induction | T | 0.55 |
| Pole width | m | 0.08 |
| Nominal Ampere*turns/pole (@ 154 T/m$^2$) | A | 2230.1 |
| Conductor (Copper) | mm*mm | 7*7 |
| Conductor coolant hole | mm | Ø 3 |
| Number of turns | | 20 |
| Nominal Current Density | A/mm$^2$ | 2.72 |
| Nominal Current (@ 154 T/m$^2$) | A | 111.5 |
| Magnet Resistance | mΩ | 37.9 |
| Nominal Voltage per magnet | V | 4.23 |
| Nominal Power per magnet | kW | 0.471 |
| Number of hydraulic circuit in parallel per coil | | 1 |
| Number of hydraulic circuit in parallel per magnet | | 2 |
| Temperature increase (max) | °C | 9 |
| Total Water Flow Rate | m$^3$/s | 1.25*10$^{-5}$ |
| Water speed | m/s | 0.89 |
| Pressure drop | Pa | 265250 |
| Yoke Weight per Magnet | kg | 261.9 |
| Coil Weight per Magnet | kg | 33.2 |
| Total Weight of 1 Magnet (inc. ancillary) | kg | 354 |
| Iron Longitudinal Mechanical Length | m | 0.225 |
| Overall Magnet Length | m | 0.275 |
| Overall Magnet Width | m | 0.58 |
| Overall Magnet Height | m | 0.58 |

**Table 4.6.11 - Magnetic field harmonic content of the sextupole magnet**

**Magnetic Field Harmonic Analysis** @ = O (0.0;0.0) – Code: POISSON   Field coefficients
Normalization radius =  1.0 cm
Interpolation radius = 1.0 cm
(Bx - iBy) = i[sum n*(An + iBn)/r * (z/r)**(n-1)]

| n | n(An)/r | n(Bn)/r | Abs(n(Cn)/r) | Units |
|---|---|---|---|---|
| 3 | 1.9500E+02 | 0.0000E+00 | 1.9500E+02 | Gauss/cm$^2$ |
| 9 | 1.9685E-02 | 0.0000E+00 | 1.9685E-02 | Gauss/cm$^8$ |
| 15 | -2.3662E-01 | 0.0000E+00 | 2.3662E-01 | Gauss/cm$^{14}$ |
| 21 | 8.0547E-01 | 0.0000E+00 | 8.0547E-01 | Gauss/cm$^{20}$ |

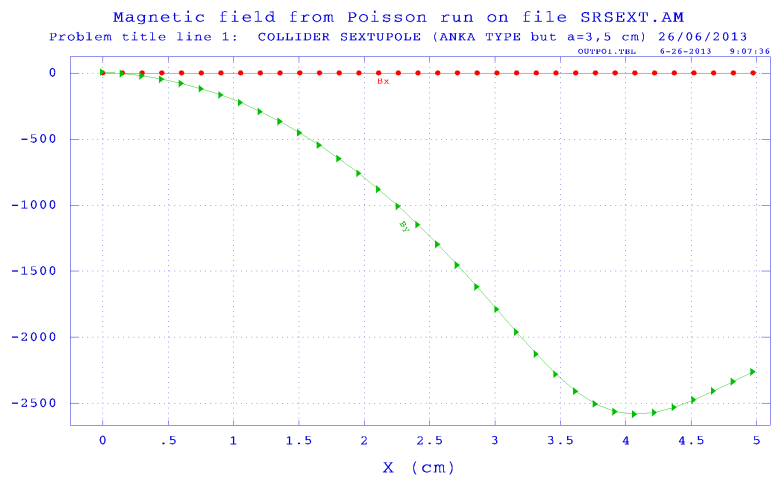

**Figure 4.6.13 - Magnetic field profile on the mid-plane of the sextupole magnet (POISSON).**

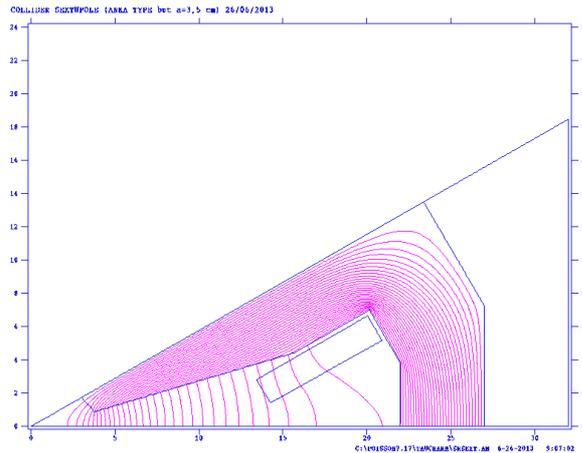

**Figure 4.6.14 - Geometry and magnetic flux distribution on 1/12 of the sextupole magnet (POISSON).**

### Wiggler magnet

For low energy running, at 2 GeV c.m. and below, it is foreseen to use 8 wiggler magnets to reduce the damping times and increase the emittance. The DAΦNE wiggler magnets are a good example of wigglers to re-use or to re-build. In this hypothesis Table 4.6.12 reports the parameter list of such a magnet. Figure 4.6.15 shows the magnetic field profile on the wiggler axis on the mid-plane and finally Figure 4.6.16 shows a picture of one of eight existing DAΦNE wigglers.

**Table 4.6.12 The DAΦNE wiggler magnet parameter list**

| Parameter | Units | |
|---|---|---|
| Nominal Magnetic Field | T | 1.8 |
| Wiggler number | | 8 |
| Nominal Gap (@ pole center) | mm | 42 |
| Nominal Gap (@ pole edge) | mm | 40 |
| Wiggler period length | mm | 640 |
| Number of period | | 3 |
| Number of full poles | | 5 |
| Number of half poles | | 2 |
| Wiggler length (incl. end clamps) | m | 2.098 |
| Nominal Ampere*turns/pole | A | 54000 |
| Number of turns/pole | | 80 |
| Conductor (Copper) | mm*mm | 7*7 |
| Conductor coolant hole | mm | Ø 4 |
| Nominal Current Density | A/mm$^2$ | 18.53 |
| Nominal Current (@ 2.0 GeV) | A | 675 |
| Nominal Power | kW | 254 |
| Water circuits/coil in parallel | | 5 |
| Water flow/circuit | l/min | 2.3 |
| Water flow/magnet | l/min | 161 |
| Pressure drop/circuit | atm | 4.5 |
| Water temp. rise | °C | 30 |

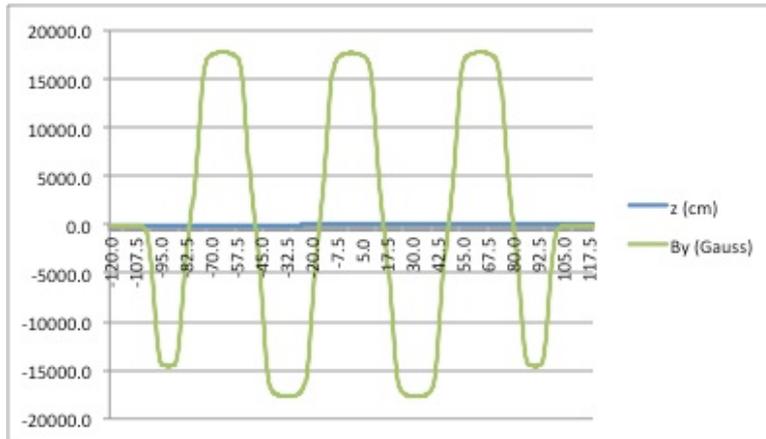

**Figure 4.6.15 - Magnetic field profile along the wiggler axis on the mid-plane.**

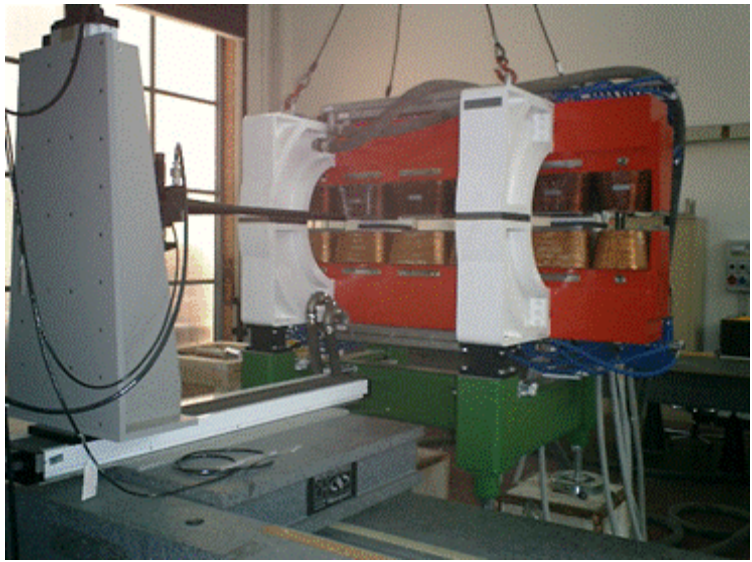

**Figure 4.6.16 - The DAΦNE wiggler magnet under measurement.**

## 4.7    Mechanical engineering

The Main Ring girders in an Arc cell will provide common mounting platforms for different sets of magnets, as shown in Figure 4.7.1. Multipolar magnets are mounted on girders number 2 and 4. Dipoles are installed on separate girders, numbered 1 and 3, because of their height difference and less stringent alignment and stability requirements. Gradient dipoles have the same stability requirements like multipolar magnets.

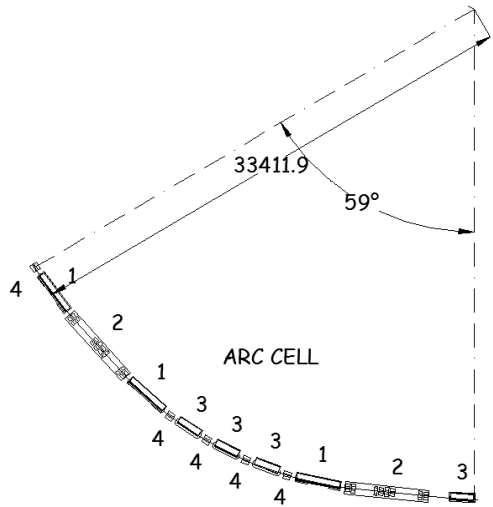

**Figure 4.7.1. – Main Rings Arc cell layout.**

General functional requirements of the magnet–girder support system are given as follows:

- Raise the centers of the magnets to the nominal beam height of 1.2 m. This height was chosen based on stability considerations and as usual choice at LNF. Give the beam height and tilt angle for the inclined positron ring with respect to the electron ring of about 11 mrad.
- Provide a stable platform for assembling and aligning the magnets outside the tunnel. The stringent alignment tolerances can only be met by precision alignment techniques requiring out-of-tunnel assembly and alignment. The magnet alignment must remain unperturbed during the transportation and installation of the magnet–girder assemblies in the tunnel.
- Meet girder-to-girder alignment requirements, both during the initial alignment and subsequently to compensate for long-term floor settlement.
- Meet dynamic stability requirements under expected ambient floor motion, flow-induced vibrations, and temperature fluctuations of the tunnel air and process water. In addition, the overall width of the magnet–girder support system must be less than 0.8 m, for ease of transportation and assembly in the tunnel. The support design must also be cost effective without sacrificing speed of installation and alignment.

### 4.7.1 Conceptual Design Features

In many recent synchrotron light sources the girders have been precisely fabricated with very stringent top surface tolerances (~15 µm flatness) and with T-slot type alignment features. Magnets, built with equally tight tolerances, are fastened directly to the girder's top surface without an interface of alignment hardware. After a careful examination of this approach and taking into account the experience done on DAΦNE, SPARC, CNAO and CTF3, it was decided to design Tau/Charm girders and magnets with conventional tolerances, and to use a vibrating-wire alignment technique for aligning the multipolar magnets up to about 30 µm precision. A typical girder with its mounting pedestals is shown in Figure 4.7.2. Girders are approximately 0.8 m wide and 0.8 m high. They are fabricated by welding commercially available plates of thicknesses ranging from 20 to 30 mm. After welding, the girders are stress-relieved by commercial thermal treatment. Girders are mounted on three pedestals that are grouted to the floor with non-shrinking epoxy grout. For mounting and height adjustment, 30 mm-diameter bolts with spherical washers are used. The girder will be over-constrained in order to minimize static deflection and raise the first natural frequency of the magnet–girder assembly.

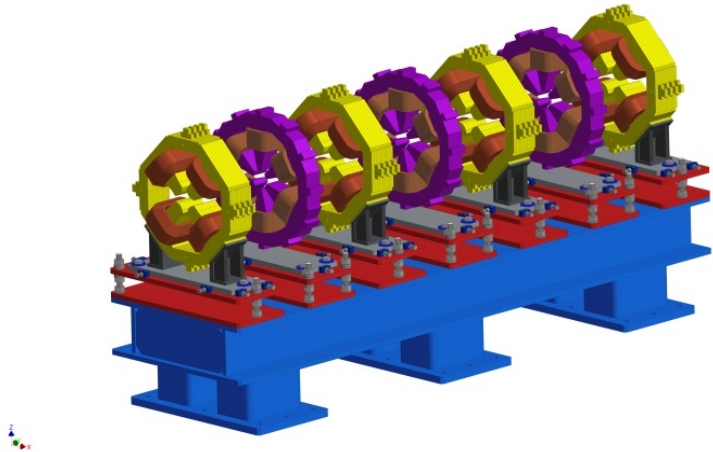

**Figure 4.7.2 – Example of girder with magnets.**

### 4.7.2    Magnet–Girder Assembly and Alignment

The Tau/Charm lattice magnets have magnetic alignment tolerances that exceed mechanical assembly tolerances and the capability of conventional alignment techniques to locate the magnetic components within the required tolerances. Therefore, a vibrating wire alignment technique, originally developed at Cornell University and subsequently adopted at SLAC and other place all around the world, will be adopted. It has been shown that this technique is capable of aligning magnets on the same girder to within 10 μm.

In this alignment technique, a current generator supplies an alternating current to the wire and the field of the magnet causes the wire to vibrate in case the wire is out of magnetic axis. Adjusting the magnet position can be found the true position where the wire does not vibrate and hence the wire is on the axis of the magnet. This standing waves on the wire are detected by optical wire sensors. In this way, the null center of the magnet can be located to within a few microns. Laser trackers are then used to transfer the position of the wire to the reference marker on the girder. Initially, the magnets will be installed and aligned on the girder with a laser tracker. The top-half of the multipolar magnets will then be split and the vacuum chamber will be installed. The ends of the chamber will be sealed with plastic caps. The caps will have small holes in either end to allow the ends of the vibrating wire to protrude through while a positive purge of dry nitrogen gas is maintained. A clean wire will be installed into the vacuum chamber prior to bake out and conditioning. Vibrating wire support brackets will be attached to either end of the magnet girder assembly and the wire will be secured to X–Y translation stages mounted on these brackets. The magnet is aligned by moving it to a "null" position that stops the wire from vibrating. The core of the magnet is then fastened to its support frame.

### 4.7.3    Installation of the Magnet–Girder Assembly

A transporter system with low pressure tires will be used to transport the girder–magnet assemblies from the alignment laboratory to the storage ring tunnel for final installation.

During the early phases of girder installation, the dipole girders will be installed first. The vibrating wire support brackets will be left attached for the entire installation process. An air pad system can be used for the final transportation stage inside the tunnel in order to damp any shocks to the girder and assuring accurate magnet location. The girder transporter will locate the girder assembly over the pedestal studs that will constraint the girder to the tunnel floor.

Instrumented torque wrenches will be used in conjunction with laser trackers to precisely offload the girder from the air pads onto the pedestals' studs. Once the girder is fixed to the floor, in-situ vibrating wire measurements will be repeated to confirm alignment of the magnets.

### 4.7.4    Mechanical Stability of the Magnet–Girder Support System

Noise sources that can influence the mechanical stability of the girder assembly are ground settlement, "cultural noise" floor motion, flow-induced vibrations, and thermal transients. These sources can be categorized in terms of the frequency range: fast when greater than a few Hz or slow when operating at frequencies lower than one Hz. Noise sources are also categorized based on the time-scale of the excitation, as being short (<1 hour), medium-term (<1 week), or long-term (>1 week). Short-term noise sources include natural and "cultural noise" vibrations, flow induced vibrations, and power supply jitters. Thermal transients due to temperature changes of the cooling water or the tunnel air, as well as gravitational and tide effects, constitute medium-term sources. Floor settlement or seasonal temperature changes, which may have direct impact on the alignment of components, are considered to be long-term effects. The high peak of "cultural noise" or human activity in the Tor Vergata area or LNF site is typically observed in the frequency range from 5 to 25 Hz. Ground motion from ocean waves is centered at about 0.2 Hz.

### 4.7.5    Short-Term Stability – Ambient Ground Motion Measurement

Two ground motion measurement campaigns were performed at LNF site in 2009-2010. The first campaign of ground motion measurements was performed on the Tor Vergata site on April 2011 with the collaboration of experts from the LAPP laboratory (Annecy France) and from CERN (Geneva, Switzerland). Seven different points have been measured in five days in order to characterize the site and to compare the influence of various vibration sources. The locations were referred to critical spots of the SuperB accelerator complex: IP (1-6), storage rings and spin rotators (4-7), electron source (2-5), SR laboratory (3). Short term measurements have been performed at points 1-2-3-5-6 while long term measurements were performed at points 4 and 7, see Figure 4.7.3.

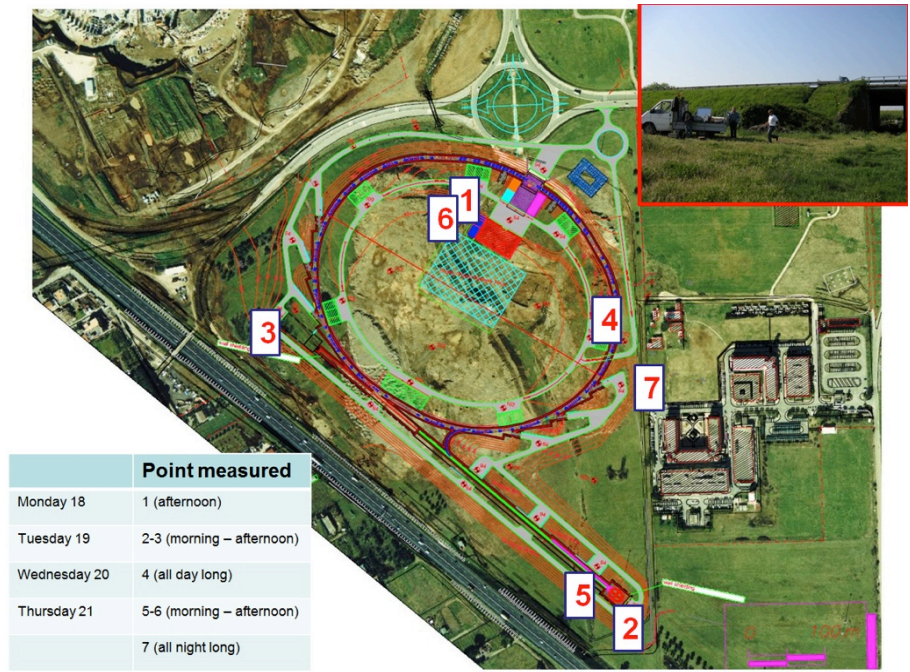

| | Point measured |
|---|---|
| Monday 18 | 1 (afternoon) |
| Tuesday 19 | 2-3 (morning – afternoon) |
| Wednesday 20 | 4 (all day long) |
| Thursday 21 | 5-6 (morning – afternoon) |
| | 7 (all night long) |

**Figure 4.7.3 - Vibrations measured points at Tor Vergata site.**

Ground Motion (GM) measurements have been performed in the frequency range of [0.1;100] Hz. Below 1 Hz, GM is due to earth motion, mostly to the micro seismic peak, while above 1 Hz GM is due to "cultural noise" due to human activities. However a beam-based feedback is usually used in accelerators to stabilize directly the beam below 0.1 Hz (and often at higher frequencies). Measurements showed that the amplitude of GM is very low above 100Hz, a level sufficient for the SuperB accelerator. In order to measure vertical GM in this wide frequency range, geophones (model Guralp CMG-40T from Guralp company) and accelerometers (model Endevco 86 from Brüel & Kjaer company) have been used. Point 2 was measured first, see Figure 4.7.3. Instruments were located at about 10 m from the highway and about 6 m below the highway asphalt floor. Figure 4.7.4 shows the power spectral density (PSD) measured at this point. The frequency range [5; 25Hz] of the high peak corresponds exactly to the traffic noise. The amplitude of PSD is almost the same versus time in the three directions. The corresponding integrated RMS of vertical GM has been calculated. The GM in the range [0.2; 100Hz] is almost the same as the data from 5Hz to 25Hz show. As a consequence, most of the noise is coming from the highway that can be considered a very high source of vibrations. The corresponding vertical displacement varies from 73 to 94 nm in the frequency range [1; 100Hz].

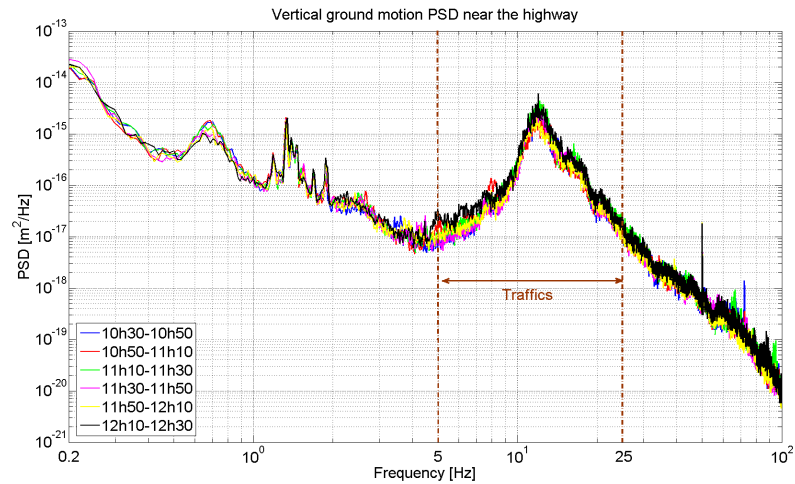

**Figure 4.7.4 - Vertical power spectral density at point 2.**

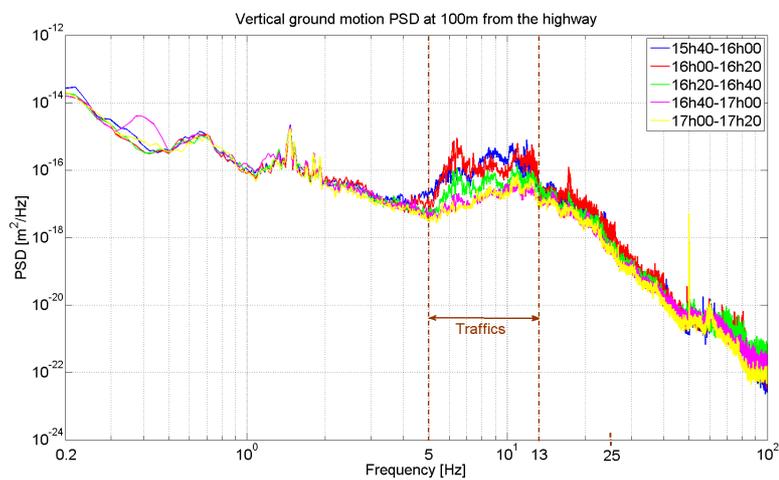

**Figure 4.7.5 - Vertical power spectral density at point 3.**

Even though the highway is very close to the Tor Vergata site, considering that a region of about 100m from the highway must be left free. At this latter distance the high vibration peak disappears (see Figure 4.7.5) and it turns out that the Tor Vergata site is acceptable from a vibrations point of view. Measurements were performed also at points 1 and 6 (see Figure 4.7.3) where the SuperB collider hall and final focus magnets was planned. In the vertical direction the amplitude and the frequency range of the traffic noise are very small, moreover the PSD amplitude does not change as a function of the time and day (see Figure 4.7.6). The amplitude of GM is very small on average and even in transient (sigma): around 20nm above 1Hz and 40nm above 0.2Hz. Points 4 and 7 are very close to each other. Point 4 was measured during the day while point 7 was measured during the night because of logistic reasons. Amplitude variations are small in average and transient between 10 nm and 30 nm above 1Hz in the three directions over the 24 hours data taking period.

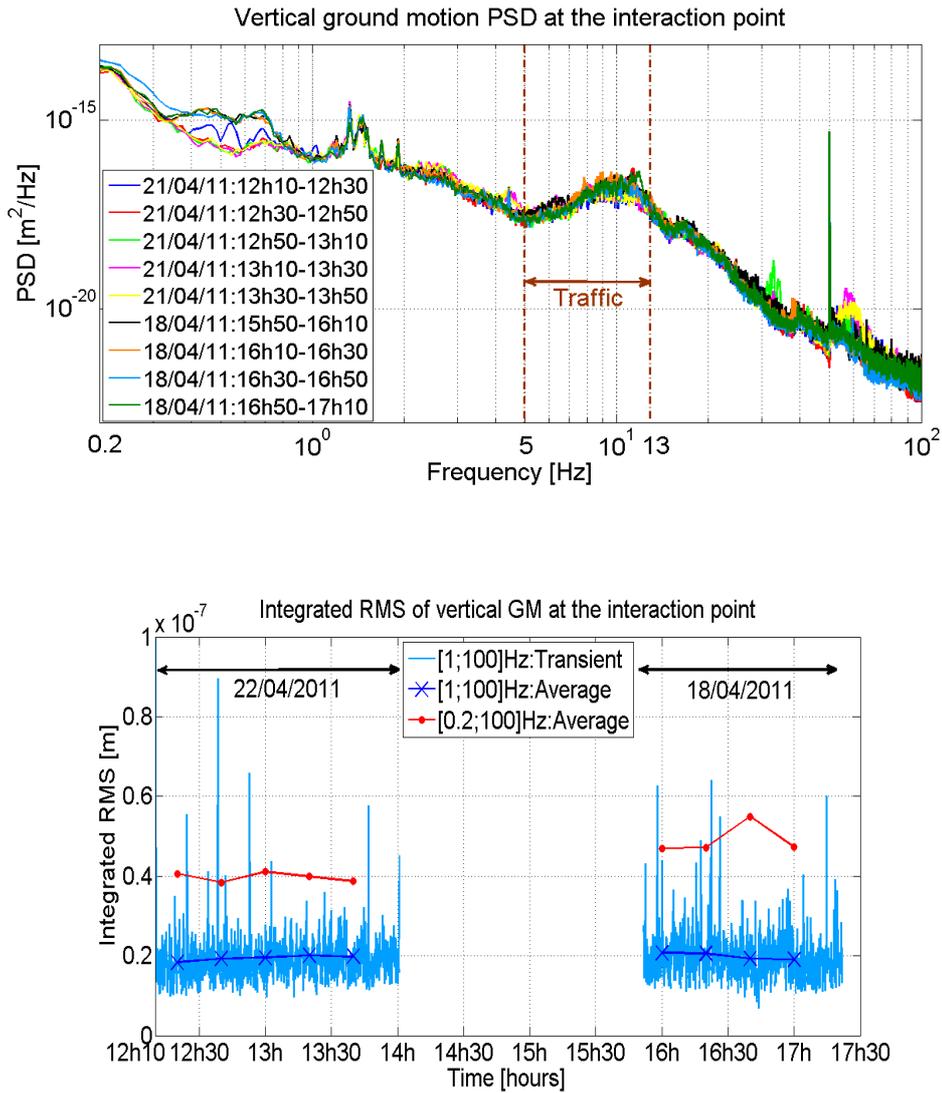

**Figure 4.7.6 - Vertical PSD at points 1 and 6.**

The minimum value is reached during the night at 2h50 and the maximum at 9h30 in the morning mainly due to increased traffic. The vertical RMS integrated in the frequency range [1; 100Hz] is shown in Figure 4.7.7 for all the points. Note that the blue curve corresponds to the measurements performed directly near the shoulder of the highway (Point 5). Amplitude of ground motion decreases with the distance from the highway and is almost the same for all the points located at a minimum distance of 100m from the highway.

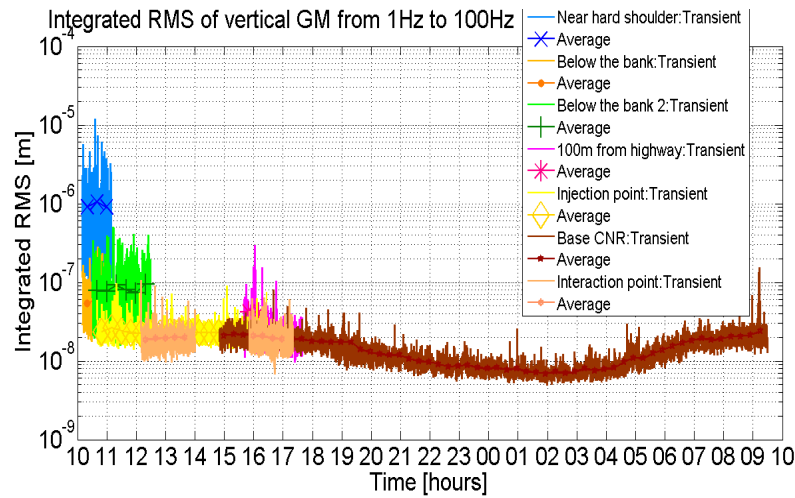

**Figure 4.7.7 - Vertical RMS of ground motion for all the points integrated in the frequency range [1; 100Hz].**

## Cooling Flow-Induced Vibrations

Cooling flow-induced vibrations of the water headers can be transmitted to the magnets and the vacuum chambers by flexible hoses and pipes. The effects of flow-induced vibrations will be mitigated by paying attention to several useful design guidelines, in particular:

all rotating equipment including fans, blowers, compressors, and pumps should be outside the storage ring tunnel, preferable tens of meters away from the tunnel floor and ceiling. The pumping station pad foundation must be disconnected from the main ring tunnel foundation and preferably passively damped.

The flow velocities in the cooling water headers should be kept less than 2 m/s. The header supports should be designed to minimize their vibration, such as by integrating viscoelastic dampers in the headers hangers.

## Thermal Stability

Ambient temperature variations will result in displacements of both the magnets on the girders and the BPMs on the vacuum chambers. To ensure acceptable thermal deformations of the ring components, cooling water and tunnel air temperatures must be maintained to within ±0.1ºC of their nominal values, 32ºC and 25ºC, respectively. Air-conditioning temperature cycling of ~1-hour duration will be maintained in order to take advantage of the thermal inertia of the support system. Lowering the beam height from 1.2 m to 1 m would reduce the vertical thermal expansions of the assembly proportionately. The tilted main ring (positrons) of about 11 mrad in order to get a vertical separation in the three overlap regions, will vary from zero to a maximum of about 0.9 m and will introduce more sensitivity to the thermal stability.

## 4.8 Survey and alignment

The required alignment tolerances are defined primarily by the physics requirements of the accelerator. At this stage of design, these tolerances have been defined on the base of simulation made from physicists. The methodological approach, for survey and alignment system follows what has already been done for other challenging accelerators or FEL around the world. Survey and alignment provides the foundation for positioning the beam-guiding magnet structures in all 6 degrees of freedom within the required tolerances. Although the tools and instrumentation available for this task have changed over the years and faster and more accurate measurements are possible, only limited control of the environmental conditions is possible. This ultimately sets an upper limit for the achievable measurement and subsequent control network accuracy.

## Tolerances

The required positioning tolerances are an essential part of the survey and alignment design. Those tolerances dictate the instruments and methods necessary to obtain the positioning goals. Table 4.8.1 provides the required global tolerances obtained from the optimization of the collider optic while Table 4.8.2 outlines the tolerated values for specific components. Relative tolerances of multipolar magnets facing each other are illustrated in Table 4.8.3. These tolerances represent the most stringent requirements for the storage ring and have been taken into account for the Tau/Charm survey and alignment network design.

**Table 4.8.1 - Required Global Tolerances**

| Global tolerances | ± 3 mm |
|---|---|
| Horizontal positioning | ± 3 mm |
| Vertical positioning | ± 3 mm |

**Table 4.8.2 - Tolerated values**

| Error kind | Studied | Reduced for Vertical Emittance | Accepted |
|---|---|---|---|
| Dipole rotation | 400μrad | 300μrad | 150μrad |
| Girder rotation | 400μrad | 350μrad | 175μrad |
| Girder DX | 200μm | 200μm | 100μm |
| Girder DY | 200μm | 130μm | 65μm |
| Quadrupole rotation | 400μrad | 150μm | 75μm |
| Quadrupole DX | 100μm | 100μm | 50μm |
| Quadrupole DY | 100μm | 90μm | 45μm |
| Sextupole DX | 100μm | 100μm | 50μm |
| Sextupole DY | 100μm | 100μm | 50μm |
| BPM offset X and Y | 100μm | 100μm | 50μm |

**Table 4.8.3 - Girder-to-Girder and magnet-to-magnet positioning tolerances**

| Relative tolerances | Girder to Girder | Magnet-to-Magnet |
|---|---|---|
| Horizontal positioning | ± 0.10 mm | ± 0.050 mm |
| Vertical positioning | ± 0.065 mm | ± 0.045 mm |
| Longitudinal | ± 0.50 mm | ± 0.1 mm |
| Roll angle | ± 0.175 mrad | ± 0.075 mrad |

## Control Network Design

To obtain certain tolerances it's important to provide an appropriate alignment network and choose all the instruments and equipment needed. State of the art equipment and procedures can assure proper mechanical positioning. However, the position tolerances of the machine components are not achievable with standard procedures alone. The mechanical alignment will be realized using instruments such as laser trackers, Total Station in combination with stretched wires and optical levels. A network of reference nodes will be built and will be qualified by referring the coordinates of each node to a properly chosen coordinate system.

All components will be accurately fiducialized by means of either laser trackers. A stable site is obviously a crucial starting point for the alignment of a particle accelerator. For this reason maximum attention has been given to this aspect in the design phase in order to obtain good ground stability. Optimization algorithms will be used to achieve the best possible configuration for machine operation by analyzing "as-built" geometry at each stage of the assembly and by modifying the alignment criteria to suit. This is especially important when defining the as-built magnetic axis of the machine and the subsequent alignment of components aligned to it. During the assembly phase and beyond, metrology processes will ensure that the machine and its supporting systems are dimensionally compliant for the successful operation of the machine. According to the size of the accelerator to be built, a Primary and Secondary control network are required to achieve the preview tolerances. The Primary alignment network is an outside network that spans the entire site, it is made of concrete pillars and their basement are well deep in the ground in order to assure a very stable position in the time. The relative position of each pillar can be determined by means of optical instruments or can be referred to a Global Position Satellite that can help to monitor the strain of the network itself versus time. The secondary alignment network is a cloud of reference targets well distributed along the inner volume of the accelerator tunnel as well as all the other building housing accelerator components. Specific elements or holes said Sight-Risers and well distributed in the alignment area are used to interconnect, by means of optical interface, the two separated networks. These elements are simply holes (like chimney) foreseen in tunnel ceiling concrete walls with the inner bore of about 500mm. Sight Risers, as we will see later, constrains the error propagation of the Secondary Network to the level achieved by the Primary reference network. Least square software like STARNET[1] are commonly used to simulate a complex alignment network and to optimize the targets distribution and minimum number of references to be foreseen. The Tau/Charm complex will spans an area of about 10 hectares with a Main Rings circumference of about 360 m and a Linac of about 200m.

## Primary Control Network

In Figure 4.8.1 is a footprint of the Tau/Charm complex, where is possible to see 9 monuments in orange (Primary Network) located at 103 meters above sea level and 15 Sight Risers in blue.

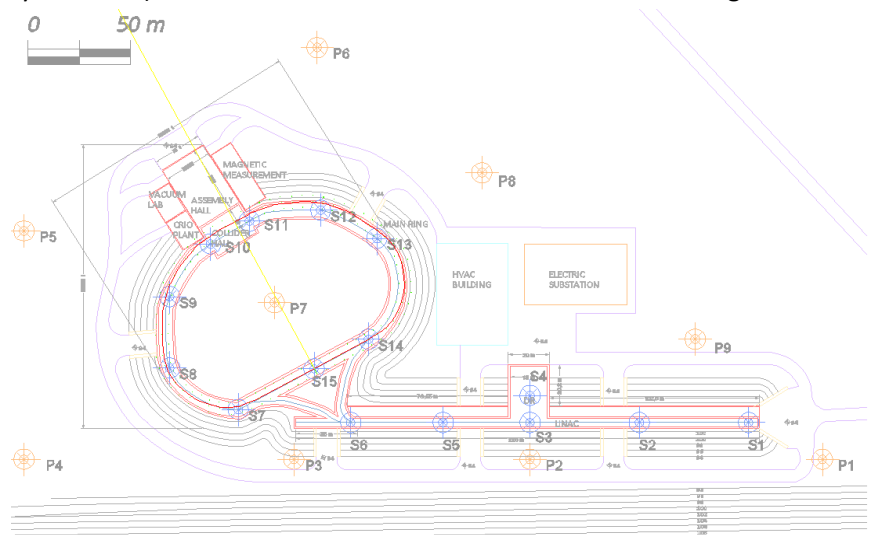

**Figure 4.8.1 – Tau/Charm Primary alignment network.**

The Primary Survey Network consists of permanent survey pillars positioned on the worksite area. These pillars will be used as fixed reference points to define the global coordinate system for civil engineering works, and to provide a stable reference for monitoring purposes. The network will evolve as the project develops and in the future will provide the global data for an enhanced reference system to be installed within the Building.

---

[1] http://www.microsurvey.com/products/starnet/

The primary control network spans the entire accelerator facility and ties the accelerator enclosures into one reference system. Therefore, they require a deep foundation and a secondary outer shell for temperature stability. Concrete pillar monuments can vary greatly in design, but are generally simple monuments consisting of reinforced concrete set within a tubular concrete form. The leveling mount and GPS antenna are secured to a stainless steel pin which is anchored within the top of the pillar. The foundation of the pillar can be coupled to exposed bedrock or be a larger mass of concrete set within a pit in soil. The pillar's ultimate design may vary depending upon availability of building materials, location, site conditions, and project requirements. These monuments can also be used by the construction companies for layout and construction surveys. In the Tau/Charm complex, accordingly to the geology of the site, pillars should have the basement at about 30 m below the ground surface and the outer diameter of the pillar should be not less than 50 cm.

The primary and secondary networks must be established, measured, and analyzed before accelerator equipments are installed. However, sufficient time has to elapse for the concrete to cure before the control network monuments can be considered stable. Most accelerator tunnels are constructed by the open cut method (cut and cover tunneling method). So, the foundation will be displaced because of the change in the load and the release of the stress.

### Pre-analysis

By means of the Star*Net least square code we have simulated the alignment primary network. Preliminary results of the computed error distribution are shown in Figure 4.8.2. The maximum error of about 38 μm was obtained at pillar S2. A list of node results is shown in table 4.

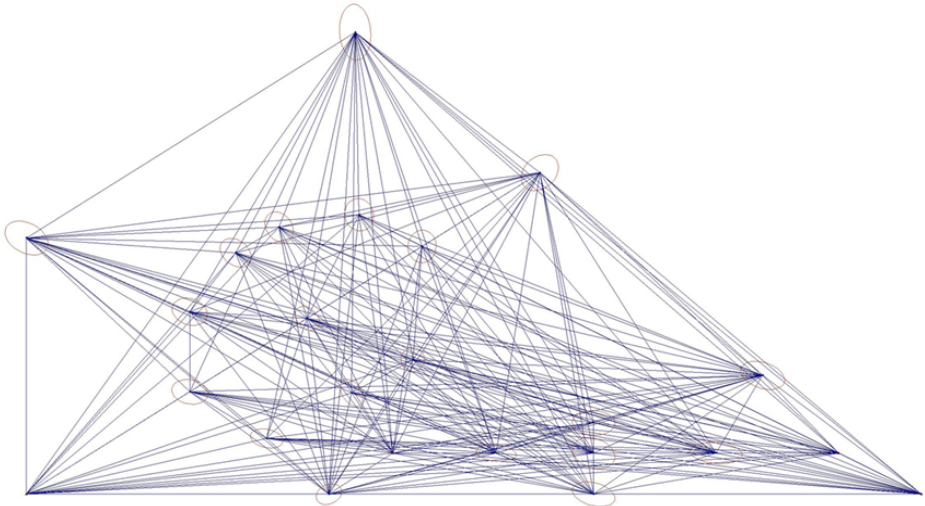

**Figure 4.8.2 - Primary network optimization with error ellipses for every monument.**

To obtain the results shown in the Table the following hypotheses have been set:

1. Measurements made with the same parameters obtained from the laser instrument datasheet (Total Station Leica TDA5005)

2. Seven virtual GPS receivers (Leica GNSS GS15) have been set for long term static observations. (Word Zone UTM 33A, Ellipsoid WGS-84)

**Secondary control Network**

The Laser tracker will be used for measuring the secondary control network, followed by a least squares analysis of the data to produce the final control point coordinates prior to setting out the collider components. Current laser tracker systems obtain point accuracy on the order of ±0.05 mm in a spherical volume with a radius of 10/25 m around the instrument measurement head. For measuring the secondary control network with laser trackers, the primary control points are included in the measurement process by means of about 15 sight-risers that are part of the data analysis. Sight risers constrain the error propagation of the secondary control network to the level achieved by the primary reference network. In Figure 4.8.3 is shown the 3D distribution of targets along the tunnel used to simulate with Star*Net the Secondary alignment Network in the Main Ring Tunnel. According to the geometrical target distribution used for this application, we have chosen to set monument section every 5 meters along the beam line in the tunnel. The total targets in the tunnel are 408 and the number of measuring station are 65. The floor monuments every 5m of the storage ring will necessitate core drilling to recess the target fixtures. The six wall-mounted targets every 5m of the storage ring, LINAC and Damping Ring, will be grouted to concrete wall by Hilti (HIT-RE 500) resin inserts.

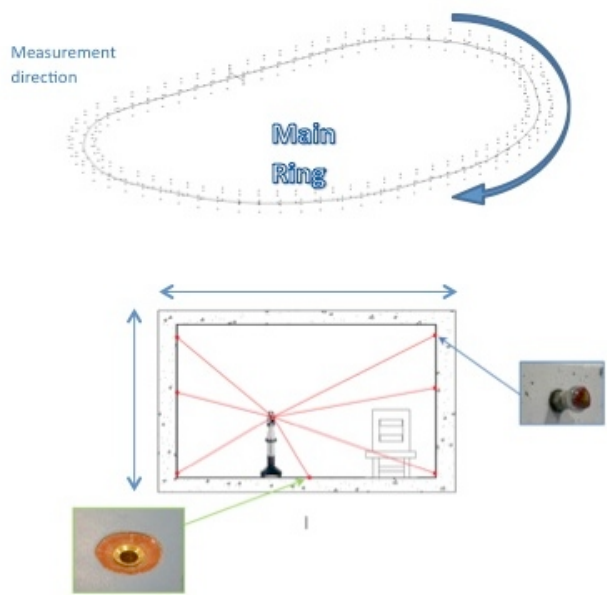

**Figure 4.8.3 - Target distribution in the tunnel and in the typical cross section.**

The secondary alignment network survey was simulated with Laser Tracker Leica LTD500. In Figure 4.8.4 is shown a plot of the error propagation without Sight Risers interconnection with the primary control network. As it can see the maximum error occurs in P1 and is about 987micron. The same traverse simulation was made adding the primary control network contribution by means of 6 sight risers, Figure 4.8.5. The maximum error, that occurs at point P442 located in the straight section of the ring, is about 289 microns, see the sight riser results in Figure 4.8.6.

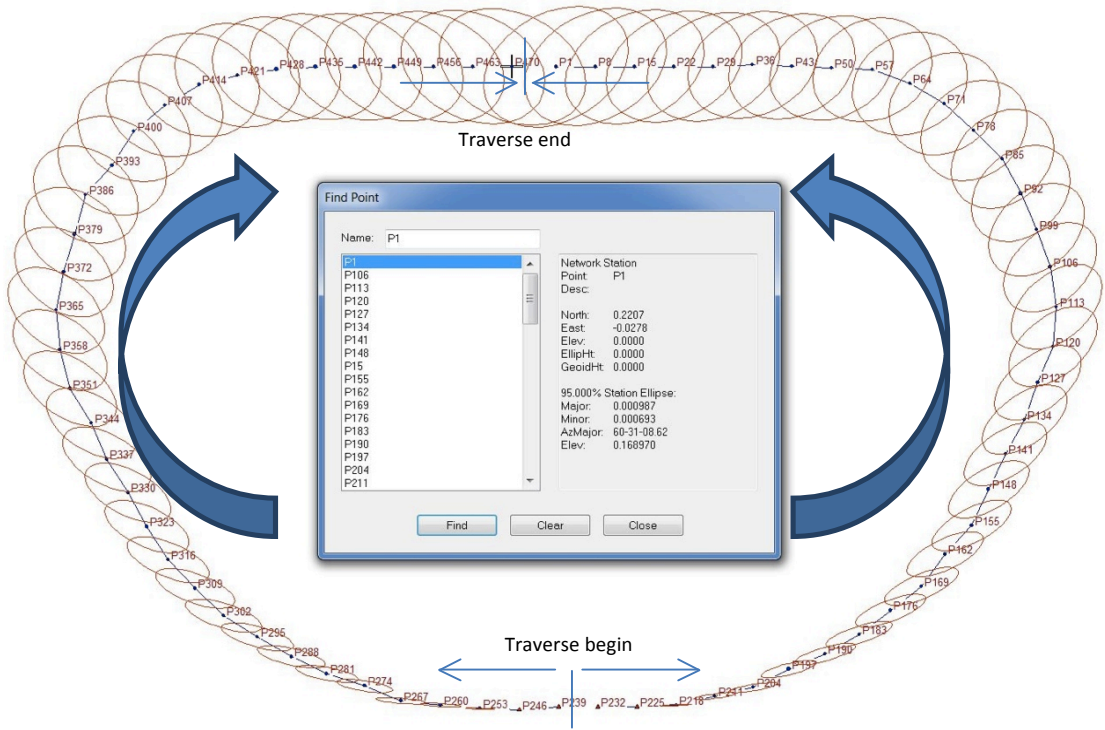

Figure 4.8.4 - Traverse simulation study to define the max virtual error

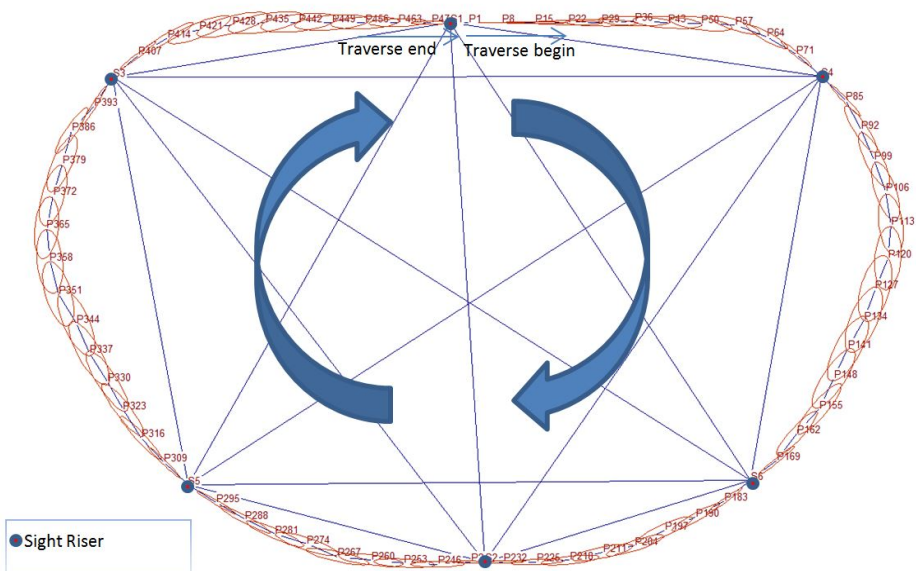

Figure 4.8.5 - Traverse simulation with 6 Sight risers to see their contribution in preliminary alignment results.

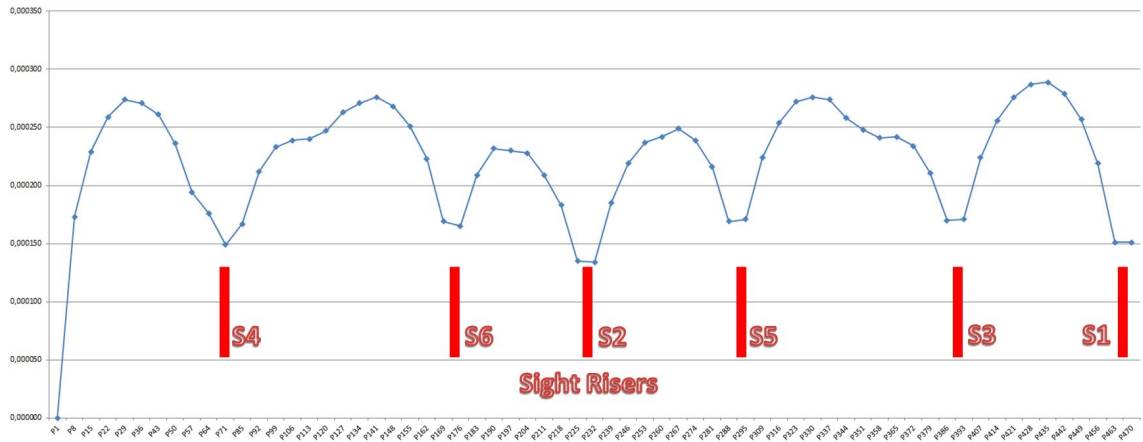

**Figure 4.8.6 - Semi-Major axis of error ellipses vs points with Sight Risers contributions.**

### Smoothing of the Magnets

This phase is the final alignment of the magnets. For the Tau/Charm the final alignment must be applied to all quadrupoles and dipoles with index field which have nearly the same sensitivity to misalignments. The process can only start once the magnets are connected and beam pipe under vacuum, so that all the mechanical forces are taken into account. The objective is to obtain a relative radial and vertical accuracy of 0.03 mm over a distance of 20 m.

As with the first alignment, the accuracy mentioned is applied at the fiducials. The vertical smoothing is performed with direct optical levelling measurements while the radial one is done by wire offset measurements or Laser Tracker. For this latter operation, access to the tunnel is required with the ventilation system regulated to give minimum air-flow. This smoothing process initially corrects both residual errors in the pre-alignment and ground motion. As various geo-mechanical and structural forces are acting on the tunnel, the reference network mainly tends to move vertically, but magnets may also become tilted by a transverse component of this motion and by the strain of the floor thus also generating a radial displacement. Repeated measurements of the network are very expensive and in fact useless if, on the other hand, tilt, radial and vertical measurements are made directly on the magnets and then processed with respect to a local trend curve within a sliding window along the machine. This efficient method allows an optimal and minimal detection of the magnets which need to be realigned. At the end of the process, the misaligned magnets are moved, while keeping a contingency for relative movement (within the tolerance) at interconnection level.

### Planned alignment procedures:

1. Procurement of alignment equipment and people training.
2. Multiple survey campaigns of the primary and secondary control network and link between them by means of sight-risers.
3. Magnet fiducialization by means of vibrating wire technique
4. Alignment of magnetic components inside the tunnels
5. Alignment smoothing of magnetic components.

Some specific tools or techniques may also be taken into considerations in order to fulfill the alignment requirements, such as Hydrostatic alignments, Stretched Wire Systems, Vibrating wires fiducialization, Micro/nano positioning.

## 4.9    Power Electronics

As far as the power converters are concerned, the main activity regarded the evaluation of the total power consumption of magnetic loads to establish the quantity and the electrical specifications of power supply systems (PSS) to be installed. The power estimating procedure, with reference to the V55 version of the lattice and based on data provided by magnets group, refers to datasheets of magnetic devices (Dipoles, Quadrupoles, Sextupoles, Octupoles and Correctors) distributed along the Main Ring (MR) and the Damping Ring (DR) sections of the accelerator. This document has to be intended as a preliminary evaluation of the total power requirements and subsequent sizing of the Power Supply Systems (PSS) and wirings.

At the time of writing, the evaluation procedure omits Octupoles and Correctors, being their electrical data not available yet. They will be included in a forthcoming revision.

### 4.9.1    Loads power consumption evaluation and sizing of Power Supply System (PSS)

In this section, a brief discussion on the method applied for the evaluation of the total electrical loads and the related sizing of power converters is presented, followed by a list of achievements. All computations are based on electrical datasheets of magnetic components (lattice V55) available at the time of drafting.

In the estimating procedure the following constraints were considered:

- Maximum current absorbed by each magnet was increased by a 20% compared to reported nominal current. As a result of increased value of the maximum current, maximum voltage drop on each component grows up by a 20%, bringing the maximum power rating of each magnet to a 44% higher value, compared to the nominal maximum power.
- To perform a realistic evaluation of total electrical power required by the PSS, is essential to take into account not negligible power losses on wirings, due to the very high currents flowing in most of the magnets and due to the cables length (cable resistance is not irrelevant in this application).
- To calculate the total power absorbed by the whole power supply system from the main power distribution, an hypothetical and safe value of the power converters efficiency (90%) was chosen, as it's not known, at the moment, which of the devices available on the market will be installed.

### 4.9.2    Damping Ring

A schematic diagram of the Damping Ring (DR) is shown in Figure 4.9.1.

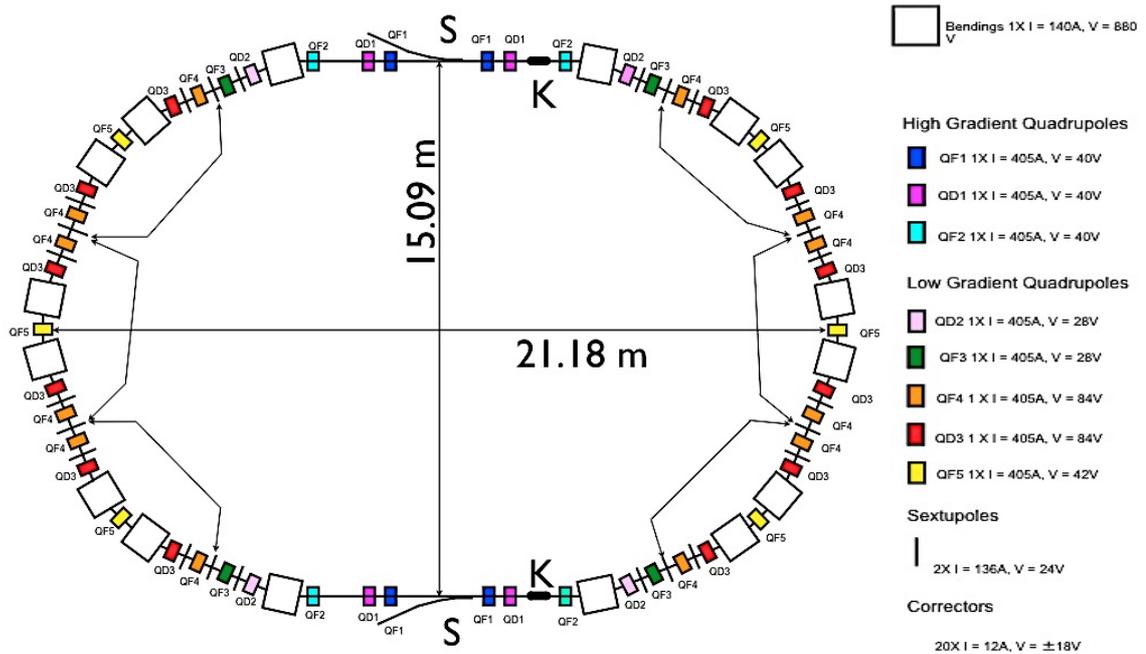

**Figure 4.9.1 – DR magnetic layout.**

The coloured blocks represent groups of magnetic components belonging to specific families. Magnets of each family are series-connected and fed by a single unit of dc power converter. This particular approach in connecting and feeding magnetic components best meets the needs of optimization, reducing costs and volume of Power Supply System (PSS) and wirings.

Starting from previous considerations, the estimation procedure provides the results listed in Table 4.9.1. These results are also based on the following assumptions:

- Damping Ring perimeter (approximated): **58m**
- Power converters are hosted in a dedicated technical room adjacent to the Damping Ring room.
- Each series of magnetic components is connected to its own dc power converter by a couple of parallel unipolar cables. A parallel (and close) cable arrangement is preferred to avoid EMI phenomena that could be generated by wide "loops". This approach critically increases the amount of installed wirings.

Considering the Damping Ring dimensions and the specific layout of (PSS), it's reasonable to allocate the power converters in a single dedicated technical room adjacent to the Damping Ring room. Correctors (20 units) are not included, but with reference to the SuperB project, a whole power consumption of about 10 kW can be supposed, spread on 20 dc power supply.

**Table 4.9.1 - Damping Ring - Power Consumption Evaluation**

| Number of Devices (Power Supplies) | Equipment | Equipment Label | Magnet Quantity (per Power Supply) | Max Voltage (Series) | Max Power Consuption (Series/Steady State) | Cable Section | CABLE Lenght (Total) | Max Cable Overall Voltage Drop | Power Supply DC Output Voltage | Power Supply DC Output Power (Estimated) | Power Suppy Equipment AC Nominal Power (Estimated) |
|---|---|---|---|---|---|---|---|---|---|---|---|
| | | | n° | [V] | [kW] | Sez[mm²] | [m] | [V] | [V] | [kW] | [kW] |
| 1 | BENDING MAGNET | BENDING MAGNET | 16 | 1337,79 | 195,2110 | 1x300 | 150,00 | 1,40 | 1339,20 | 195,42 | 214,96 |
| 1 | QUADRUPOLE MAGNET HG | QF1 | 4 | 37,46 | 15,1666 | 1x300 | 150,00 | 3,89 | 41,35 | 16,74 | 18,42 |
| 1 | QUADRUPOLE MAGNET HG | QF2 | 4 | 37,46 | 15,1666 | 1x300 | 150,00 | 3,89 | 41,35 | 16,74 | 18,42 |
| 1 | QUADRUPOLE MAGNET HG | QD1 | 4 | 37,46 | 15,1666 | 1x300 | 150,00 | 3,89 | 41,35 | 16,74 | 18,42 |
| 1 | QUADRUPOLE MAGNET LG | QD2 | 4 | 25,36 | 10,2684 | 1x300 | 150,00 | 3,89 | 29,25 | 11,84 | 13,03 |
| 1 | QUADRUPOLE MAGNET LG | QD3 | 12 | 76,09 | 30,8053 | 1x300 | 150,00 | 3,89 | 79,98 | 32,38 | 35,62 |
| 1 | QUADRUPOLE MAGNET LG | QF3 | 4 | 25,36 | 10,2684 | 1x300 | 150,00 | 3,89 | 29,25 | 11,84 | 13,03 |
| 1 | QUADRUPOLE MAGNET LG | QF4 | 12 | 76,09 | 30,8053 | 1x300 | 150,00 | 3,89 | 79,98 | 32,38 | 35,62 |
| 1 | QUADRUPOLE MAGNET LG | QF5 | 6 | 38,04 | 15,4027 | 1x300 | 150,00 | 3,89 | 41,94 | 16,98 | 18,68 |
| 2 | SEXTUPOLE MAGNET | SEXTUPOLE MAGNET | 12 | 18,04 | 1,9088 | 1x300 | 300,00 | 1,02 | 19,05 | 2,02 | 2,22 |
| | | | | | | | | | | Total Power [kW] | Total Power [kW] |
| | | | | | | | | | | 365,11 | 399,82 |

### 4.9.3   Main Rings

Applying the same procedure used for the DR to estimate the power consumption of the magnetic components installed in the Main Ring, the results listed in Table 4.9.2 were obtained.

It should be noted that magnets of only one storage ring, of the two constituting the Main Ring, were included in the present analysis, since, actually, electrical specs of just one ring are available at the time of writing. Nevertheless it is reasonable to suppose almost equal overall power consumption for both of the rings. Hence, a realistic value of total power absorbed by all the magnets of the MR can be obtained doubling the results depicted in Table 4.9.2, where Octupoles (14 units) are not included. These results are based on the following assumptions:

- Main Ring approximated circumference: 360 m
- At the time of writing not sufficient data are available in order to define a series-interconnection strategy for magnets feeding, as it has done for DR magnets. Thus the analysis proceeds supposing each magnet powered by a dedicated dc power converter. This approach critically increases the amount of installed wirings and the quantity of implemented dc power converters.
- Power converters hosted in four dedicated technical room adjacent to the MR area, arranged along the orthogonal axes of the MR, in order to minimize cables length.
- Each magnetic component is connected to its own dc power converter by a couple of parallel unipolar cables. A parallel (and close) cable arrangement is preferred to avoid EMI phenomena that could be generated by wide "loops". This approach critically increases the amount of installed wirings.
- Total Power consumption of MR can be obtained considering twice the value provided in Table 4.9.2.

**Table 4.9.2 - Main Ring (1of 2) - Power Consumption Evaluation**

| Number of Devices (Power Supplies) | Equipment | Equipment Label | Magnet Quantity (per Power Supply) | Max Voltage (Series) | Max Power Consuption (Series/Steady State) | Cable Section | CABLE Lenght (Total) | Max Cable Overall Voltage Drop | Power Supply DC Output Voltage | Power Supply DC Output Power (Estimated) | Power Supply Equipment AC Nominal Power (Estimated) |
|---|---|---|---|---|---|---|---|---|---|---|---|
| | | | n° | [V] | [kW] | Sez[mm²] | [m] | [V] | [V] | [kW] | [kW] |
| 8 | BENDING MAGNET | BQDM | 1 | 35,64 | 8,7577 | 1x300 | 1200,00 | 2,36 | 36,33 | 9,34 | 10,27 |
| 4 | BENDING MAGNET | BQDMA | 1 | 33,59 | 7,7261 | 1x300 | 600,00 | 2,21 | 34,57 | 8,23 | 9,06 |
| 4 | BENDING MAGNET | BSUP | 1 | 18,89 | 5,3281 | 1x300 | 600,00 | 2,71 | 30,10 | 6,09 | 6,70 |
| 12 | BENDING MAGNET | BARC | 1 | 32,99 | 9,3043 | 1x300 | 1800,00 | 2,71 | 35,71 | 10,07 | 11,08 |
| 12 | BENDING MAGNET | BS/B4/B2 | 1 | 21,07 | 5,9187 | 1x300 | 1800,00 | 2,70 | 30,38 | 6,68 | 7,35 |
| 4 | BENDING MAGNET | B3 | 1 | 17,44 | 4,0563 | 1x300 | 600,00 | 2,24 | 27,36 | 4,58 | 5,03 |
| 4 | BENDING MAGNET | B1 | 1 | 13,16 | 2,3084 | 1x300 | 600,00 | 1,69 | 23,79 | 2,60 | 2,86 |
| 2 | BENDING MAGNET | BSB1 | 1 | 0,39 | 0,0033 | 1X120 | 300,00 | 0,21 | 52,83 | 0,01 | 0,01 |
| 101 | QUADRUPOLE MAGNET LG | MRQM | 1 | 7,28 | 2,2900 | 1x300 | 15150,00 | 3,03 | 47,66 | 3,24 | 3,57 |
| 22 | QUADRUPOLE MAGNET HG | MRQM | 1 | 10,42 | 3,2800 | 1x300 | 3300,00 | 3,03 | 37,75 | 4,23 | 4,66 |
| 38 | SEXTUPOLE MAGNET | MRSM | 1 | 5,07 | 0,6800 | 1X300 | 5700,00 | 1,29 | 29,60 | 0,85 | 0,94 |
| | | | | | | | | | Total Power [kW] | Total Power [kW] | |
| | | | | | | | | | 814,80 | 896,08 | |

### 4.9.4    Control system

The local control of the power supply must be done via a controller, positioned inside the cabinet containing the power supply, or close to it. This controller allows commands, initializations and reset execution giving an overview of status information of the addressed power supply and resumes the most important status information. The controller must communicate with the remote computer system by a communication standard.

The amount of power needed by control systems constitutes a very small fraction of overall power budget, therefore it is not included in the estimating procedure at the present stage of the preliminary study.

### Conclusions on power electronics

Although in the previous analysis some components were not included (Octupoles and Correctors), it's still possible to provide a realistic evaluation of the whole power load enforced by the Main Ring and the Damping Ring.

Summarizing the results presented above, for Damping Ring and Main Ring, total power consumptions (magnetic loads) equal to approximately 360 kW dc (400kW ac) for DR and 1.8 MW dc (2MW ac) for MR, respectively, are predictable. Both values were rounded up to take into account contribution of Octupoles and Correctors not included in the evaluation procedure.

Considering the particular connection strategy chosen for DR and MR magnets, previously illustrated, quantities and electrical characteristics of dc power converters were identified.

For the Damping Ring, it's reasonable to suppose a number of 31 dc power converters with voltage comprised in the range [30.. 1500] V and dc power in the range **[**2.. 200] kW.

For the Main Ring, it's reasonable to suppose a number of 450 (225 for each of the two rings of MR) dc power converters with voltage comprised in the range [30.. 60] V and dc power in the range [1.. 12] kW.

The matter of connecting methods (series or stand alone connections) to energize the magnetic components of the Main Ring is still under discussion. Thus the quantities and characteristics of dc power converters (MR), presented above, may significantly change in the future.

As pointed out at the beginning, this document has to be intended only as a preliminary study. Therefore it could be subjected to significant further updates on the basis of the new data, available in the future, coming from overall progress of the Tau/Charm project.

PART 3 *Conventional Facilities*

# Contents

# 5   Conventional Facilities

Conventional facilities depend on the site choice. We refer in the following to the Tor Vergata University campus, where a slot was assigned for the SuperB project and could accommodate easily the Tau/Charm complex.

## 5.1   Site overview

The site proposed for the construction of the Tau/Charm factory is part of the campus at the University of Rome "Tor Vergata". It is in the South-Eastern area of Rome. It is located (see Figure 5.1.1) to the West side of the CNR research area, at the East of City of Sport facility, currently under construction and near to the Rome-Naples highway that runs from West to South. This location is reasonably close to the INFN Frascati National Laboratory LNF (about 4 km).

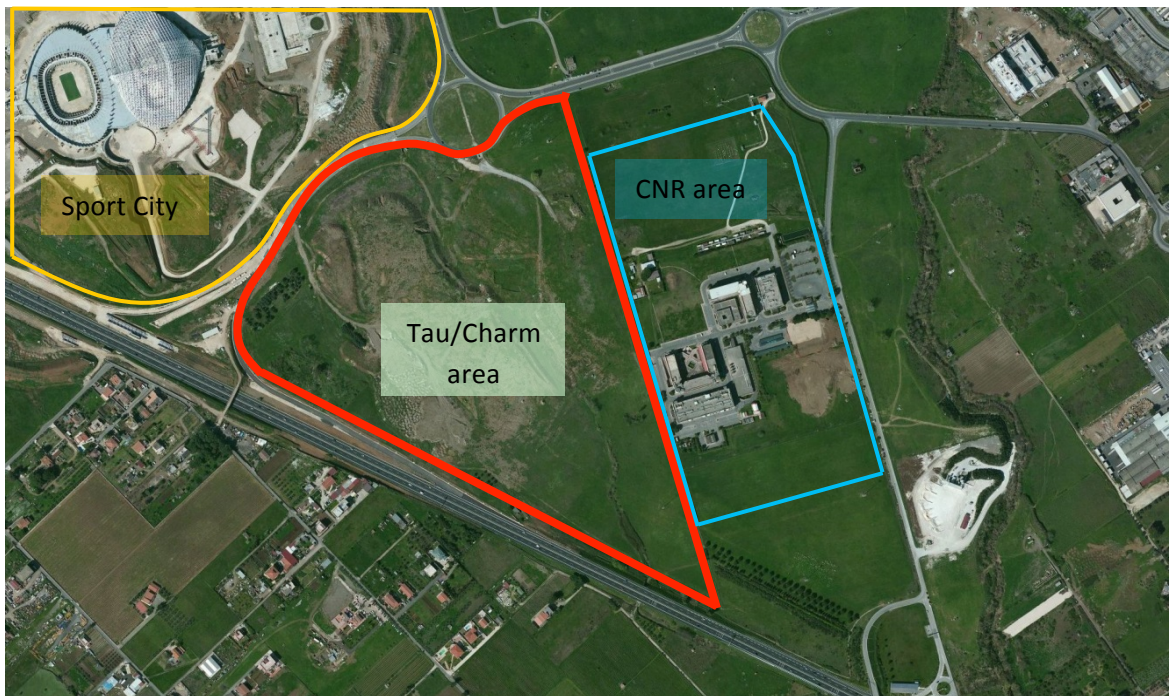

Figure 5.1.1 – Aerial view of the Tor Vergata University site.

Its position is strategic because of its vicinity to the CNR centre and to the university complex also in view of future collaboration projects. The lot, with a triangular shape, has long sides which are parallel to the highway and to the CNR area. It covers approximately an area of 28 hectares and it is located at an elevation ranging from 94 to 108 m above sea level. At present there are no buildings, but only wells and an underground power line of 20 kV.

From a urban point of view, the site is described in Tor Vergata detailed plan that identifies the site as green spaces (shown in Figure 5.1.2 with the initials VA4). Moreover, we need to remind that this area is at high development potential and make sure that future urban projects will not interfere with our experiment.

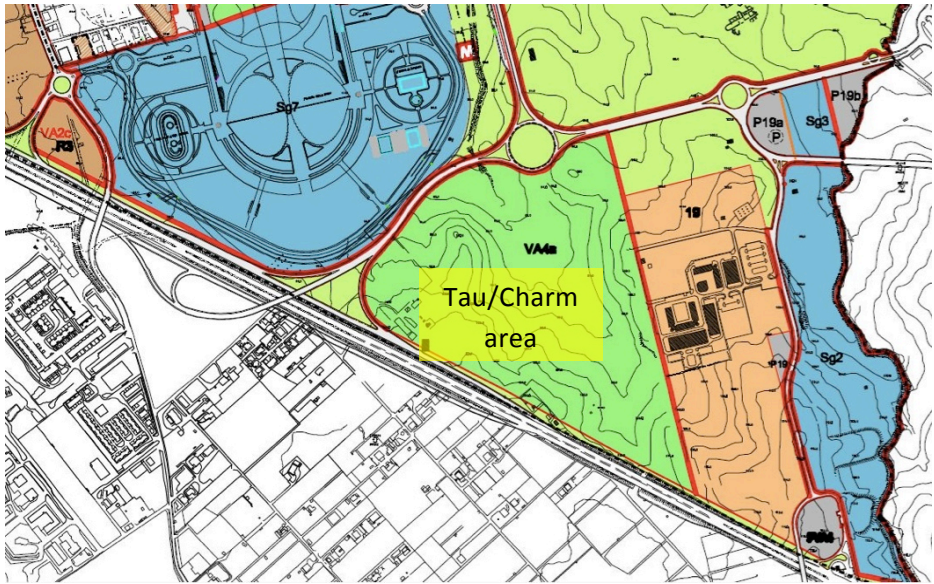

**Figure 5.1.2 - Map extract of the detailed plan of the University campus of Tor Vergata, highlighting the project area.**

### 5.1.1    Geology and hydrogeology

At the moment no dedicated geological and hydrogeological study of the site has been performed yet, but we know preliminary data of the adjacent area thanks to 2005 geological campaign survey before the construction of the City of Sport. We only present a brief introduction about geology and site layout, but geological investigations and site assessment have to be reviewed in detail.

The morphology is typical of plano-altimetric sub-flat trends of the Colli Albani area, with an elevation that ranges from 94 m and 108 m above sea level. From a geological point of view, the Tau-Charm site is related to the Latium volcano and the Lombardo creek crosses this area from South to North. From a hydrogeological point of view, we are in the presence of formations slightly permeable for primary porosity due to the degree of cohesion, sometimes of alteration and rearrangement reached after its lithification. We do not observe underground filtration, but it is believed that there is a groundwater at about 40 m depth from the ground level. For this reason, it is necessary to carry out a more detailed analysis to verify the presence of underground water and to study its variation over time.

### 5.1.2    Description of the soil strata found

The soil consist of volcanic products with different physical-mechanical characteristics. It has a fairly good degree of stability and is classified as grained soils from medium to large, weakly or moderately cemented. Data are known up to a depth of about 30 m.

Starting from the ground surface, the soil includes the following main stratigraphic levels (see Figure 5.1.3):
- • a first layer with a thickness of about 1 m of vegetable soil and fill material consisting of sediment and debris of various kinds;
- • below there is a 3-4 m layer of uncompact brown pyroclastic material;
- • a layer of about 15 – 20 m of more compacted grey pyroclastic material: it includes pockets of scoria, lava, tuff elements;
- • at 30 m below the surface, there is a very thick layer of tuff rock.

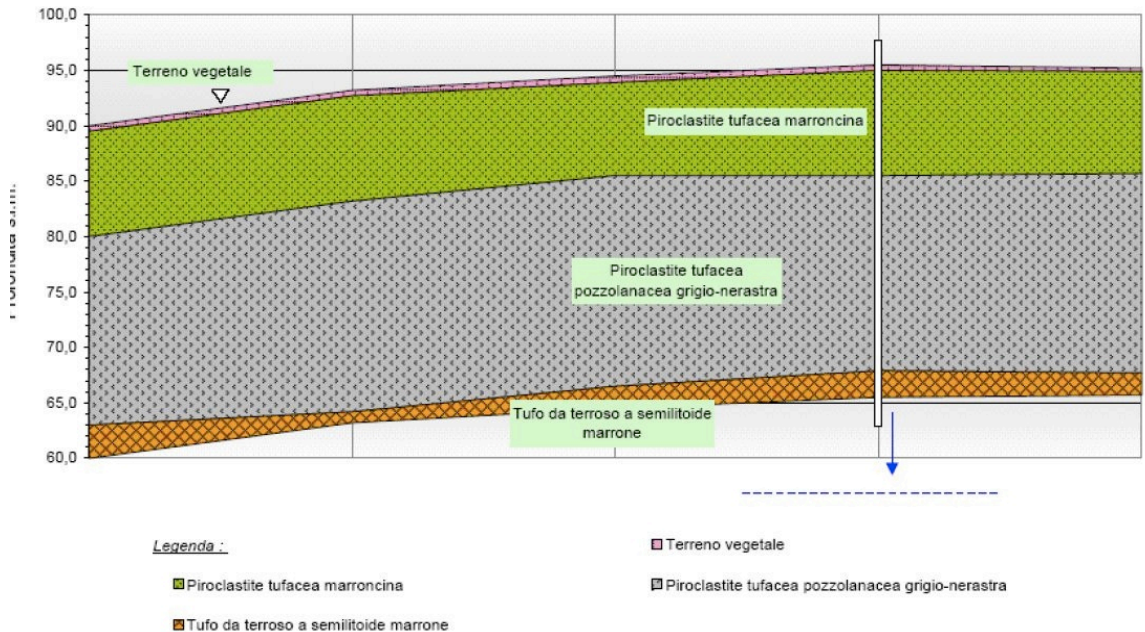

**Figure 5.1.3 - Schematic stratigraphic section of the study area.**

The pyroclastic layers consist of compact material of sand and clay that is able to support heavy loads and exhibits excellent damping properties. Moreover, they have good draining, in fact the underground water can be found at about 40 m depth well below the pyroclastic layers.

Over the site there is currently a wide and thick backfill from the excavations for the construction of the City of Sports; it is necessary, so, carry out a land surveying in order to consider the recent changing of the area and remove this layer of uncompacted materials (about 10 m), to found the buildings on better ground.

A thorough campaign of geological surveys will be predisposed to obtain the necessary data for the dimensioning of the foundation of the buildings and the parameters that must be used for the calculations of all the structures.

The survey will be targeted to the precise definition of the geotechnical features and, in particular, to the aspects of deformation of the land affected by the mains building (collider hall, damping ring building, undulators, experimental hall).

## 5.2   Mechanical layout

The layout of the Tau/Charm complex has been designed for all its parts, that is the Injector, the Transfer lines and the Main Rings and the Damping Ring. In the following a short description of these components is given, the details are reported in Part 1, where the various components are described. Figure 5.2.1 shows the Gun to Positron Source part. The electron gun is positioned at an angle with respect to the Linac1, used to accelerate the electrons up to the positron target. In line with the LINAC1 it is foreseen room to accommodate eventually the FEL gun. Modulators and klystrons are housed in the building placed beside the LINAC tunnel and each modulator and klystron drive three accelerating structures. Room at the end of LINAC1 was allocated to house the positron converter.

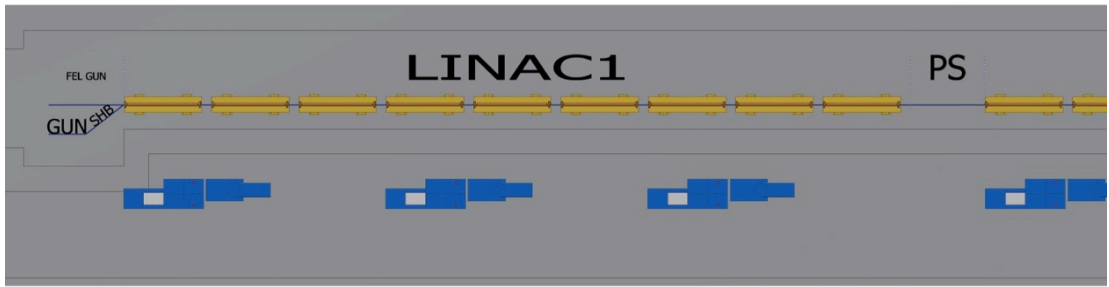

**Figure 5.3.1 – Layout of the Electron Source up to the Positron Source.**

Figure 5.2.2 shows the layout of the Injection complex from the Electron Source up to the entrance of Linac3 which accelerates the beams from 1 GeV to the final injection maximum energy (2.3 GeV at present). Positrons are produced at 600 MeV, accelerated in Linac2 up to 1 GeV, and then injected and extracted at 1 GeV from the Damping Ring. On the extraction TL from the DR a bunch compressor is planned. Electrons not used for positrons production will continue in Linac2 and Linac3. LINAC2 consists of 18 accelerating structures driven by six klystrons.

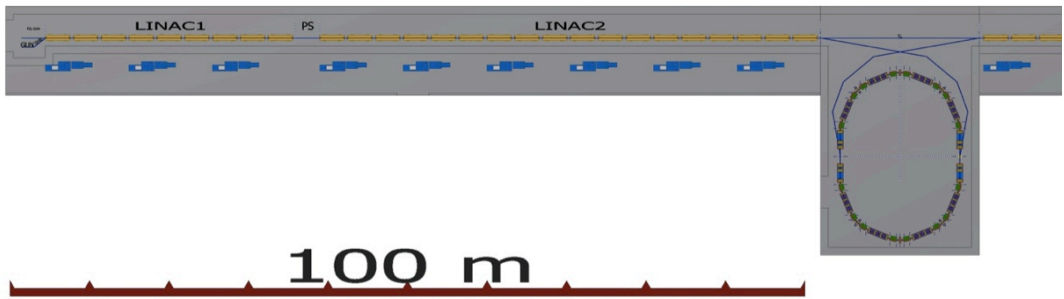

**Figure 5.2.2 – Layout of Linac1, Linac2 and Damping Ring.**

Linac3 and Transfer Lines layout are shown in Figure 5.2.3 LINAC3 consists of twenty one accelerating structures driven by seven klystrons. At the end of LINAC3 beams are split and injected into two separate transfer lines and eventually injected into the main rings.

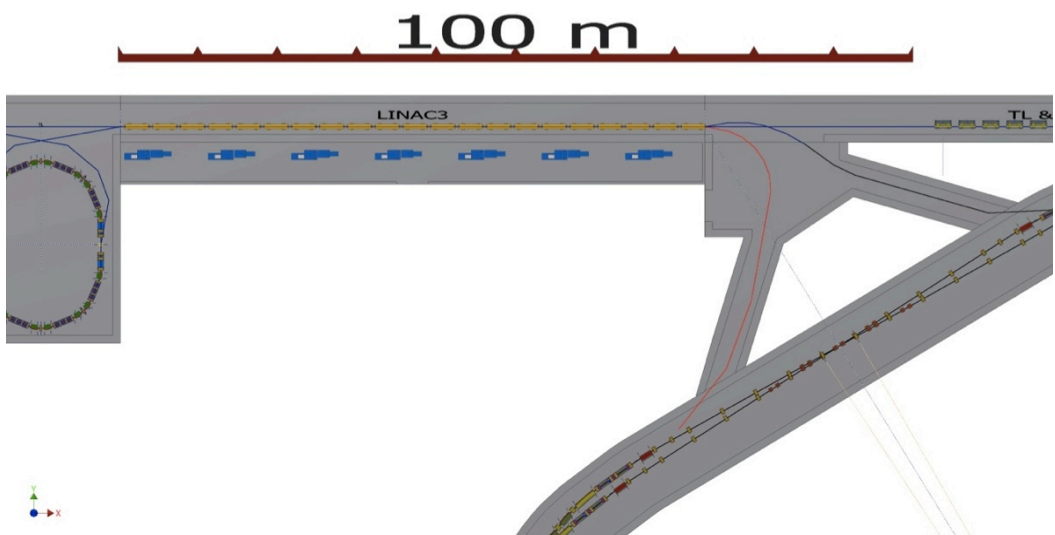

**Figure 5.2.3 – Layout of Linac3 and Transfer Lines to the Main Rings.**

Figure 5.2.4 shows a possible layout of the SASE-FEL facility which can be installed after the Linac3, including the hall for experiments. The undulator numbers and type are only indicative. The undulator tunnel and the experimental hall can be extended up to about 700 m from the GUN.

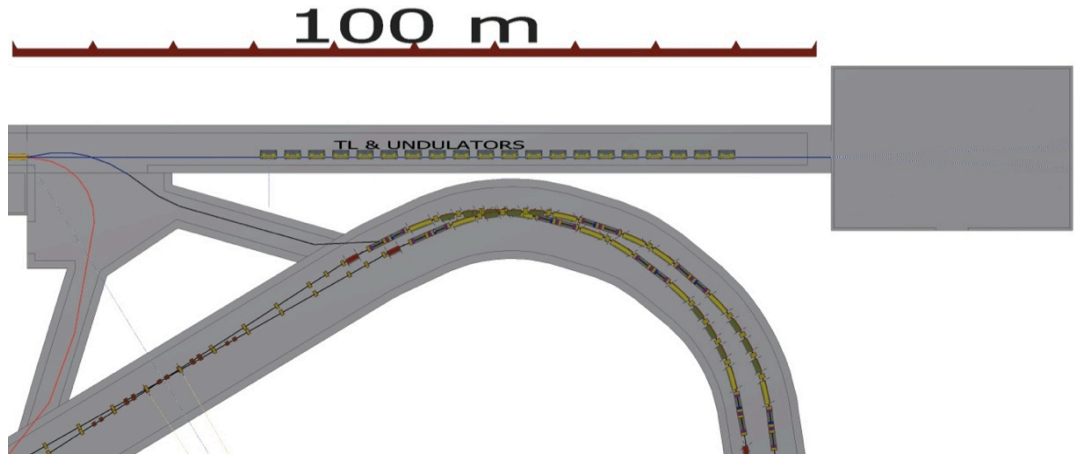

**Figure 5.2.4 – Layout of Linac4, undulators and experimental hall for a possible SASE-FEL facility.**

The Damping Ring hall and layout are shown in Figure 5.2.5. Only positrons will be stored in the DR, injected and extracted at 1 GeV. The extraction transfer line house the bunch compressor.

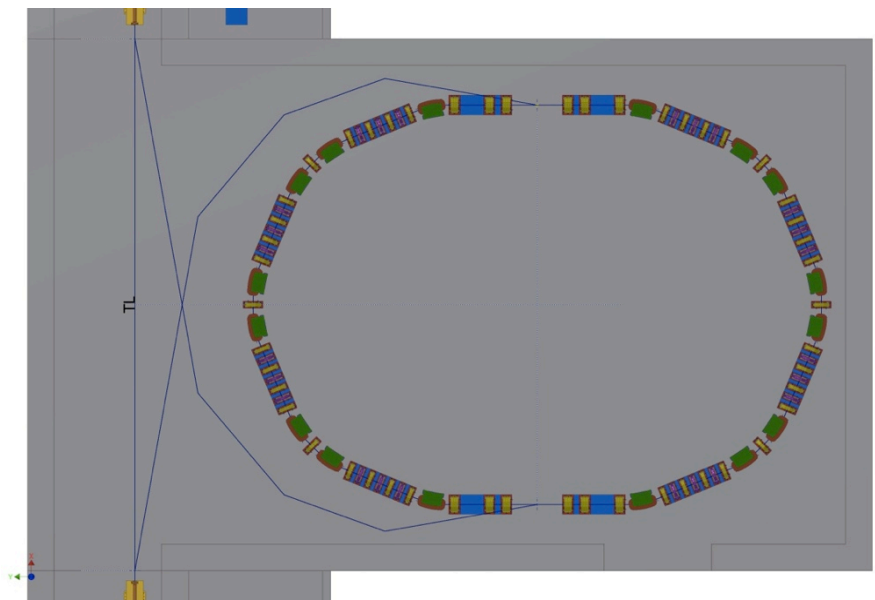

**Figure 5.2.5 – Layout Damping Ring and IN and OUT positrons lines.**

Figure 5.2.6 shows the layout of Main Rings tunnel with the Transfer Lines and the Experimental Hall. Finally, a detail of the Arcs overlap is shown in Figure 5.2.7. The Main Rings will be tilted one with respect to the other by 11 mrad, the tilt angle will be provided by very small solenoids close to the doublets in the Final Focus. Finally Figure 5.2.8 shows a detail of the two rings inside the tunnel.

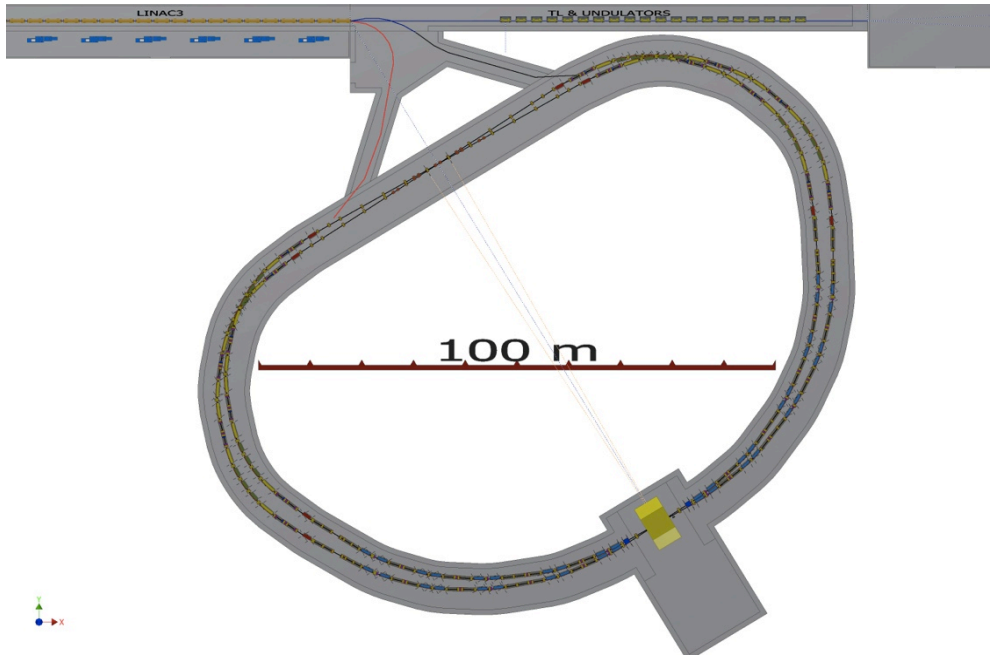

**Figure 5.2.6 – Layout of Main Rings tunnel with Transfer Lines and Experimental Hall.**

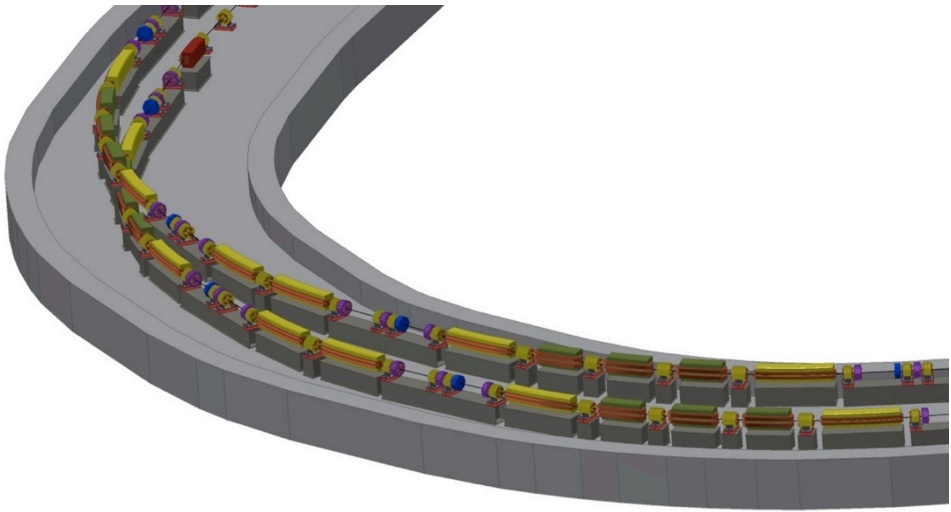

**Figure 5.2.7 – Detail of Main Rings Arcs overlap in the tunnel.**

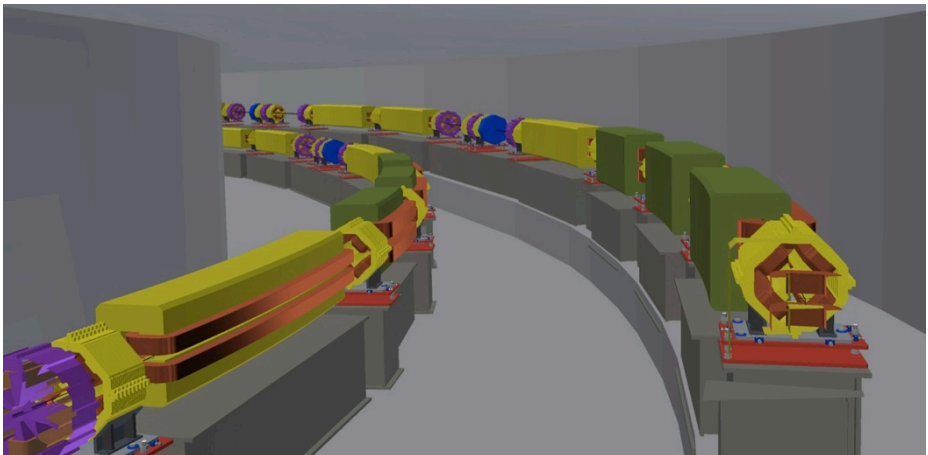

**Figure 5.2.8 – Detail of the two rings inside the tunnel.**

### 5.2.1 Damping Ring mechanics

The Damping Ring is about 15 m wide and 21 m long. The magnet layout is almost complete, we have 16 dipoles, 12 long quadrupoles, 38 short quadrupoles and 24 sextupoles. Figure 5.2.9 shows the pictorial tridimensional view of the CAD model. The injection and extraction devices are foreseen in the two straight section even though the solution of having injection and extraction devices concentrated on the same straight section is also possible. The RF cavity will be installed in the straight section as well.

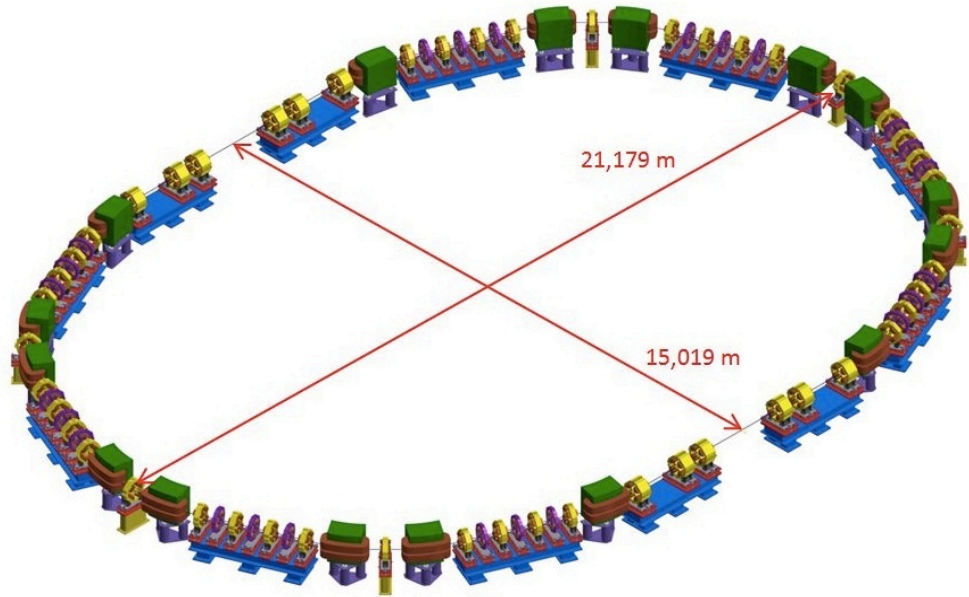

**Figure 5.2.9 - Damping Ring 3D pictorial view.**

The completion of 3D CAD magnetic model puts the basements for the engineering of the vacuum and diagnostic components, the alignment network, electric power bus bar and refrigeration pipe distribution etc. Magnets can be grouped in eight magnetic cells composed of four short quadrupoles and three sextupoles plus two dipoles at the beginning and the end of the cell itself. All the multipolar magnets in the straight section between dipoles are installed on a single girder support that provide the capability of assembling and pre-alignment of the component outside the damping ring hall. The latter choice optimizes the standardization of the design, ease the mechanical installation as well as the manufacturing and design cost, see Figure 5.2.10.

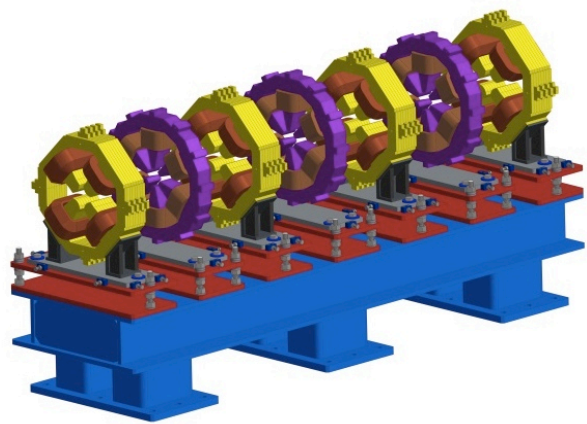

**Figure 5.2.10 - Multipolar magnets and girder in the magnetic cell.**

The remaining quadrupoles and dipoles are installed on standalone girders, see Figure 5.2.11.

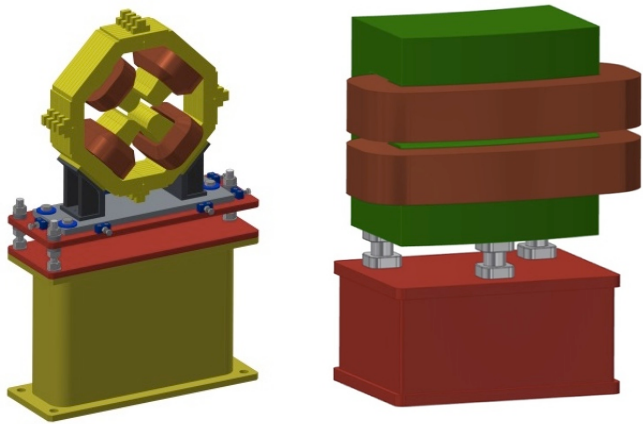

**Figure 5.2.11 - Assembly of short quadrupole and dipole and girder supports.**

The final geometry of all mechanical devices supporting the magnets will be performed to minimize the vibration sensitivity response. Taking into account the preliminary magnetic design, the Zeroth mechanical design of the magnets was performed: overall dimensions, cross section, poles section, magnetic coils geometry, poles gap.

## 5.3   Infrastructures and Civil Engineering

The conventional facilities have been designed to provide all underground and above ground buildings, services and infrastructure needed to support the experiment of the Tau/Charm factory.

Before starting to build the machine, it is necessary to perform some preliminary activities:

- Fence of the site;
- Explosive ordinance disposal;
- Archeological digging;
- Power line deviation;
- Ground motion measurements;
- Utilities connections;
- Sewerage;
- Artesian well sinking.

A preliminary layout (shown in Figure 5.3.1) indicates the preliminary architectural design of the facility. It includes a tunnel that will accommodate two accelerator rings, an experimental hall that will include the collider hall, housing the Detector, and the assembly hall, a tunnel for the Linac, a building for the Damping Ring, a building housing the modulators and the klystrons. The complex includes also a Vacuum Lab, Cryo Lab and a Magnetic Measurements building. Furthermore, there will be service buildings that will house mechanical and electrical equipment supporting: a HVAC building and a main electrical station located East of the main ring and other secondary electrical substation distributed around the outer side of ring building.

Suitable access will be needed during both the construction phase (during which a great deal of excavated material will be removed) and the operational phase. Trucking routes and deposit locations will need to be identified. For the installation of components, shipping by road is likely the main delivery option and the roads to the site must be able to accommodate both the length and weight requirements of the major components.

The Figure 5.3.1 shows the entrance to the site from the roundabout, the gatehouse and, following the route parallel to the fence, the accelerator complex buildings. An internal road network of driveways and pedestrian ways allow connect all buildings.

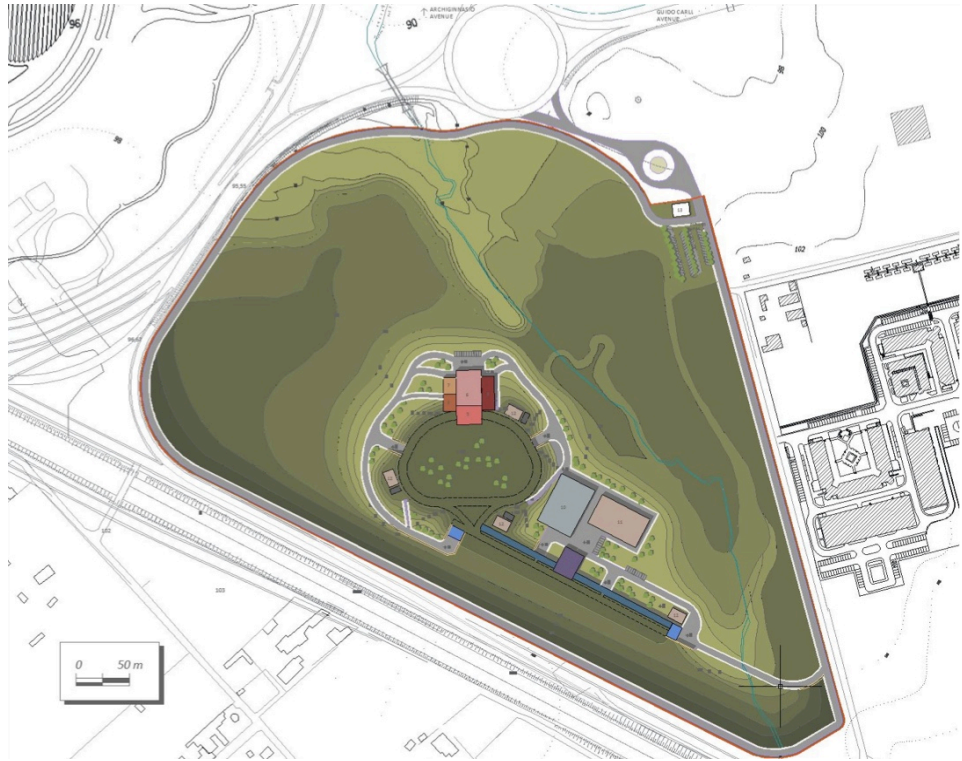

**Figure 5.3.1 - Preliminary layout of the site.**

The Main Rings and Linac tunnels will be underground and be built in a cut and cover method. For reasons of radiation safety, they have boundary walls of about one meter thickness and are covered with filling material. The soil rising above them must have a thickness of about 4 m, depending on detailed radioprotection calculations. Special attention has to be devoted to the shielding of the Collider Hall, Linac and other building, interested by the beam, according to the Radioprotection Safety Guidelines.

For the design of the structures, three aspects are of fundamental importance: alignment, ground vibration and seismicity of the site.

Alignment and stability are very important for reliable accelerator operation. Even more critical is the stability of the Collider Hall, that must be able to accommodate the detector mounting and movement and allow its repositioning without unsatisfactory deflection or settlement over time.

The vibration limits are associated with the user-supplied research instruments. A first campaign of detailed ground motion measurements has been performed at different locations of the site (See Section 4.7). For reference, Table 5.3.1 shows the tolerance acceptable defined for Synchrotron Soleil.

**Table 5.3.1 - Long time settlement defined for Synchrotron Soleil**
**(Extract from SuperB Site Commitee report)**

| OPERATING SPECIFICATION | |
|---|---|
| Long term settlement (vertical): | 100 μm over 10 m per year |
| | 10 μm over 10 m on a diurnal cycle |
| | 1 μm over 10 m in short-term (about 1 hour) |
| Punctual static load of 500 kg: | Δz<6 μm under the load |
| | Δz<1 μm at 2 m |
| Dynamic load of 100 kg: | Δz<1 μm (ptp) at 2 m |
| Vibrations (0,1 – 70Hz) due to all effects induced by the facility, added to the external effects: | Δz<1 μm peak to peak |
| | Δz<4 μm peak to peak |

From a seismic point of view, the Tau/Charm complex is located in an area classified as "seismic of 2B category". This classification is in accordance with the new classification of the Latium region in force since 22 may 2009. In this category the maximum ground acceleration is between 0,17 g to 0,15 g, where g is the gravitational acceleration.

The regulation in force (NTC 2008) allows to define all those operational parameters and the input data to be taken into account at the moment of planning and carrying out of any manufactured article, both in reinforced concrete and in metallic carpentry (pillars dimensions, earth anchorage, anchoring bolts, building materials, etc.). For buildings and structures that need particular stability, we could study more stringent safety standards about resistance to seismic actions. While for others buildings we can adopt the general regulation in force.

### 5.3.1 Architecture

The facilities will be located at the center of the site, below the Lombardo creek (see Figure 5.3.2).

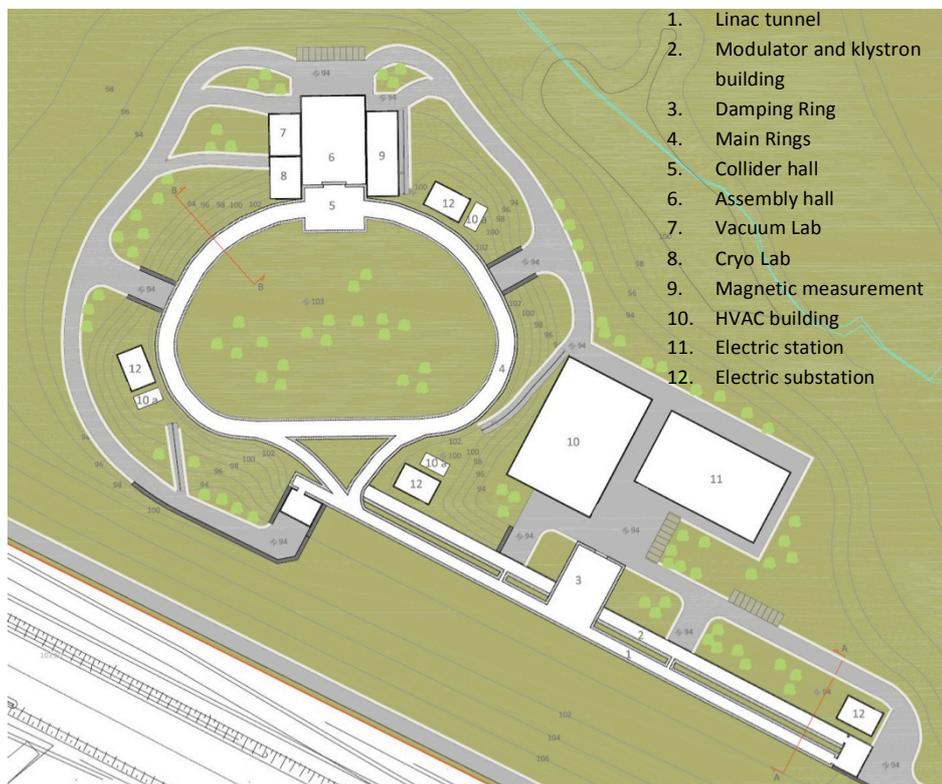

1. Linac tunnel
2. Modulator and klystron building
3. Damping Ring
4. Main Rings
5. Collider hall
6. Assembly hall
7. Vacuum Lab
8. Cryo Lab
9. Magnetic measurement
10. HVAC building
11. Electric station
12. Electric substation

**Figure 5.3.2 – Tau/charm complex.**

The approximate areas for each of building are listed in Table 5.3.2.

**Table 5.3.2 - Area and cubature of buildings**

| BUILDING | | AREA (m$^2$) | CUBATURE (m$^3$) |
|---|---|---|---|
| Linac tunnel | underground | 1360,00 | 8980,00 |
| Mod. and Klystr. building | surface | 1135,00 | 8512,50 |
| Damping Ring | surface | 470,00 | 3525,00 |
| Main Rings tunnel | underground | 2725,00 | 14715,00 |
| Collider hall | surface | 432,00 | 8000,00 |
| Assembly hall | surface | 800,00 | 15340,00 |
| Vacuum Lab | surface | 190,00 | 1710,00 |
| Cryo Lab | surface | 190,00 | 1710,00 |
| Magnetic measur. building | surface | 384,00 | 3456,00 |
| HVAC building | surface | 1750,00 | 15750,00 |
| Electric station (HV/MV) | surface | 1500,00 | 13500,00 |
| Electric substation (for n. 4) | surface | 150,00 | 600,00 |
| Gatehouse | surface | 150,00 | 450,00 |

The Linac runs parallel to the highway; it is a fully underground tunnel of 230,0 m in length. Its cross section has a rectangular shape with a width of 4,0 m and a free height of 3,6 m (see Figure 5.3.3). At intervals of 70 m, there is a connection passage between the linac and the klystron building, which can be used for evacuation in case of emergency. The Linac building has two entrances, one from the East side and the other from the South-West side. Two overhead crane with 20 t of capacity will be installed above of 3,6 m of height. Cooling water pipes and waveguide are installed in the lower part of the tunnel (below the planking level) and electric power lines are installed on the ceiling. The building housing the modulator and the klystron plants (see Figure 5.3.3) is parallel to the Linac tunnel, but above ground. It will have a rectangular cross section and it will be far from the Linac about 2,0 m for reasons of radioprotection.

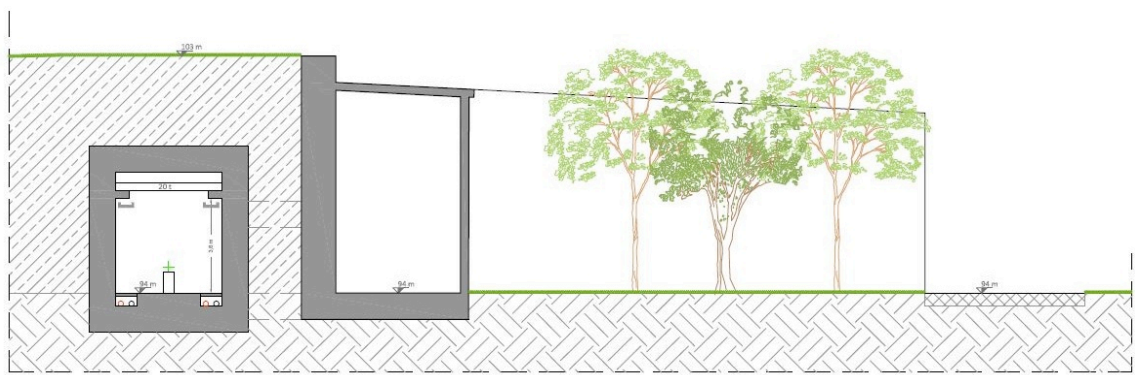

**Figure 5.3.3 – Cross section type of Linac and klystrons building (section A-A).**

At about 100 m from the beginning of Linac, it is foreseen the Damping Ring building. It is a surface building that has a rectangular plan with a size of 20 x 23,5 m. The longer side is perpendicular to the direction of the Linac. The driveways and pedestrian entrance to the damping ring is in the North-East side. At the end of the Linac (West side), two underground tunnel will be built for the transfer line to inject the beams in the Main Rings.

The two Main Rings will be placed side by side in a single tunnel. This tunnel looks like an ellipsoid flattened in the transfer line side. It will have a length of about 340 m and it will be fully underground. Two large and symmetric driveways and pedestrian entrances are foreseen. As the tunnel for the Linac, it will have a rectangular shape, with a width of 6,0 m and a free height of 3,6 m. It will accommodate also the ancillary equipments like trays for the power and control cables, cooling water pipes (see Figure 5.3.4).

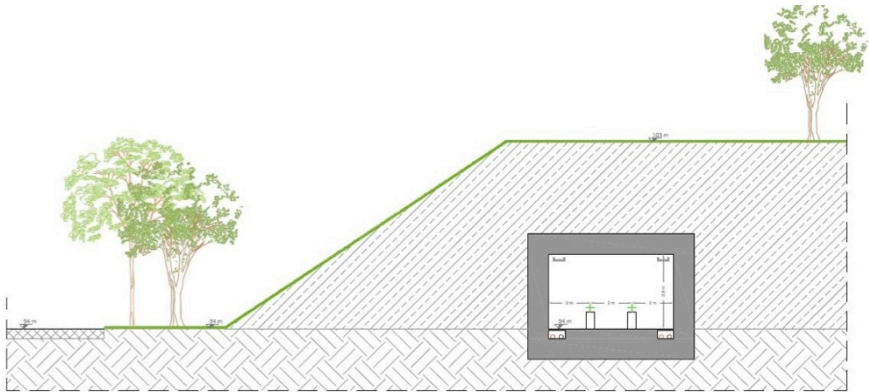

**Figure 5.3.4 – Cross section type of Main Rings tunnel (section B-B).**

In the underground buildings, a floor drainage system will be provided to contain, collect and treat any free-running water.

The experimental hall will be on the North side of the Main Rings; it will include the collider hall, housing the Detector, and the assembly hall (see Figure 5.3.5).

The collider hall has a rectangular plans with dimensions 16 x 22 m. It consists of a main hall that has enough central space host the detector on the beamline. It also has several work areas on either side and a large door that put it in connection with the Assembly hall. Its walls perform also radiation shielding.

The assembly hall consists of a wide room with dimensions of 33 x 23,5 m; it has an electronic house and the tracks to move the detector. Inside the building it will be installed an overhead crane with 20 t of capacity.

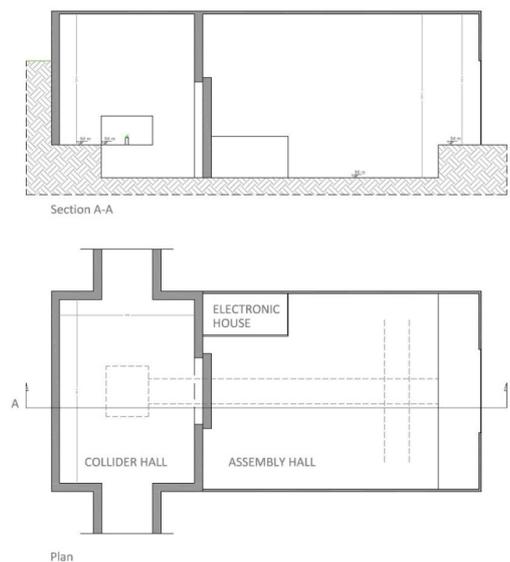

**Figure 5.3.5 - Plan and Section A-A of the experimental hall.**

After construction, all disturbed areas will be re-vegetated with a combination or indigenous plant materials, seeding, sods and/or wildflowers/groundcovers to minimize the effects of soil erosion.

## 5.4  Fluids

In this section we describe the whole fluid mechanical systems required for the Tau/Charm factory. HVAC systems are part of the infrastructure serving the accelerator and, they are one of the components that are most sensitive to the operation of the entire project, determining the efficiency and performance of the whole system. The main fluid systems to be achieved are:

- cooling systems;
- water cooling systems with very low conductivity;
- ventilation and air conditioning systems;
- water treatment and water source distribution systems;
- compressed air systems;
- gas fluid systems.

The fluid mechanical systems are necessary to ensure the control of environmental and technology parameters inside the laboratories. The main services are the following: checking the research equipment temperatures control, thermo-hygrometric parameters control in the buildings, thermal loads disposal, optimal ventilation and pressurization of all buildings (as required by law), air filtering and sanitization in order to limit the presence of pollutants, compliance with fire regulations, remote control and storage environmental parameters. This design has been developed in a single technological station operated by a remote control room; the heat transfer fluid used for thermal energy dissipation is water.

**Main devices**

The main devices chosen to ensure the successful operation of the accelerator are:

- dry cooler: the optimization of energy performance has directed the choice towards the use of machines operating in free-cooling mode. The heat transfer cooling fluid for such devices is optimal in case of low external temperature and humidity (for systems with water temperature  t = 32 ° C).
- chillers and cooling towers: the designed system will provide a value of COP not less than 6. The adopted machine to reach under certain load conditions, up to a value of COP 10.

Only components with high performance and durability such as high-efficiency pumps, inverters and steel heat exchangers will be used. Table 5.4.1 reports not only the kind and nominal quantity to be installed, but also the nominal and operation values of thermal power to be adopted.

**Table 5.4.1 – Main devices**

| central | component | Make & Model | circuit | nominal quantity | NOMINAL VALUES | | | OPERATION VALUES | | |
|---|---|---|---|---|---|---|---|---|---|---|
| | | | | | thermal power | electric power unitary consumption | DT [°C] | thermal power | attenuation | electric power consumption |
| | | | | | [kWt] | [kWel] | | [kWt] | [%] | [kWel] |
| CONFIGURATION | | | | | | | | | | |
| SC00 | Dry Cooler | BAC DFCV-AD | main dry-cooler | 2 | 800 | 26,4 | | 1600 | | 52,8 |
| SC00 | pump | grundfoss con Inverter | main dry-cooler | 2 | 800 | 15 | 5 | | | 30 |
| SC00 | heat exchanger | | main dry-cooler | 2 | 800 | 0 | | | | 0 |
| SC00 | chiller centrifugal water/wate | YORK YK R4R4K4 5DG ( | main chiller | 2 | 7500 | 1262 | | 10000 | 67% | 841,3 |
| | | | | | | | | 8400 | 56% | 706,7 |
| SC00 | transformer 20/6kVolt 1'300kW | | electric chiller | 2 | | | | | | 0 |
| SC00 | evaporative cooling tower | BAC 2 x S3-D 1056 L in | main tower | 2 | 7500 | 122 | | | | 244 |
| SC00 | evaporative cooling tower | Cillichemie | treatment tower | 2 | | 2 | | | | 4 |
| SC00 | pump | grundfoss con Inverter | main tower | 2 | 7500 | 90 | 5 | | | 180 |
| SC00 | pump | grundfoss con Inverter | main chiller | 2 | 7500 | 90 | 5 | | | 180 |

## Principle of operation of the technological station

Each unit is equipped with a primary circuit in which it is expected the interaction between the free-cooling system and the water-water-air device, for the maximum obtainable efficiency. The secondary circuits are characterized by water at 12° C, water at 32° C and with demineralized or simply softened water. All systems are remote controlled by and must operate in stand-alone mode as well as being controlled by the remote control center.

The calculations were performed considering a thermal load of 10 MW; about 1,5 MW is disposed of by machines operating in free-cooling mode; the remaining from the chillers with high COP. It has been taken into account the environmental parameters characterizing the host location of the accelerator by the attenuation factor of machine operation: in summer months, in fact, even if the dry-cooler are equipped with a precooling adiabatic systems (that manages all in all to lower the temperature of the heat transfer fluid), it was considered an additional cooling systems (chillers) to meet the needs of the users that are served by the dry-cooler (utilities to 32 ° C). Considering summer periods made up of four months, it has been assumed for the chillers an attenuation factor of 56% for 8 months a year (depending on the installed thermal power) and an higher attenuation factor (67%) for the remaining 4 months. The optimization of energy performance has directed the choice towards the use of machines operating in dry-coolers. This mode will lower electricity consumption for 8 months a year.

## Layout of the technical rooms

The dimensions of the technological station allow the operation of machines for cooling and for the fluid circulation. The station has two floors: first floor is at the ground floor (see Figure 5.4.1), the second floor is semi-terrace (see Figure. 5.4.2). In a nearby buildings there is a technical room dedicated to treatment of cooling towers water to avoid incrustations (Figure 5.4.3).

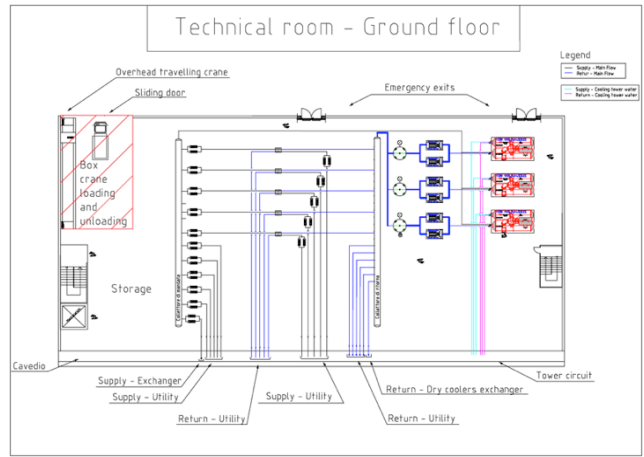

**Figure 5.4.1 - Station ground floor.**

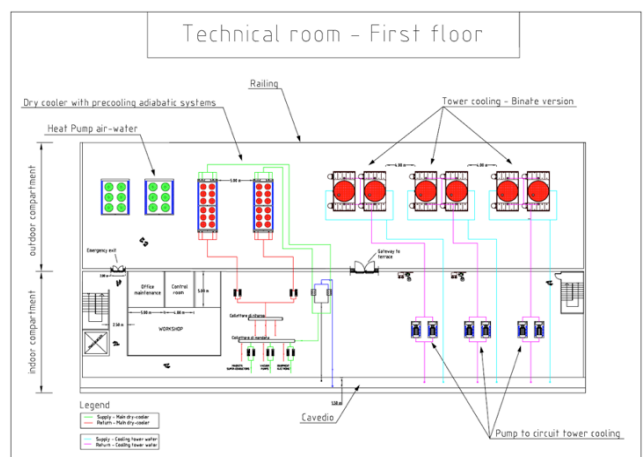

**Figure 5.4.2 - Station first floor.**

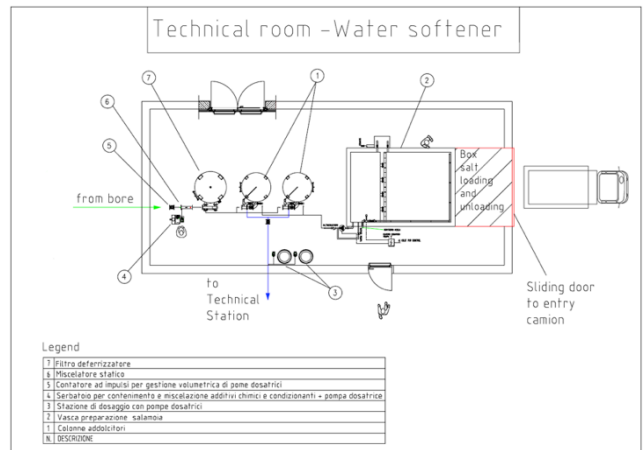

**Figure 5.4.3 - Water softener.**

Given the thermal loads (DR, collider hall and detachment point of the beam), the positioning of the technological station is based on the thermal barycenter.

**Layout of the ventilation systems**

The ventilation system guarantees the disposal of thermal loads in the air, the air circulation in all buildings and depressurization of the buildings. There are n°3 inlet air treatment station and three for extraction, placed so as to ensure an efficient air exchange.

Its value of 3 vol/h has been set taking into account the thermal load of research devices and ventilation to comply the environmental thermo-hygrometric parameters, according to Italian law. The depressurization of the buildings is ensured by three air extraction stations, according to the diagram below (see Figure 5.4.4).

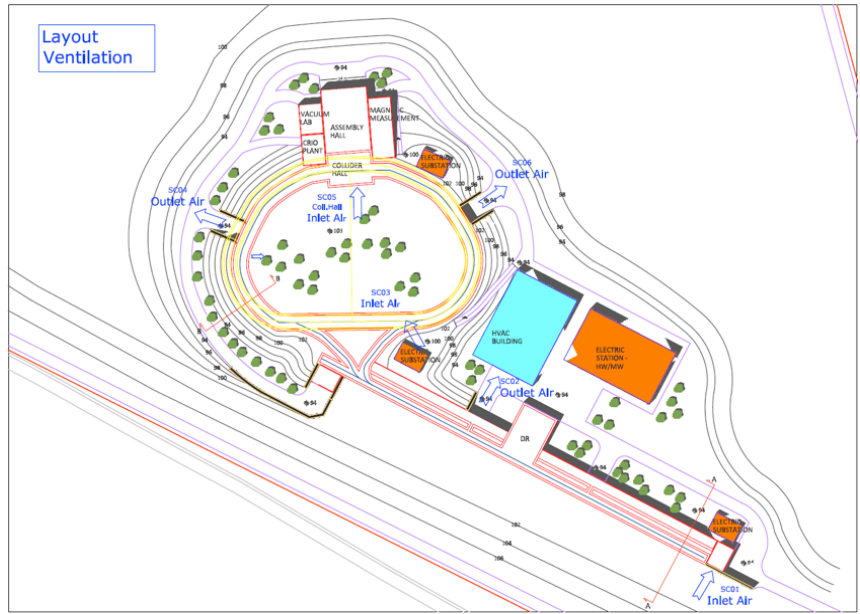

**Figure 5.4.4 - Sketch of the air extraction stations.**

## Load analysis

Table 5.4.2 contains all the values of the thermal power dissipated by the equipment (magnets, power supplies, wiring...) which are present in the accelerator. The thermal power consists of two parts: one to be dissipated in water (with dry-cooler) and the other in air (with precision air conditioning units).

**Table 5.4.2 – Thermal power dissipated**

| Device | Number of Devices (Power Supplies) | Equipment | Equipment Label | Plant Location | Power supply | Thermal power TOTAL for device <TP> | POWER CONTROL Thermal dissipation in air <TPa> | TOTAL in air | MAGNET Thermal dissipation in water <TPw> | TOTAL in water | Temperature increase (max) | nominal Water flow <Qw> | | nominal pressure drop <DP> | |
|---|---|---|---|---|---|---|---|---|---|---|---|---|---|---|---|
| | | | | | | [kW] | [kW] | [kW] | [kW] | [kW] | [°C] | [mc/s] | [mc/h] | [Pa] | [bar] |
| | | | | | | | | 301 | | 701 | | | | | |
| Power Supply | 1 | BENDING MAGNET | BENDING MAGNET | C. T. Damping Ring | | 195,21 | 59,3837 | 59,38 | 8,469 | 135,83 | 6 | 0,0000338 | 1,2980 | 283250 | 2,83 |
| Power Supply | 1 | QUADRUPOLE MAGNET HG | GF1 | C. T. Damping Ring | | 15,17 | 4,6179 | 4,62 | 2,637 | 10,55 | 9 | 0,00007 | 0,25200 | 260700 | 2,61 |
| Power Supply | 1 | QUADRUPOLE MAGNET HG | GF2 | C. T. Damping Ring | | 15,17 | 4,6179 | 4,62 | 2,637 | 10,55 | 9 | 0,00007 | 0,25200 | 260700 | 2,61 |
| Power Supply | 1 | QUADRUPOLE MAGNET HG | QD1 | C. T. Damping Ring | | 15,17 | 4,6179 | 4,62 | 2,637 | 10,55 | 9 | 0,00007 | 0,25200 | 260700 | 2,61 |
| Power Supply | 1 | QUADRUPOLE MAGNET LG | QD2 | C. T. Damping Ring | | 10,27 | 3,1522 | 3,15 | 1,779 | 7,12 | 5 | 0,000085 | 0,30600 | 250000 | 2,5 |
| Power Supply | 1 | QUADRUPOLE MAGNET LG | QD3 | C. T. Damping Ring | | 30,81 | 9,4567 | 9,46 | 1,779 | 21,35 | 5 | 0,000085 | 0,30600 | 250000 | 2,5 |
| Power Supply | 1 | QUADRUPOLE MAGNET LG | QF3 | C. T. Damping Ring | | 10,27 | 3,1522 | 3,15 | 1,779 | 7,12 | 5 | 0,000085 | 0,30600 | 250000 | 2,5 |
| Power Supply | 1 | QUADRUPOLE MAGNET LG | QF4 | C. T. Damping Ring | | 30,81 | 9,4567 | 9,46 | 1,779 | 21,35 | 5 | 0,000085 | 0,30600 | 250000 | 2,5 |
| Power Supply | 1 | QUADRUPOLE MAGNET LG | QF5 | C. T. Damping Ring | | 15,40 | 4,7284 | 4,73 | 1,779 | 10,67 | 5 | 0,000085 | 0,30600 | 250000 | 2,5 |
| Power Supply | 2 | SEXTUPOLE MAGNET | SEXTUPOLE MAGNET | C. T. Damping Ring | | 3,82 | 0,5827 | 1,17 | 0,111 | 2,65 | 4 | 0,0000066 | 0,02376 | 278000 | 2,78 |
| Power Supply | 8 | BENDING MAGNET | BQDM | C. T. M.R. | | 70,06 | 2,6796 | 21,44 | 6,078 | 48,62 | 11 | 0,000132 | 0,47520 | 265900 | 2,66 |
| Power Supply | 4 | BENDING MAGNET | BQDMA | C. T. M.R. | | 30,90 | 2,2843 | 9,14 | 5,442 | 21,77 | 10 | 0,0001300 | 0,46800 | 253000,0 | 2,53 |
| Power Supply | 1 | BENDING MAGNET | BStP | C. T. M.R. | | 21,31 | 1,9444 | 6,58 | 3,884 | 14,73 | 8 | 0,0001100 | 0,39600 | 269750,0 | 2,7 |
| Power Supply | 12 | BENDING MAGNET | BARC | C. T. M.R. | | 111,65 | 2,5230 | 30,28 | 6,781 | 81,38 | 10,0 | 0,00009000 | 0,32400 | 303115,0 | 3,03 |
| Power Supply | 12 | BENDING MAGNET | B5/B4/B2 | C. T. M.R. | | 71,02 | 1,8164 | 21,80 | 4,102 | 49,23 | 10,0 | 0,00008000 | 0,35280 | 248120,0 | 2,48 |
| Power Supply | 4 | BENDING MAGNET | B3 | C. T. M.R. | | 16,23 | 1,2936 | 5,17 | 2,763 | 11,05 | 6 | 0,0001100 | 0,39600 | 313250,0 | 3,13 |
| Power Supply | 4 | BENDING MAGNET | B1 | C. T. M.R. | | 9,23 | 0,6340 | 2,54 | 1,674 | 6,70 | 4 | 0,0001000 | 0,36000 | 237850,0 | 2,38 |
| Power Supply | 2 | BENDING MAGNET | BStB | C. T. M.R. | | 0,01 | 0,0033 | 0,01 | 0,000 | 0,00 | 0,0 | 0,00000000 | 0,00000 | 0,0 | 0 |
| Power Supply | 101 | QADRUPOLE MAGNET LG | SRQM(?) | C. T. M.R. | | 231,29 | 0,6826 | 68,94 | 1,607 | 162,35 | 6,0 | 0,00004000 | 0,23040 | 219900,0 | 2,2 |
| Power Supply | 22 | QUADRUPOLE MAGNET HG | SRQM(?) | C. T. M.R. | | 72,16 | 1,0196 | 22,43 | 2,260 | 49,73 | 10,0 | 0,00005400 | 0,19440 | 241400,0 | 2,41 |
| Power Supply | 38 | SEXTUPOLE MAGNET | SRSM(?) | C. T. M.R. | | 25,84 | 0,2091 | 7,94 | 0,471 | 17,90 | 9,0 | 0,00001250 | 0,04500 | 265250,0 | 2,65 |

## 5.5    Cryogenics

The superconducting magnets of the accelerator require refrigeration for their operation. The detector magnet and the two Final Focus cryostats will be cooled at an operating temperature respectively of T = 4.5 K and T = 1.9 K by means of a liquid helium refrigeration plant, while the Siberian Snake solenoids will be individually cooled by cryocoolers (Pulse Tube refrigerators) at T ≃ 4 K. The cryogenic plant will basically consist of a screw compressor with purification system, a cold box including two turbine expanders, a distribution feed box, a set of transfer lines and a buffer volume. It will be provided with a control system to be remotely operated. A process flow diagram of the plant is reported in Figure 5.5.1 (proposal from Linde Kryotechnik AG). The total cooling power required by the plant users is: 50W of isothermal refrigeration at 4.5K, 40W of isothermal refrigeration at 1.9K, 550W at 60 K for the radiation shields, 0.56 g/s of liquefaction rate for the current leads. The plant will operate without $LN_2$ precooling.

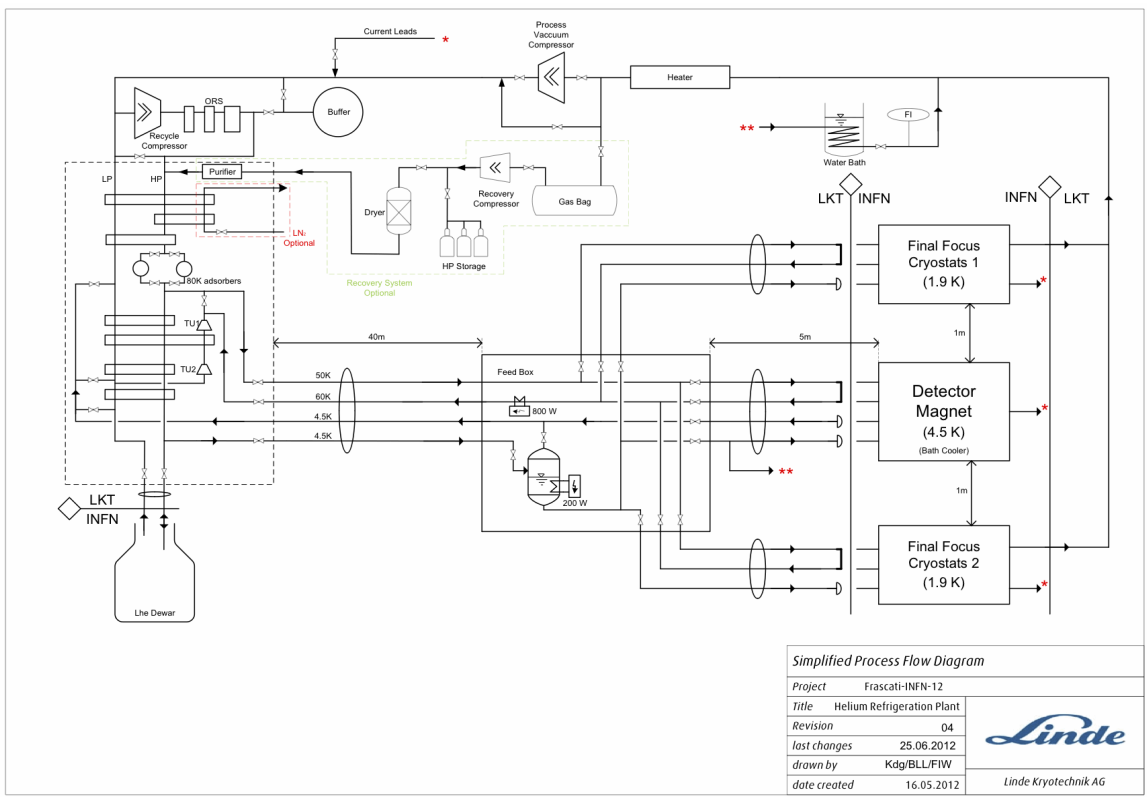

**Fig. 5.5.1 – Flow diagram of the LINDE cryogenics plant.**

## 5.6    Electrical engineering

### Load analysis

On the basis of the electrical loads, acquired or estimated during previous experiences, a preliminary design of the power supply system has been elaborated. The main electrical loads are detailed in Table 5.6.1. The expected power demand, at the moment is about 11,5 MW. The connection to the power grid, managed by the company Terna S.p.A., is at the voltage level of 150 kV.

Table 5.6.1 – Main electrical loads in the Tau/Charm complex

|  | Total Power (kW) |
|---|---|
| **MR (2 rings)** |  |
| RF HVPS | 2.320 |
| Power electronics to supply MR magnets | 2000 |
| **Linac** |  |
| Gun |  |
| RF | 2.250 |
| **DR** |  |
| Power electronics to supply DR magnets | 400 |
| RF cavity power supply |  |
| Transfer line |  |
| Detector Cryogenics | 350 |
| Detector PSU | 200 |
| FF Cryogenics + PSU | 188 |
| Computing |  |
| Civil building |  |
| Cooling and HVAC | 3.728 |
| FEL magnets and devices |  |
| Experimental facilities |  |
| **Total** | **11.436** |

At a first glance, the main components taken into account are:

- High Voltage station 150/20 kV, equipped with 2 High Voltage/Medium Voltage transformers having an apparent power of 20 MVA;
- General Medium Voltage Switch Board;
- Medium Voltage power distribution cables;
- Five Medium Voltage sub-stations 20/0,4 kV, each one equipped with 4 Medium Voltage/Low Voltage transformers having an apparent power of 1,6 MVA;
- Medium and low voltage power distribution cables;
- Two emergency electric generators, each one having an apparent power of 800 kVA;
- Uninterruptable Power Supply (UPS) to guarantee the supply of particular loads that need stability, high quality of voltage and continuity; the number and nominal power of UPS has to be evaluated;
- Tunnel lighting;
- Grounding and equipotential connections.

The MV substations have been designed using standardization criteria. Attention has been paid to specific problems related to voltage stability, harmonics pollution, electromagnetic compatibility, continuity of service, energy efficiency, operating costs and maintenance and the maintainability of the systems. The plants are designed to tolerate a significant number of fault conditions without compromising the functionality of the system. The design uses logic of modularity and is tailored to a first optimization in order to avoid unnecessary criticality or oversizing. Figure 5.6.1 shows the power distribution diagram. Based on experience in the management of DAΦNE accelerator plant, criteria against the radiation have been carefully evaluated for installations inside the accelerator, in order to avoid the use of materials and equipment sensitive to ionizing radiation and reducing the level of maintenance required.

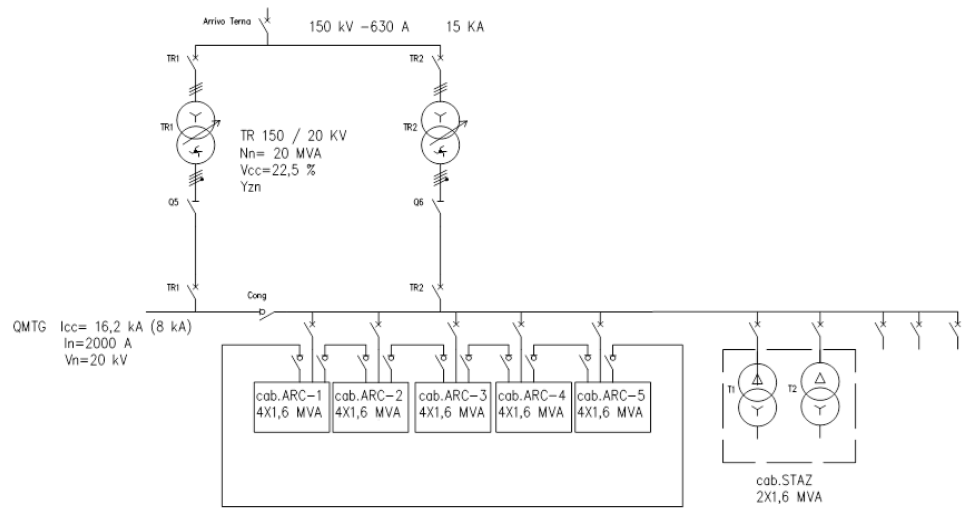

**Figure 5.6.1 – Power distribution diagram.**

## Main characteristics of tunnel and technical rooms lighting

Inside the tunnel, the choice is fluorescent lights with painted sheet-steel casing and electromechanical low-loss ballast (class B1), without capacitor. In all other environments, it is more convenient the use of electronic ballasts in A2 class and T5 lamps instead of T8. The power distribution in tunnels and in all large rooms is better using at least two bus-bars with 3 different circuits and dedicated plugs with fuse. To avoid the installation of components inside the area exposed to radiation, the emergency lighting power sources are UPS located outside of the area exposed to radiation; the distribution cables are fire-resistant FGT10OM1 for service continuity reasons. In the accelerator areas the equipment is Atex type. The lights chosen are 1 x 18 W to have a lighting level of at least 10 lux for 3 hours, to the floor, in all environments To check the respect of lighting quality needed (UNI EN 12464) for every activity in the different areas, there have been done simulations using specific software. In the simulations it has been considered a tunnel 6 m wide and 3,4 m high. The minimum values are:

- for the ways out, 150 lux;
- for working and assembling area, 300 lux;
- for emergency exit way, 10 lux.

The defined configuration is lighting sources using 2 lamps with 58 W nominal power (130 W total), installed every 2 metres of tunnel length, in two different positions (corridor and work area), so it means an average power of 65 W/m. The Figure 5.6.2 shows the result of a lighting simulation considering the work plan at a height of 1,2 m, while the Figure 5.6.3 shows the result of a lighting simulation considering the work plan at floor height.

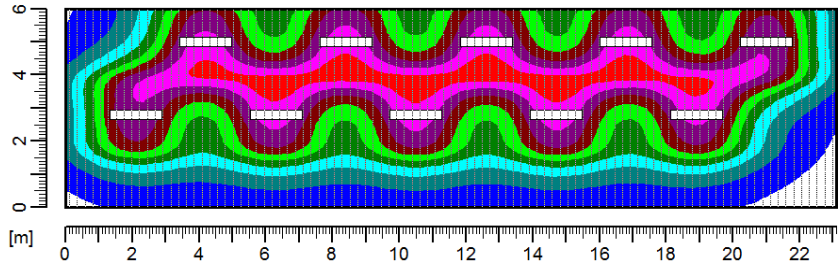

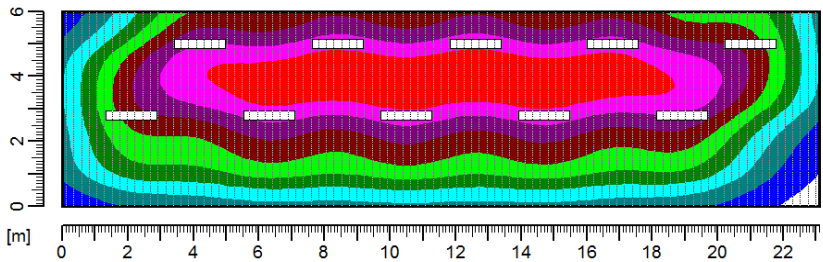

**Figure 5.6.2 - Lighting simulation considering the work plan at a height of 1.2 m.**

**Figure 5.6.3 - Lighting simulation considering the work plan at floor height.**

The Figure 5.6.4 shows the result of a lighting simulation considering emergency lighting at floor height.

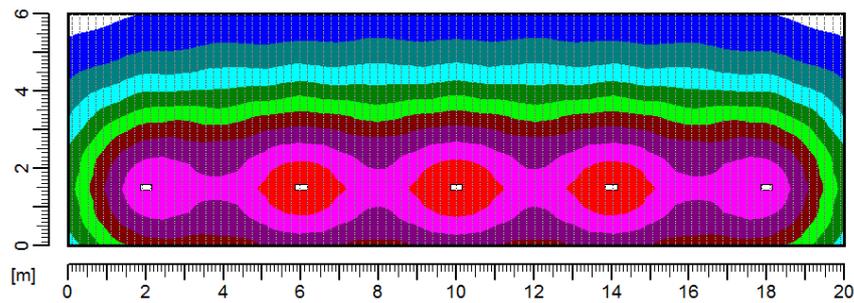

| Valori delle sezioni [lux] | | |
|---|---|---|
| 4,0 | 10,0 | 16,0 |
| 6,0 | 12,0 | 18,0 |
| 8,0 | 14,0 | 20,0 |

**Figure 5.6.4 - Lighting simulation considering emergency lighting at floor height.**

### Ground and equipotential network

The grounding system, in addition to its protective purposes in case of ground faults, performs functional requirements for accelerator devices. It is realized by copper cable that will be integrated with the armatures made sinks of foundations having connections at least every 20 metres. The equipotential bonding network of laboratory will be a strongly meshed network.

Inside the tunnel and the machinery space, there are connections from the ground network, every 20 metres, to allow direct access to it; such connections will be made from lengths of copper cable, 95 sqmm, 2 meters long. The task of these connections is the interconnection with the Common Bonding Network, the accelerator equipotential network. These grounding network access points will also be used for grounding of distribution boards.

The transformer cells equipotential bonding has to be installed inside the cabin, on the back of every transformer cell, in accessible position for measurements and have to be connected, using copper cables 95 sqmm, with:

- the ground main collector;
- the equipotential node;
- the grounding of the neutral conductor;
- the EQS and any metal frames and grids;
- the 20 kV cables socks.

## 5.7    Health, Safety and Environment (HSE)

### Introduction

The Consorzio Laboratorio Nicola Cabibbo is committed to the success of the mission objectives of the Tau-Charm Factory and to the safety of its users, staff, and the public. The Cabibbo-Lab Safety and Health and Environmental Manager will be responsible for ensuring that an HSE system is established, implemented, and maintained in accordance with requirements. The HSE Manager will provide oversight and support to the project participants to ensure a consistent Health Safety and Environment program, in this scenario safe working conditions and practices are an absolute requirement for all staff and contractors: in accordance with Italian Laws, European codes and International Technical Rules. We expect all design and work will be performed with this goal in mind.

For reaching this very high level target, we are already working on different aspects related to safety. One very important issue for the HSE Managers is to identify Authorities, Institutions responsible for the Authorizations process and Laws to be followed. As minimum requirements we expect our HSE plan to:

1. Contain a program that will protect the environment and the safety of workers and the general public by assuring that:
   a. Facilities, systems, and components needed to meet mission requirements are fully defined and are designed, constructed, and operated in accordance with applicable Italian and European laws and International Technical codes and requirements;
   b. Potential hazards to personnel associated with Tau-Charm facilities, structures, and components are identified and controlled through the timely preparation of safety assessment documents and with the help of dedicated Risk Analysis (i.e. Hazop, FMAE, ...);
   c. Potential risks to the environment are addressed through the timely and comprehensive preparation of appropriate documents (environmental risks assessment), the possibility to build the facility with a zero environmental impact must be investigated;
   d. International standards (i.e. ISO 14001 and OHSAS 18001) will be implemented to assure that all HSE risks are identified and addressed as well the responsibility and management roles.
2. Implement an effective construction safety program to ensure worker safety on the site during construction (see in Titolo IV of D.Lgs. 81/08) .
3. Provide appropriate training to ensure that project staff is adequately trained and qualified to perform their assigned work safely.

Policies and requirements to ensure implementation of these expectations will be established and communicated to all staff, contractors, and vendors.

### Preliminary and final hazard analysis (PHA - FHA)

One of the main components of our HSE program is to ensure that all hazards have been properly identified and controlled through design and related procedures. To ensure that these issues are understood at the preliminary design phase, a Final Hazard Analysis will be conducted to identify the hazards that will be encountered during the project construction and operational phases. This analysis is an update of the Preliminary Hazard Analysis that will be developed before, during the Conceptual Design Phase.

A Hazards List is going to be developed as the first step in identifying the potential hazards; it will also include preliminary (pre-mitigation) risk assessments that identified risk categories before incorporating the HSE related design and operational controls that are postulated to mitigate those risks. The identified hazards then will be further developed in the PHA, where the proposed HSE design enhancements will be taken into consideration. The FHA will analyze again the risks, including these enhancements and, in certain cases, operational controls, to establish a post- mitigation risk category.

### Fire protection & extinguishing system

In agreement with the Italian Law the Tau/Charm Factory is classified as High-Risk for the Fire Hazards point of view; for this category of activity (high-risk) it is necessary to be authorized by the competent Fire Brigade. The design for the high-risks facility must be realized with the so-called "Fire Safety Engineering (FSE)" method, this means that it is necessary to adopt a Fire Safety Management System.

Fire Safety targets for the reduction of fire hazards in the design phase of the facilities are:
- Restricting the probability of fire hazards
- Infrastructure must resist to fire for an assigned time
- Restricting Fire and Smoke propagation
- Restricting Fire propagation to the near structures and materials
- Easy accessibility of Emergency Exits – fast evacuation of personnel
- In case of emergency, Rescue Teams must work in a safe way

In the activity of Tau/Charm, in case it is not possible to fully respect the fire safety norms because of architectonic or structural hindrances, compensative safety measures will have to be applied, some of which are listed below:
- Protection of emergency exits and paths
- Realization of further emergency paths
- Emergency paths must be as short as possible
- Reduction of number of people
- Automatic detection of smoke and fire (see below)
- Reinforcement of the Fire systems (see below)
- Smoke discharge control
- Reinforcement of Emergency light
- Installation of further safety labeling
- Improvement of rescue teams
- Surveillance (see below)

In the present phase of the Conceptual Design particular emphasis has been given to the Fire Hazard (and site surveillance), in order to proceed with a preliminary design of the Fire Protection systems (and TVCC cameras system).

### Surveillance

The complex Tau/Charm Factory will be covered by a network of TVCC cameras. The cameras are speed-dome type and they will be controlled inside a control room equipped with video server, records units, security systems, and soft control software for the management of the whole laboratory

### Automatic detection of smoke and fire

Two main systems can be installed into the experimental halls of Main Ring, Damping Ring and Linac:

- one is an automatic detection system that will be devoted to analyze the quality of the air inside the experimental halls;
- the second is an Optical Fiber system that could be realized for controlling at an early stage a possible increase of temperature (before fire starts).

For technical facilities rooms an automatic detection system for smoke and/or heat can be installed.

## Fire extinguishing systems

At the moment two systems are under study: high-pressure water mist system and inert gas system. A final decision will be taken once all the risk analysis has been completed. There is also the possibility to have a combination of the two systems according to the different use and scope of the facilities building and experimental halls.

## Inert gas fire extinguish system

The inert gas system works by saturating the air with inert gas, it can be realized thanks to pressurized bottles (up to 300 bar) filled with this "clean agent" (see for example Fig. 5.7.1).

Advantages:

- The gas can be easily displaced thru Air Management system, no effect for the machines and equipment;
- Gas discharge can happen also with personnel inside the rooms;
- The system does not need active power for working, the pressure in the bottles guarantee that it is working properly.

Disadvantages:

- The system is "one shot", once discharge ends there is not possibility to extinguish anymore;
- It is impossible to act the system only for a subarea of the protected facility if fire compartments are not installed;
- The rooms must be properly sealed;
- Maintenance is very expensive.

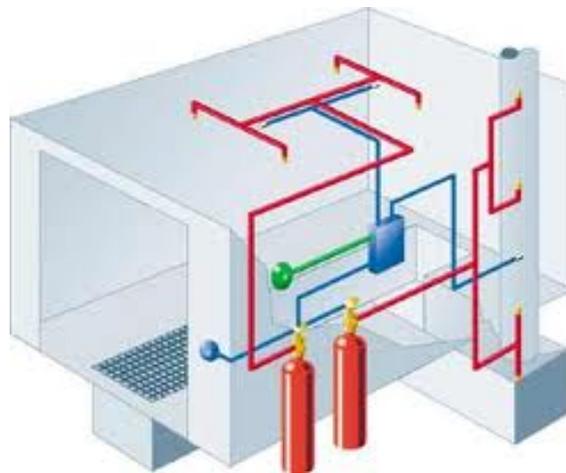

**Figure 5.7.1 - Artistic view of room protected with Inert Gas (in blue/green the detection system and in red the discharging pipes and nozzles).**

## Water mist system

The second system is a high-pressure water based system; the water is sprinkled in very small drops forming a fog. For this kind system the pumping units are very important, they are self-sustain units and the can be combined for reaching the requested flow rate (see Fig. 5.7.2). It is crucial for the availability of the system the use of back up pumping units and the reliability of the electrical network. In this preliminary stage we had already some contacts with Companies in order to have a quote and preliminary design of the system.

Advantages:

- The system can run until the fire is extinguished;
- The system can work also locally where fire is starting.

Disadvantage:

- The water discharged must be collected;
- The pumping unit must have a very high reliability;

For this system in under study the possibility to configure the nozzles in order to realize a water curtain to confine the smoke and the soot as it is going to be implemented in the European XFEL project - DESY. Several water mist fire extinguish systems are installed and successfully tested at the Gran Sasso Laboratory (LNGS) of INFN (see Fig. 5.7.3).

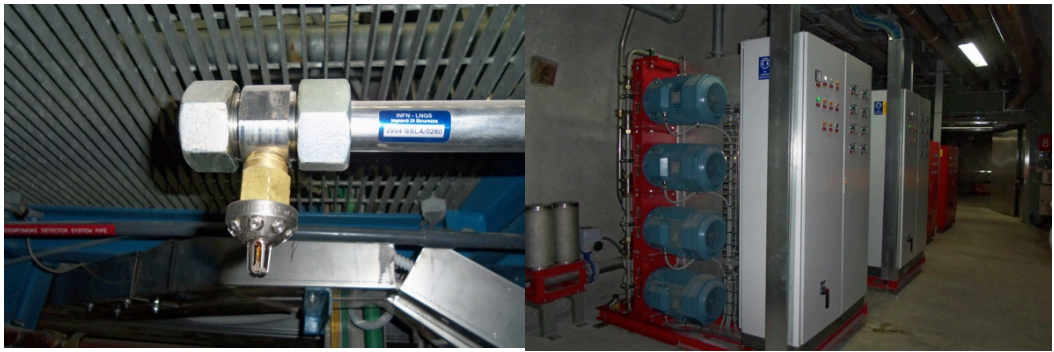

**Figure 5.7.2 - Detail of water mist nozzle and pumping units @ INFN–LNGS.**

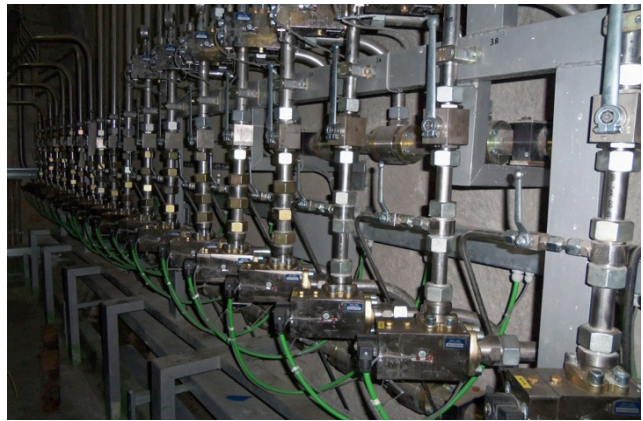

**Figure 5.7.3 - Detail of Inox piping and valves of water mist extinguish system @ INFN-LNGS.**

### Fire hydrants network

Another plant under study is the Fire hydrants network for the whole facility. In our mind, this plant will cover all the area where the facility will be built. The construction of the plant will include also the water reserve (underground pit) and the pumps unit.

## 6    Costs and Schedule

The cost of a Tau/Charm accelerator complex has been scaled down by the cost of the SuperB project, that was based on:

- SuperB Progress Report Accelerator [arXiv: 1009.6178 – Lattice version V 12];
- BINP preliminary quotations for Linac, Damping Ring and some Storage Ring magnets and girders;
- quotations from more than 30 firms from all around the world;
- experience from the most important world particle accelerators;
- more than 1000 pages of documents collected. All the documentation is available on the web repository of the Nicola Cabibbo Laboratory.

As a consequence, a WBS (Work Breakdown Structure) of about 1400 entries was prepared were the following systems were analyzed in detail:

- Normal conducting magnets;
- Power supplies;
- Vacuum;
- Mechanics;
- Electrotechnics;
- HVAC and fluids;
- Cryogenics;
- Civil engineering;
- Safety;
- Radioprotection;
- Superconducting magnets and solenoids;
- Final Focus/Interaction Region;
- Injection system;
- Instrumentation;
- Feedbacks;
- Controls;
- Radio Frequency.

Starting from this basis, different alternatives were considered in the transition from SuperB to the Tau/Charm complex. More in detail:

1. A linac for about 1.6 GeV $e^-$ and 2.5 GeV $e^+$, having a length around 280 m, and two asymmetric storage rings having a length of about 1200 m, based on the SuperB design but where some magnets were missing to allow the Tau/Charm physics.
2. As in 1., but with dedicated Tau/Charm magnetic structure and rings.
3. A linac for about 2.9 GeV $e^-$ and 2.3 GeV $e^+$, having a length around 200 m, and two symmetric storage rings having a length of about 600 m.
4. As in 3. but with storage rings having about 362 m length.
5. As in 3. but with storage rings having about 326 m length.

Some systems were left unchanged as the low energy part of the injector (gun, positron converter, low power accelerating structures), the damping ring, some system like beam diagnostics, feedbacks, controls, etc., but many others were scaled down according to the following laws and rules:

- SuperB/Tau-Charm Energy ratio (e.g. the high energy section of the linac);
- quantity unchanged but cost reduced (e.g. the storage ring multipole magnets in 1);
- building length ratio (mainly for linac and storage rings tunnels);
- concrete shielding thickness reduction;
- new component costs (e.g. the storage ring dipoles).

Even if the final lattice will be subject to some refinements and a true cost evaluation will be done in a successive phase when all the complex components will be correctly evaluated in quantity and cost and an updated, dedicated WBS will be produced, one can have a projection of the cost of a dedicated Tau/Charm accelerator on the base of the said considerations. The Table 6.1 summarizes the cost of the 26 systems taken into consideration.

**Table 6.1 - Summary of the costs of the various accelerator complex systems**

| COST EVALUATION SUMMARY (VAT Excluded) | VAT (21%) 100% (21%) TOTAL | | |
|---|---|---|---|
| | k€ | k€ | k€ |
| LINAC SYSTEM | 29614,54 | 6219,05 | 35833,59 |
| LINAC - DAMPING RING TRANSFER LINE | 4285,40 | 899,93 | 5185,34 |
| DAMPING RING | 12150,00 | 2551,50 | 14701,50 |
| ELECTRON BEAM TRANSFER LINE | 4428,17 | 929,92 | 5358,09 |
| POSITRON BEAM TRANSFER LINE | 4428,17 | 929,92 | 5358,09 |
| STORAGE RINGS | 58756,23 | 12338,81 | 71095,04 |
| POLARIZATION | 1991,00 | 418,11 | 2409,11 |
| INTERACTION REGION | 8187,06 | 1719,28 | 9906,34 |
| SYNCHROTRON LIGHT SOURCES | 0,00 | 0,00 | 0,00 |
| PHOTON LINES | 0,00 | 0,00 | 0,00 |
| GENERAL FACILITIES | 4816,42 | 1011,45 | 5827,86 |
| ELECTRIC SERVICES | 4992,19 | 1048,36 | 6040,55 |
| CRYOGENICS | 4018,00 | 843,78 | 4861,78 |
| CIVIL ENGINEERING | 35551,88 | 3555,19 | 39107,07 |
| ARCHEOLOGICAL DIGGING AND VERIFICATION | 2000,00 | 420,00 | 2420,00 |
| GEOLOGICAL PROSPECTION | 89,22 | 18,74 | 107,96 |
| GAS PIPELINE CONNECTION | 200,00 | 42,00 | 242,00 |
| WATER DUCT CONNECTION | 200,00 | 42,00 | 242,00 |
| ELECTRIC DISTRIBUTOR CONNECTION | 10200,00 | 2142,00 | 12342,00 |
| FIRE DETECTION SYSTEM | 227,69 | 47,81 | 275,50 |
| FIRE EXTINGUISHING | 736,86 | 154,74 | 891,61 |
| CRANE & LIFTING SYSTEMS | 995,32 | 209,02 | 1204,34 |
| RADIATION PROTECTION | 1083,35 | 227,50 | 1310,85 |
| CONVENTIONAL SAFETY SYSTEM | 252,00 | 52,92 | 304,92 |
| PRELIMINARY EXTERNAL AREA MAKE-UP | 3559,37 | 747,47 | 4306,84 |
| FINAL EXTERNAL AREA MAKE-UP | 1000,00 | 210,00 | 1210,00 |
| TAU-CHARM COMPLEX COST | 193762,88 | 36779,50 | 230542,38 |

The costs reported on Table 6.1 refer to the "bare" cost of the accelerator. For completeness, some other costs should be considered to have the total, all comprehensive cost of the project. Among these, should be added the cost for personnel, the cost for the hardware components integration, the contingency on civil works, the contingency for the various components and the spare parts cost, etc. These cost are strongly dependent from how the realization will proceed. Just as an example, if the Cabibbo Lab will be transformed in an ERIC Consortium, the VAT will not be paid, with a cost reduction of more than 43 M€.

Figure 6.1 shows how the cost is distributed among the different systems. Obviously, the storage ring cost is the highest one, followed by the civil engineering cost.

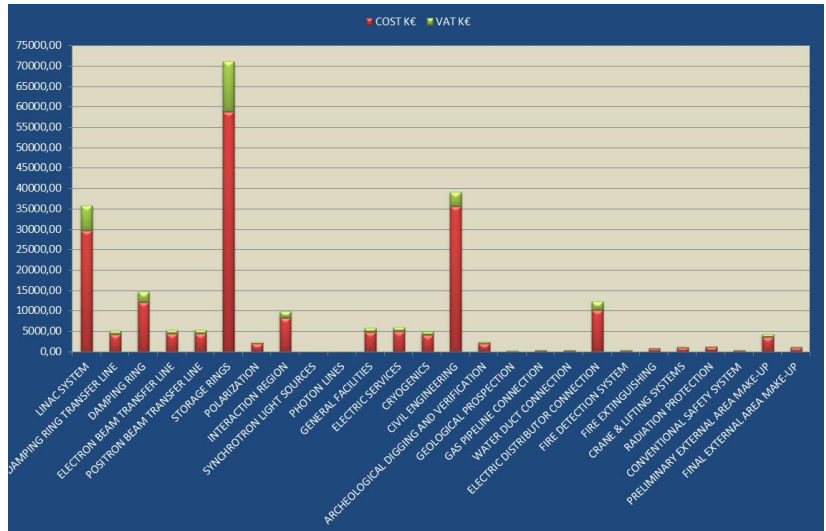

**Figure 6.1 - Cost distribution for the various systems.**

In the hypothesis that the project be approved before the end of the 2013, Figure 6.2 shows the expected spending profile, assuming that the construction of the accelerator complex can be done in six years, as shown in Figure 6.3.

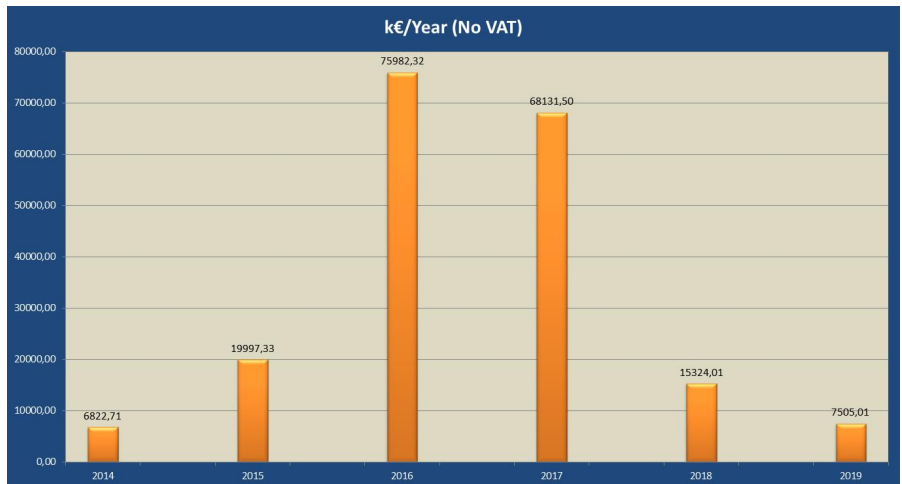

**Figure 6.2 - The spending profile figure.**

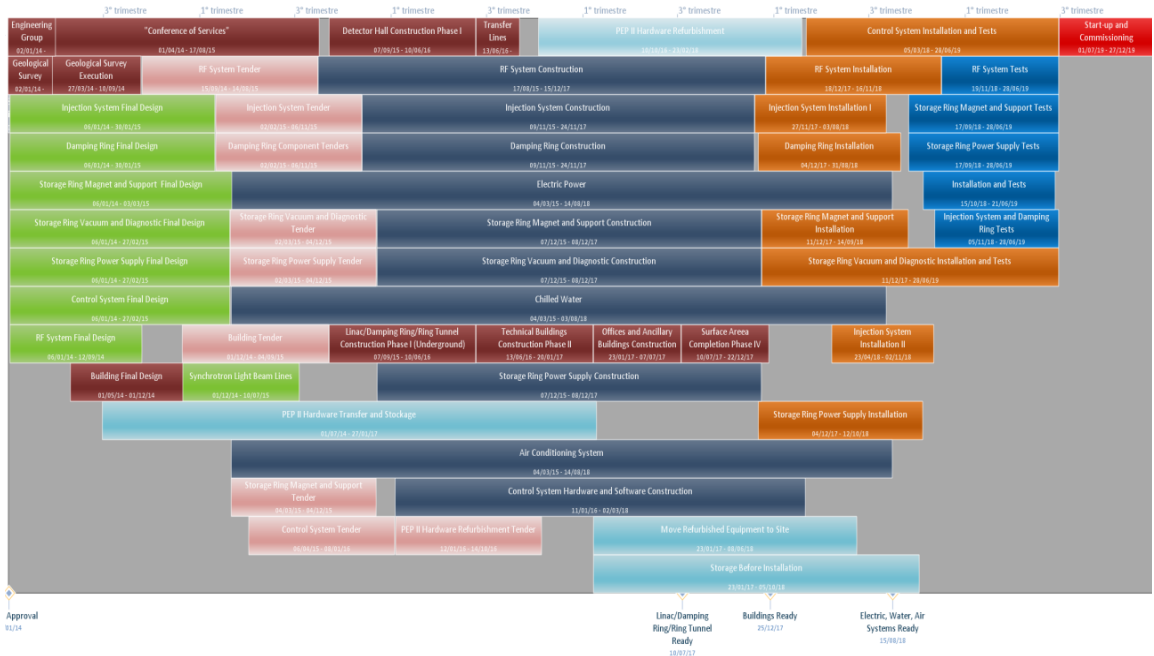

**Figure 6.3 - The crono-planning foreseen for the Tau/charm realization.**

The crono-planning shown in Figure 6.3 is also based on another assumption: the available personnel. To maintain the construction schedule is necessary that the needed personnel be available since the beginning assuming a strong collaboration with Italian and foreign Institutions in addition to a consistent number of hirings by the Cabibbo Lab. Figure 6.4 shows the number of people needed during the construction period per each year, grouped in macro-systems.

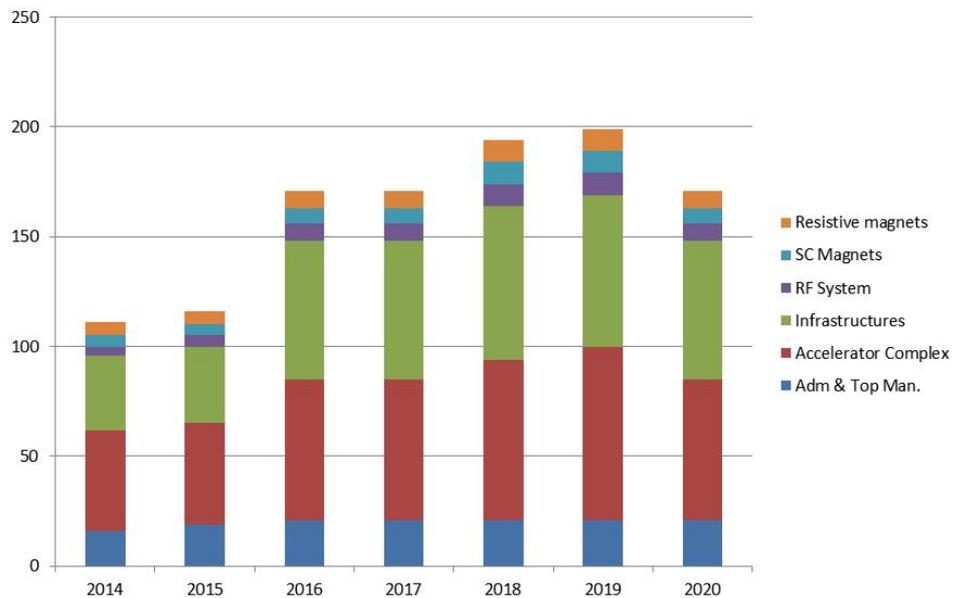

**Figure 6.4 - Personnel needed during the Tau/Charm realization.**

All the projections on the costs made for the SuperB and the various possible Tau/Charm configurations, listed at the beginning, can be plotted and Figure 6.5 shows the results of these considerations. As can be seen, the bare cost is nearly linear (taking away the first point on the left that refers to SuperB) and a rough rule in the range between 300 and 1200 m can be obtained as function of the storage ring length.

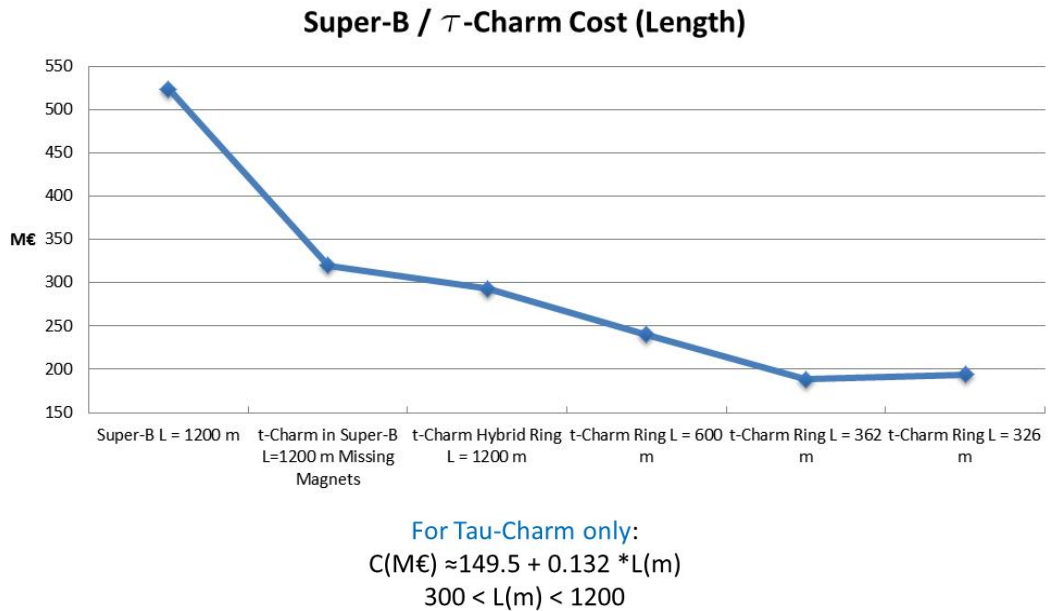

**Super-B / $\tau$-Charm Cost (Length)**

For Tau-Charm only:
C(M€) ≈149.5 + 0.132 *L(m)
300 < L(m) < 1200

**Figure 6.5 - The cost (M€) of the different Tau/Charm options, compared to SuperB.**

# 7    Tau/Charm as a SASE-FEL Facility

After the successful demonstration of exponential gain in a Self Amplified Spontaneous Emission (SASE) Free Electron Laser (FEL) and the operation up to saturation at FLASH (5 nm) and LCLS (1 Angstrom), a number of short wavelength SASE-FEL projects have been funded or proposed world wide [7.1], oriented as user facilities. Free electron lasers are poised in fact to take center stage as the premier source of tunable, intense, monocromatic photons of either ultra-short time resolution or ultra-fine spectral resolution. The choice of FEL radiation wavelength ranges from infrared down to hard X-ray, and the adopted linac technology is based on normal conducting (S-band or C-band) or superconducting accelerating structures (L-band). In this context the possibility to drive a SASE X-ray FEL using the Tau/Charm Linacs can be considered, as was done for the 6 GeV electron linac of the SuperB project [7.2]. We refer in the following to the work done for SuperB and published in [7.3].

The Tau/Charm injection system layout is shown in the Figure 3.1.1 (Part 1). The Linac sections L1, L2 and L3 are based on S-band (f = 2.856 GHz) structures equipped with the SLED system. The total length of the 3 Linacs is about 220 m. To achieve an energy of 6 GeV to obtain the photon wavelegth (between 1.5 and 3 Angstrom) proposed for the SuperB-FEL project [7.3], additional Linac sections can be installed at the end of L3. We make here the hypothesis to use the C-band (f = 5712 MHz) technology, which is being developed at LNF in the framework of the EU-TIARA project, and will be soon mounted at SPARCLAB. Assuming an accelerating gradient of 40 MV/m, additional 80 m of Linac sections (about 40) should be added.

The injection repetition cycle is 40 ms for each beam, corresponding to 25 Hz. Operating the linacs at a repetition frequency of 100 Hz, the timing scheme allows to accelerate two beam pulses for a SASE FEL facility, during the store time of the positrons in the DR, without affecting the injection rate for the Tau/Charm.

The beam for the SASE FEL would be produced by a dedicated high brightness photo-injector similar to that used at SPARC-LAB at LNF. A 50 Hz pulsed magnet will be used to combine the FEL beam with the Tau/Charm injection beams. The maximum linac energy for the electron beam is 2.9 GeV, a long space is available for the FEL extension: Linac extension, transfer lines, undulators and experimental halls.

The FEL injection system (S-band, 2.856 GHz) is composed by one 1.6 cell RF photo-injector followed by 2 TW structures embedded in a solenoid magnetic field as required to operate in the Velocity Bunching mode. It is a copy of the SPARC-LAB photo-injector, 8 m long.

The linac can be operated for the FEL in single or multi-bunch mode with a pulse length lower than 800 ns, to be compatible with SLED system, and with a repetition rate of 50 Hz. The charge per bunch can be chosen to better match the emittance and peak current requirements for the FEL operation.

After the photo-injector the beam is accelerated up to 2.9 GeV in Linac L1, L2 and L3. Two pulsed magnets are needed to separate the FEL bunches from the Tau/Charm bunches in the region of the positron converter and other two can be used in the region of Damping Ring injection and extraction, between linac L2 and L3. In this regions two magnetic bunch compressor systems can be installed, suitably designed to increase the peak current.

A layout of the Tau/Charm complex with the FEL facility is shown in Figure 7.1.

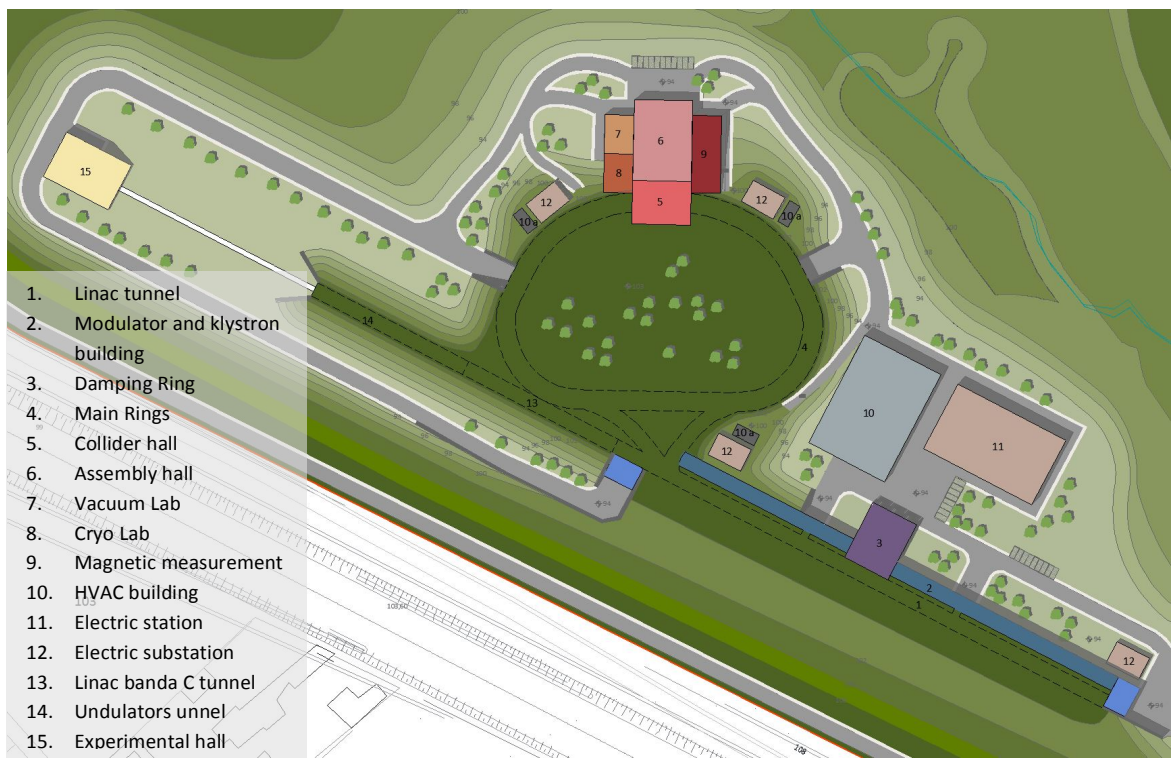

1. Linac tunnel
2. Modulator and klystron building
3. Damping Ring
4. Main Rings
5. Collider hall
6. Assembly hall
7. Vacuum Lab
8. Cryo Lab
9. Magnetic measurement
10. HVAC building
11. Electric station
12. Electric substation
13. Linac banda C tunnel
14. Undulators unnel
15. Experimental hall

**Figure 7.1 - Tau/Charm complex with the SASE-FEL option.**

To estimate the photons wavelength we consider an electron beam that traverses an undulator, emitting electromagnetic radiation at the resonant wavelength:

$$\lambda_r = \frac{\lambda_u}{2\gamma^2}\left(1 = a_u^2\right)$$

(7.1)

where $\lambda_u$ is the undulator period, $\gamma$ the beam relativistic factor and $a_u = K/\sqrt{2}$ for a planar undulator with undulator parameter given by $K = 0.934\ \lambda_u[cm]\ B[T]$, being $B$ the peak magnetic field. Figure 7.2 shows the achievable resonant wavelength versus $\lambda_u$ and K, assuming a 6 GeV electron beam energy ($\gamma = 11743$).

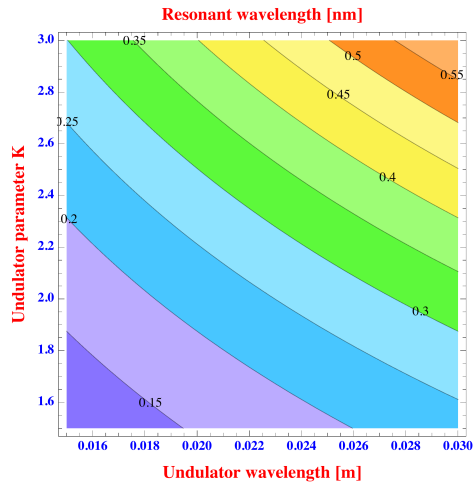

**Figure 7.2 – Resonant wavelength $\lambda_r$ versus undulator period $\lambda_u$ and parameter K as predicted by eq. (7.1).**

With this layout it is possible to obtain the same performances at 6 GeV as described in [7.3]. Using SPARC [7.4 to 7.7] like planar undulators with K = 2 and $\lambda_u$ = 2.8 cm [7.8] one can expect an output radiation wavelength of 3 Angstrom. Shorter wavelengths down to 1.8 Angstrom can be achieved by using an undulator with shorter period as the one foreseen for the future SPARC experiments with $\lambda_u$ = 1.8 cm and K = 2 (see Table 7.1 below). In Figure 7.3 the saturation length versus beam emittance and peak current as predicted in [7.3] are shown.

**Table 7.1 – Possible undulators parameters**

|  | Units | SPARC-like | Short period |
|---|---|---|---|
| Period $\lambda_u$ | cm | 2.8 | 1.8 |
| $a_u$ (=K/√2) |  | 1.51 | 1.2 |
| Section length | m | 3.36 | 2.16 |
| Gap length | m | 0.42 | 0.27 |
| $\lambda_r$ | Å | 3.16 | 1.525 |

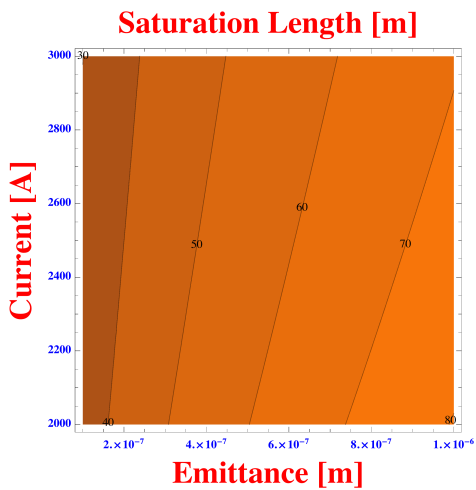

**Figure 7.3 – Saturation length versus beam emittance and peak current as predicted in [7.3].**

## Conclusions

Preliminary considerations for a possible X-ray FEL source making use of the Tau/Charm electron linacs have been presented. A systematic study on the design and beam dynamics has to be done.

The cost of a 6 GeV SASE-FEL facility addition to the Tau/Charm complex, VAT excluded, is listed in Table 7.2.

**Table 7.2 - Summary of the costs of a SASE-FEL facility at 6 GeV (keuro)**

|  | (kE uro) |
|---|---|
| LASER SYSTEM | 1500 |
| S-BAND INJECTOR | 3500 |
| C-BAND LINAC | 54000 |
| MAGNETIC COMPRESSOR 1 | 500 |
| MAGNETIC COMPRESSOR 2 | 1500 |
| X-BAND SYSTEM | 2000 |
| DIAGNOSTIC UPGRADE | 2500 |
| SYNCHRONIZATION UPGRADE | 1000 |
| UNDULATOR CHAIN | 20000 |
| ADDITIONAL BUILDINGS | 14000 |
| X-FEL | 100500 |

More advanced FEL scheme than SASE could be also considered, for example: "Self-seeding" [7.9], "High-gain harmonic generation", "Seeded harmonic cascade" or the possibility to produce ultra-short pulses at the attosecond level. To extend the user opportunity, the electron beam could be also extracted at Linac L3 energy (~3 GeV) and used to produce radiation in the soft X-ray range with a rich user program, as in the case of the SPARX project [7.10]. Such a high brightness Linac might drive several other applications, such as powerful THz radiation sources and also advanced accelerators concepts, like the Plasma acceleration or Dielectric wake field acceleration, to increase the final beam energy.

Among the possible new applications of a X-ray facility at 6 GeV are:

- the protein bio-imaging with high resolution, a new field useful for both scientific and pharmaceutic uses: proteins structure (like proteins in the cell membrane) cannot be determined with storage ring based synchrotron light sources, since crystallization is impossible there and for the radiation damage induced;

- the "Matter under Extreme Conditions" field, a field very interesting for astrophysics applications. With a high power laser to compress the material at high temperature and pression and using the FEL beam as a probe, it is possible to study directly the status of the matter in these conditions, very similar to the stars body;

- the study of the matter, liquid, solid or gaseous, in the "far from the equilibrium conditions" and the dynamics (return to equilibrium) of these processes, a new field with an enormous potential. This can be done by combining the wavelength with femtoseconds pulses and coherent X rays;

- the study of the materials structure and their behavior under stress, and the "Nano-science".

All the mentioned studies will require the development of new techniques for positioning the systems to study, as well as new detectors, to obtain 3D information and to measure in the femtoseconds time lapse. This new field is in fast development and will come to maturity in the very near future.

# 8    Tau/Charm as a Beam Test Facility

The for coming scenarios of High Energy Physics (HEP) of the next 10-20 years beyond LHC (multi TeV LC, CLIC, or High Energy LHC) and for new neutrinos large scale experiments and high luminosity flavor factories will more and more demanding for the particles detectors with very high performance and radiation hardness. This require strong efforts in the research and development of electronics and it's integration, data acquisition system and analysis, and to push the actual theology of particle detectors beyond the today border. In this scenario, Test Beam (TB) and irradiation user facility, to be installed at the end of the Tau/Charm Linac (see Figure 8.1), will be fundamental tools for the European HEP community, and the extracted beam of the TauCharm accelerators can play a fundamental role.

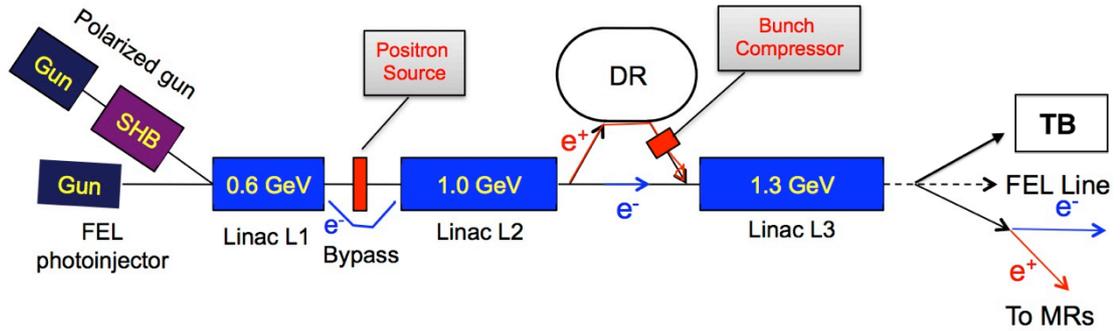

**Figure 8.1 – Tau/Charm Injection Complex.**

At the same time, photons and neutrons source obtained from the FEL line and the dumped beam in an optimized user area can host material science, biological and medical tests and application. A 50 Hz pulsed magnet will intercept the beam coming from the injection system in to the Test Beam building where 4 experimental halls and 2 control rooms can be hosted, see Figure 8.2 as an example.

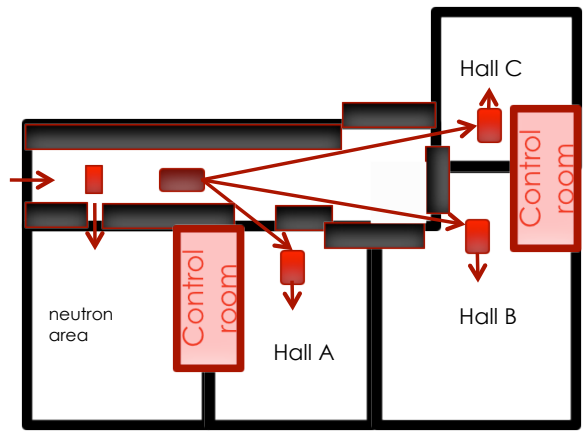

**Figure 8.2 – Beam Test Facility halls.**

The beam is dumped on an optimized target (W/Cu) producing secondary beam composed of hadrons, among which neutrons, electrons, positrons and photons (see Figure 8.3 for a typical production ratio on tungsten).

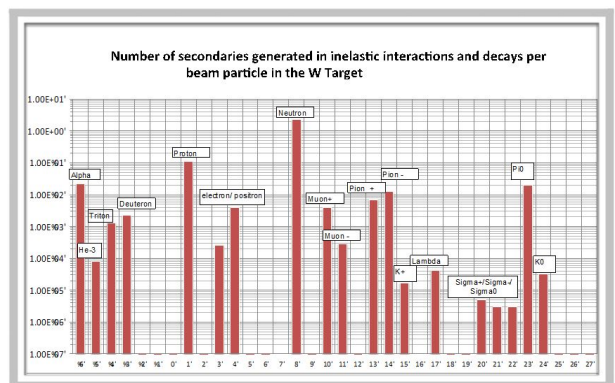

**Figure 8.3 – Number of secondaries generated on a W target.**

A hall at 90 degree with respect to the beam dumper can host an area dedicate to neutrons irradiation and tests while an electromagnetic powered magnet will select different out coming particles selected by their momentum, available in up to three different experimental hall.

The facility can be optimized for the transport of single particle per injection system spill mainly for detectors calibration purpose. All the high intensity injection system beam could be also available for high intensity electron irradiation study or accelerator diagnostics device tests.